\documentclass[12pt,a4paper,dvips]{article}
\usepackage{graphicx,epsfig}
\usepackage{hhline}

\usepackage{amsmath,amssymb}
\usepackage{times}
\usepackage[varg]{txfonts}
\DeclareMathAlphabet{\mathbold}{OML}{txr}{b}{it}

\usepackage{array,multirow,dcolumn}
\usepackage[mathlines,displaymath]{lineno}
\usepackage{rotating}

\usepackage[numbers,square,comma,sort&compress]{natbib}
\usepackage{hypernat}
\usepackage{textcomp}
\newcommand{\tablecaption}{%
\caption}

\newcolumntype{.}{D{.}{.}{-1}}
\newcolumntype{-}{D{-}{-}{-1}}

\usepackage[dvipsnames]{color}
\definecolor{rltred}{rgb}{0.75,0,0}
\definecolor{rltgreen}{rgb}{0,0.5,0}
\definecolor{rltblue}{rgb}{0,0,0.5}

\newcounter{pdfadd}    
\usepackage[hyperindex,bookmarks,bookmarksnumbered,breaklinks,a4paper,unicode]{hyperref}
\hypersetup{%
  pdftitle        = {Measurement of the inclusive ep Scattering Cross Section
 at low Q2 and x at HERA},
  urlcolor        = rltblue,       
  urlbordercolor  = 0 0 0.5,
  filecolor       = rltblue,       
  filebordercolor = 0 0 0.5,
  linkcolor       = rltred,        
  linkbordercolor = 0.75 0 0,
  citecolor       = rltgreen,      
  citebordercolor = 0 0.5 0,
  pagecolor       = rltgreen,      
  pagebordercolor = 0 0.5 0,
  menucolor       = rltgreen,      
  menubordercolor = 0 0.5 0,
  colorlinks    = true,
  pdfauthor     = {H1 Collaboration},
  pdfsubject    = { },
  pdfkeywords   = {High-Energy Physics, Particle Physics, Proton Structure, DIS}
}

\newlength{\dinwidth}
\newlength{\dinmargin}
\newcommand{\boldftwo}{\mbox{$\tilde{F}_2$}}
\newcommand{\boldxft}{\mbox{$x\tilde{F}_3$}}
\newcommand{\boldfl}{\mbox{$\tilde{F}_L$}}
\newcommand{\boldxftth}{\mbox{$x\tilde{F}^{\rm th}_3$}}
\newcommand{\boldflth}{\mbox{$\tilde{F}^{\rm th}_L$}}

\newcommand{\ncred}{\mbox{$ \sigma_{r,{\rm NC}}^{\pm}$}}
\newcommand{\ncredth}{\mbox{$ \sigma_{r,{\rm NC}}^{{\rm th,} \pm}$}}
\newcommand{\ncredne}{\mbox{$ \sigma_{r,{\rm NC~920}}^{\pm}$}}
\newcommand{\ncredre}{\mbox{$ \sigma_{r,{\rm NC~820}}^{\pm}$}}
\newcommand{\ncdd}{\mbox{$\frac{\textstyle {\rm d^2} \sigma^{e^{\pm}p}_{{\rm NC}}}{\textstyle {\rm d}x{\rm d} Q^2}$}}

\newcommand{\ccdd}{\mbox{$\frac{\textstyle {\rm d^2} \sigma^{e^{\pm}p}_{{\rm CC}}}{\textstyle {\rm d}x{\rm d} Q^2}$}}
\newcommand{\ccddne}{\mbox{$\frac{\textstyle {\rm d^2} \sigma^{e^{\pm}p}_{{\rm CC~920}}}{\textstyle {\rm d}x{\rm d} Q^2}$}}
\newcommand{\ccddre}{\mbox{$\frac{\textstyle {\rm d^2} \sigma^{e^{\pm}p}_{{\rm CC~820}}}{\textstyle {\rm d}x{\rm d} Q^2}$}}
\newcommand{\ccddneth}{\mbox{$\frac{\textstyle {\rm d^2} \sigma^{{\rm th,}~e^{\pm}p}_{{\rm CC~920}}}{\textstyle {\rm d}x{\rm d} Q^2}$}}
\newcommand{\ccddreth}{\mbox{$\frac{\textstyle {\rm d^2} \sigma^{{\rm th,}~e^{\pm}p}_{{\rm CC~820}}}{\textstyle {\rm d}x{\rm d} Q^2}$}}

\newcommand{\ccred}{\mbox{$ \sigma_{r,{\rm CC}}^{\pm}$}}
\newcommand{\ccredp}{\mbox{$ \sigma_{r,{\rm CC}}^{+}$}}
\newcommand{\ccredm}{\mbox{$ \sigma_{r,{\rm CC}}^{-}$}}

\newcommand{\Fig}{\mbox{figure}}
\newcommand{\Tab}{\mbox{table}}
\newcommand{\Eq}{\mbox{equation}}
\newcommand{\Sec}{\mbox{section}}

\newcommand{\FFig}{\mbox{Figure}}

\newcommand{\EEq}{\mbox{Equation}}

\newcommand{\Figs}{\mbox{figures}}
\newcommand{\Tabs}{\mbox{tables}}
\newcommand{\Eqs}{\mbox{equations}}

\newcommand{\FFigs}{\mbox{Figures}}

\newcommand{\asmz}{\alpha_s(M_Z^2)}
\newcommand{\msbar}{\mbox{$\overline{\rm{MS}}$}\ }

\newcommand{\bs}{\overline{s}}
\newcommand{\bc}{\overline{c}}
\newcommand{\bu}{\overline{u}}
\newcommand{\bd}{\overline{d}}

\newcommand{\bU}{\overline{U}}
\newcommand{\bD}{\overline{D}}

\newcommand{\pb}{\,pb$^{-1}$}

\begin{document}

\makeatletter \def\NAT@space{} \makeatother

\begin{titlepage}
 
\noindent
DESY 09-158 \hfill ISSN 0418-9833 \\
January 2010 \\

\vspace*{4.5cm}

\begin{center}
\begin{Large}

{\bfseries Combined Measurement and QCD Analysis of the Inclusive $\mathbold{e^{\pm}p}$ 
Scattering
 Cross Sections  at HERA}

\vspace*{2cm}

H1 and ZEUS Collaborations

\end{Large}
\end{center}

\vspace*{2cm}

\begin{abstract} \noindent
A combination is presented of the  inclusive deep inelastic cross
sections measured by the H1 and ZEUS Collaborations 
in neutral and charged current 
unpolarised $e^{\pm}p$ scattering
at HERA during the period $1994$-$2000$. The data span six orders of magnitude in 
negative four-momentum-transfer squared, $Q^2$, and in Bjorken $x$.
The combination method used takes the correlations of
systematic uncertainties into account,
resulting in an improved accuracy. The combined
data are the sole input in a NLO QCD analysis which
determines a new set of parton distributions, HERAPDF1.0, with
small experimental uncertainties.
This set includes an estimate of the model and parametrisation uncertainties of the fit result. 
\end{abstract}

\vspace*{1.5cm}

\begin{center}
\end{center}

\end{titlepage}

\begin{flushleft}
\begin{center}                                                                                     
{                      \Large  The H1 and ZEUS Collaborations              }                               
\end{center}                                                                                       
F.D.~Aaron$^{13, a8}$,
H.~Abramowicz$^{72, a36}$,
I.~Abt$^{57}$,
L.~Adamczyk$^{19}$,
M.~Adamus$^{84}$,
M.~Aldaya~Martin$^{31}$,
C.~Alexa$^{13}$,
V.~Andreev$^{54}$,
S.~Antonelli$^{9}$,
P.~Antonioli$^{8}$
% ...

{\it
\vspace{0.4cm}

% ...
$^{8}$   {\it INFN Bologna, Bologna, Italy}~$^{b17}$ \\
 $^{9}$   {\it University and INFN Bologna, Bologna, Italy}~$^{b17}$ \\
% ...
$^{13}$  National Institute for Physics and Nuclear Engineering (NIPNE), Bucharest, Romania \\
% ...
$^{19}$   {\it Faculty of Physics and Applied Computer Science, AGH-University of Science and \mbox{Technology}, Cracow, Poland}~$^{b27}$ \\
% ...
$^{31}$   {\it Deutsches Elektronen-Synchrotron DESY, Hamburg, Germany} \\
% ...
$^{54}$  Lebedev Physical Institute, Moscow, Russia~$^{b6}$ \\
% ...
$^{57}$   {\it Max-Planck-Institut f\"ur Physik, M\"unchen, Germany} \\
% ...
$^{72}$   {\it Raymond and Beverly Sackler Faculty of Exact Sciences, School of Physics, Tel Aviv University, Tel Aviv, Israel}~$^{b16}$ \\
% ...
$^{84}$   {\it Institute for Nuclear Studies, Warsaw, Poland} \\
 $^{85}$  Fachbereich C, Universit\"at Wuppertal, Wuppertal, Germany \\
 $^{86}$  Yerevan Physics Institute, Yerevan, Armenia \\
 $^{87}$   {\it Meiji Gakuin University, Faculty of General Education, Yokohama, Japan}~$^{b18}$ \\
 $^{88}$   {\it Department of Physics, York University, Ontario, Canada M3J1P3}~$^{b14}$ \\
 $^{89}$   {\it Deutsches Elektronen-Synchrotron DESY, Zeuthen, Germany} \\
 $^{90}$  Institut f\"ur Teilchenphysik, ETH, Z\"urich, Switzerland~$^{b10}$ \\
 $^{91}$  Physik-Institut der Universit\"at Z\"urich, Z\"urich, Switzerland~$^{b10}$ \\

\vspace{0.4cm}
{ \small


$^{a1}$  Also at Physics Department, National Technical University, Zografou Campus, GR-15773 Athens, Greece \\
$^{a2}$  Also at Rechenzentrum, Universit\"at Wuppertal, Wuppertal, Germany \\
$^{a3}$  Also at University of P.J. \v{S}af\'arik, Ko\v{s}ice, Slovak Republic \\
$^{a4}$  Also at CERN, Geneva, Switzerland \\
$^{a5}$  Also at Max-Planck-Institut f\"ur Physik, M\"unchen, Germany \\
$^{a6}$  Also at Comenius University, Bratislava, Slovak Republic \\
$^{a7}$  Also at DESY and University Hamburg, Helmholtz Humboldt Research Award \\
$^{a8}$  Also at Faculty of Physics, University of Bucharest, Bucharest, Romania \\
% ...
$^{a36}$   Also at Max Planck Institute, Munich, Germany, Alexander von Humboldt Research Award\\
% ...
$^{b6}$  Supported by the Deutsche Forschungsgemeinschaft \\
% ...
$^{b16}$   Supported by the Israel Science Foundation\\
$^{b17}$   Supported by the Italian National Institute for Nuclear Physics (INFN) \\
% ...
$^{b27}$   Supported by the Polish Ministry of Science and Higher Education as a scientific project (2009-2010)\\
% ...

}
}

\end{flushleft}
 


\section{Introduction \label{sec:int}}
Deep inelastic scattering (DIS) at HERA has been central to the exploration
of proton structure and quark-gluon interaction dynamics as
prescribed in perturbative Quantum Chromodynamics (QCD). 
HERA allowed inelastic $ep$ 
interactions to be studied
at high centre-of-mass energy, $\sqrt{s} \simeq 320\,$GeV, where $s = 4 E_e E_p$,
the lepton beam energy $E_e \simeq 27.5$\,GeV and the proton beam
energy $E_p = 920$\,GeV for most of the running period.
Operation of HERA proceeded
in two phases, HERA I, from 1992-2000, and HERA II, from 2002-2007.
The luminosity collected by each of the collider
experiments, H1 and ZEUS, in unpolarised $e^+ p$ and $e^- p$ scattering
during the first phase was approximately $100$\pb and $15$\pb, respectively.

The  paper presents the combination of the published H1~\cite{Collaboration:2009bp,Collaboration:2009kv,Adloff:1999ah,Adloff:2000qj,Adloff:2003uh}   
and ZEUS~\cite{Breitweg:1997hz,Breitweg:2000yn,Breitweg:1998dz,Chekanov:2001qu,
zeuscc97,Chekanov:2002ej,Chekanov:2002zs,Chekanov:2003yv,Chekanov:2003vw}  measurements
from HERA I on inclusive DIS in neutral (NC) and charged current (CC)
reactions which cover a wide range of negative four-momentum-transfer
squared, $Q^2$, and Bjorken $x$. 
The combination is performed
using a method introduced 
in \cite{glazov} and  extended in \cite{Collaboration:2009bp}. This also
 provides a model-independent check of the data consistency. The correlated 
systematic uncertainties and global normalisations are fitted such that 
one coherent data set is obtained. Since H1 and ZEUS have employed different 
experimental techniques, using different detectors and methods of kinematic reconstruction, 
the combination leads to 
 a significantly reduced uncertainty. 

The combined data set contains complete information on inclusive DIS cross sections
published by the H1 and ZEUS collaborations based on data collected in the years
$1994$-$2000$. The kinematic range of the NC data
is $6 \cdot 10^{-7} \leq x \leq 0.65$ and 
$0.045 \leq Q^2 \leq 30000 $\,GeV$^2$,
for values of inelasticity $y$ between $0.005$ and $0.95$,
 where 
$y = Q^2/sx$. The kinematic range of the CC data is 
 $1.3 \cdot 10^{-2} \leq x \leq 0.40$ and $300 \leq Q^2 \leq 30000 $\,GeV$^2$,
for values of $y$ between $0.037$ and $0.76$.
An extensive overview of the results
on HERA I data from both H1 and ZEUS is given in \cite{klein-2008}.

Analyses of the $x$ and $Q^2$ dependences of the NC and CC DIS
cross sections measured at HERA have determined sets of quark and gluon 
momentum distributions in the proton, both from H1 \cite{Collaboration:2009kv}
and ZEUS \cite{Chekanov:2005nn}. In such analyses, the lower $Q^2$ NC data 
determine the low-$x$ sea quark and gluon distributions.
The high-$Q^2$ CC data, 
together with the difference between NC $e^+p$ and $e^-p$ cross sections 
at high $Q^2$, constrain the valence quark distributions.
The use of the HERA CC data allows
the down quark distribution  in the proton
to be determined without assuming
isospin symmetry.       
In addition, the use of  HERA data alone for the determination of parton distribution functions (PDFs)
eliminates the need for heavy target corrections, which must be applied to 
DIS data from nuclear targets.
In this
paper the combined HERA  data are used to determine a new set of parton 
distributions termed HERAPDF1.0. 
Consistency of the input data ensures that the experimental
uncertainty of the HERAPDF1.0 set 
can be determined using rigorous statistical methods.  
Uncertainties resulting from model assumptions
and from the choice of PDF parametrisation are also considered, 
similarly to~\cite{Collaboration:2009kv}. 

The paper is organised as follows. In \Sec\,\ref{sec:meas} the measurements
by H1 and ZEUS and the input data are described. In \Sec\,\ref{sec:comb}
the combination of the NC and CC data sets from H1 and ZEUS is
discussed. The QCD analysis is described in \Sec\,\ref{sec:fit}.
A  summary is given in \Sec\,\ref{sec:sum}.

\section{Measurements of Inclusive DIS Cross Sections \label{sec:meas}}
\subsection{Cross Sections and Parton Distributions}
\label{xsecns}
The neutral  current deep inelastic $e^{\pm}p$ scattering cross section, at tree level,
is given by a linear combination of generalised structure functions. For unpolarised beams
it can be expressed as
\begin{eqnarray} \label{ncsi}     
 \ncred =\ncdd \cdot \frac{Q^4 x}{2\pi \alpha^2 Y_+}                                                     
  =            \tilde{F_2} \mp \frac{Y_-}{Y_+} \tilde{xF_3} -\frac{y^2}{Y_+} \tilde{F_L}~,
\end{eqnarray}                                                                  
where the electromagnetic coupling constant $\alpha$, the photon              
propagator and a helicity factor are absorbed 
in the definition of the reduced cross section \ncred, and $Y_{\pm}=1 \pm (1-y)^2$.                                                                              
The functions $\boldftwo$, $\boldfl$  and $\boldxft$
depend on  the electroweak parameters as~\cite{kleinriemann}
\begin{eqnarray} \label{strf}                                                   
 \boldftwo &=& F_2 - \kappa_Z v_e  \cdot F_2^{\gamma Z} +                      
  \kappa_Z^2 (v_e^2 + a_e^2 ) \cdot F_2^Z~, \nonumber \\   
 \boldfl &=& F_L - \kappa_Z v_e  \cdot F_L^{\gamma Z} +                      
  \kappa_Z^2 (v_e^2 + a_e^2 ) \cdot F_L^Z~, \nonumber \\                     
 \boldxft &=&  \kappa_Z a_e  \cdot xF_3^{\gamma Z} -                     
  \kappa_Z^2 \cdot 2 v_e a_e  \cdot xF_3^Z~.                                   
\end{eqnarray} 
Here $v_e$ and $a_e$ are the vector and axial-vector weak couplings of 
the electron to the $Z$ boson 
and $\kappa_Z(Q^2) =   Q^2 /[(Q^2+M_Z^2)(4\sin^2 \theta_W \cos^2 \theta_W)]$. 
The effective values for the electroweak mixing angle,
 $\sin^2  \theta_W=0.2315$,  and the $Z$ boson mass, $M_Z=91.187$~GeV, are used~\cite{PDG}.
At low $Q^2$, the contribution of $Z$
 exchange is negligible and $\sigma_{r,{\rm NC}} = F_2  - y^2 F_L/Y_+$.
The contribution of the term containing the structure function $F_L$ is only significant for large values of $y$.

In the Quark Parton Model (QPM),  
$\boldfl=0$\,\cite{PhysRevLett.22.156} and the 
other functions in \Eq\,\ref{strf} are given as
\begin{eqnarray} \label{ncfu}                                                   
  (F_2, F_2^{\gamma Z}, F_2^Z) &=&  [(e_u^2, 2e_uv_u, v_u^2+a_u^2)(xU+ x\bar{U})
  +  (e_d^2, 2e_dv_d, v_d^2+a_d^2)(xD+ x\bar{D})]~,            
                                 \nonumber \\                                   
  (xF_3^{\gamma Z}, xF_3^Z) &=& 2  [(e_ua_u, v_ua_u) (xU-x\bar{U})
  +  (e_da_d, v_da_d) (xD-x\bar{D})]~,                        
\end{eqnarray} 
where  $e_u,e_d$ denote the electric charge of up- or
down-type quarks while $v_{u,d}$ and $a_{u,d}$ are 
the vector and axial-vector weak couplings of the up- or 
down-type quarks to the $Z$ boson.
Here   $xU$, $xD$, $x\bU$ and $x\bD$ denote
the sums of up-type, of down-type and of their 
anti-quark distributions, respectively. 
Below the $b$ quark mass threshold,
these sums are related to the quark distributions as follows
\begin{equation}  \label{ud}
  xU  = xu + xc\,,    ~~~~~~~~
 x\bU = x\bu + x\bc\,, ~~~~~~~~
  xD  = xd + xs\,,    ~~~~~~~~
 x\bD = x\bd + x\bs\,, 
\end{equation}
where $xs$ and $xc$ are the strange and charm quark distributions.
Assuming symmetry between sea quarks and anti-quarks, 
the valence quark distributions result from 
\begin{equation} \label{valq}
xu_v = xU -x\bU\,, ~~~~~~~~~~~~~ xd_v = xD -x\bD\,.
\end{equation}

Defining a 
reduced cross section for 
the inclusive unpolarised charged current $e^{\pm} p$ 
scattering  as
\begin{equation}
 \label{Rnc}
 \ccred =  
  \frac{2 \pi  x}{G_F^2}
 \left[ \frac {M_W^2+Q^2} {M_W^2} \right]^2
          \ccdd
\end{equation}
gives a sum of charged current structure functions, analogous to \Eq~\ref{ncsi}, as
\begin{eqnarray}
 \label{ccsi}
 \sigma_{r,CC}^{\pm}=
  \frac{Y_+}{2}W_2^\pm   \mp \frac{Y_-}{2} xW_3^\pm - \frac{y^2}{2} W_L^\pm \,.
\end{eqnarray}
For the Fermi constant, an effective value $G_F=1.16639\times 10^{-5} $~GeV$^{-2}$
is used, the $W$ boson mass is $M_W=80.41$~GeV \cite{PDG}.   In the QPM $W_L^\pm = 0$ 
and the CC structure functions represent sums and differences
of quark and anti-quark-type distributions depending on the 
charge of the lepton beam as 
\begin{eqnarray}
 \label{ccstf}
    W_2^{+}  =  x\bU+xD\,,\hspace{0.05cm} ~~~~~~~
  xW_3^{+}  =   xD-x\bU\,,\hspace{0.05cm}  ~~~~~~~ 
    W_2^{-}  =  xU+x\bD\,,\hspace{0.05cm} ~~~~~~~
 xW_3^{-}  =  xU-x\bD\,.
\end{eqnarray}
From these equations it follows that
\begin{equation}
\label{ccupdo}
 \ccredp = x\bU+ (1-y)^2xD\,, ~~~~~~~
 \ccredm = xU +(1-y)^2 x\bD\,. 
\end{equation}
Therefore the NC and CC measurements may be used
to determine the combined sea quark distribution functions, $x\bU$ and $x\bD$,
and the valence quark distributions, $xu_v$ and $xd_v$. 
A QCD analysis in the DGLAP 
formalism~\cite{Gribov:1972ri,Gribov:1972rt,Lipatov:1974qm,Dokshitzer:1977sg,Altarelli:1977zs}
also allows the 
gluon momentum distribution, $xg$, in the proton 
to be determined from scaling violations.
\subsection{Reconstruction of Kinematics}
\label{diskine}
The deep inelastic $ep$ scattering cross section
of the inclusive neutral  and charged 
current reactions
depends on the centre-of-mass energy $s$ and on two kinematic variables,
 $Q^2$ and $x$. Usually $x$ is obtained from the measurement
of the inelasticity $y$ and from $Q^2$ and $s$ through the relationship
 $x=Q^2/(sy)$. The 
salient feature of the  HERA collider experiments is the possibility
to determine the NC event kinematics from the scattered
electron\footnote{In this paper, the term electron is used for both electrons and positrons, unless otherwise stated.} $e$, or from the hadronic final state $h$, or using a combination of the
two.  The choice of the most appropriate kinematic reconstruction method
for a given phase space region is 
based on resolution, measurement accuracy and radiative correction effects and has
been optimised differently for the two experiments.
The use of different  reconstruction 
techniques by the two experiments
contributes to the improved accuracy of the combined data set.

For NC scattering,
in the ``electron method'', the inelasticity and the negative four-momentum-transfer squared 
can be calculated using the electron kinematics as
\begin{equation}
 y_e = 1-\frac{\Sigma_e}{2 E_e}\,,~~~~~~~~~~~~ 
Q^2_e = \frac{P_{T,e}^2} {1 - y_e}\,,~~~~~~~~~~~~x_e = \frac{Q^2_e}{s y_e}\,.
 \label{eq:emeth}
\end{equation}
Here $\Sigma_e = E'_e(1-\cos\theta_e)$,
where $\theta_e$ is the angle between the scattered electron direction
and the proton beam direction\footnote{%
In the right-handed H1 and ZEUS coordinate systems,
 the $z$ axis points along the proton beam direction,
termed the forward direction.  
The $x$ ($y$) axis is directed horizontally 
(vertically).
}, 
$E_e'$ is the scattered electron energy, 
and $P_{T,e}$ is its transverse momentum.
Similar relations are obtained from the hadronic final state
reconstruction~\cite{yjb}, and are used for CC scattering,
\begin{equation}
 y_h  = \frac{\Sigma_h}{2 E_e}\,,~~~~~~~~~~~~ 
Q^2_h = \frac{P_{T,h}^2} {1 - y_h}\,,~~~~~~~~~~~~x_h = \frac{Q^2_h}{s y_h}\,,
 \label{yjb}
\end{equation}
where
$\Sigma_h = \sum_i{(E_i-p_{z,i})}$
is the hadronic $E-P_z$ variable with the sum extending over  the reconstructed hadronic final state
particles $i$, and $P_{T,h} = \left| \sum_i \mathbold{p}_{\perp,i} \right|$
is the total transverse momentum of the hadronic final state
with $\mathbold{p}_{\perp,i}$ being the transverse momentum vector of the particle $i$.
The combination of $P_{T,h}$ and $\Sigma_h$ defines
the hadronic scattering angle
\begin{equation}\label{eq:thh}
\tan \frac{\theta_h}{2} = \frac{\Sigma_h}{P_{T,h}}
\end{equation}
which, within the QPM, corresponds to the direction of the struck quark.
In the ``sigma method''~\cite{ysigma} the total $E-P_z$ variable                                                 
\begin{equation} \label{eq:sigma}
 E-P_z = E'_e (1-\cos{\theta_e}) + 
 \sum_i \left(E_i - p_{z,i}\right) =  \Sigma_e + \Sigma_h
\end{equation}
is introduced. For non-radiative events this variable equals $2E_e$ such that
  \Eqs~\ref{eq:emeth} and \ref{yjb} can be modified as
\begin{equation}\label{eq:yh}
 y_{\Sigma}  = \frac{\Sigma_h} {E-P_z}\,, ~~~~~~~~~~~~~ Q^2_{\Sigma}=\frac{P^2_{T,e}}{1-y_{\Sigma}}\,,~~~~~~~~~~~~ x_{\Sigma} = \frac{Q^2_{\Sigma}}{s y_{\Sigma}}\,.
\end{equation} 
A hybrid ``e-sigma method'' \,\cite{ysigma,Adloff:1999ah,Breitweg:2000yn} 
uses $Q^2_e$ and $x_\Sigma$ to reconstruct the event kinematics
\begin{equation} \label{eq:esigma}
  y_{e\Sigma} =  \frac{Q^2_e}{s x_{\Sigma}} = \frac{2E_e}{E-P_z}\,y_{\Sigma}\,,~~~~~~~~~~~~~ Q^2_{e\Sigma} = Q^2_e\,,~~~~~~~~~~~~ x_{e\Sigma} = x_{\Sigma}\,.
\end{equation}
An extension of the  sigma method~\cite{Collaboration:2009bp,Collaboration:2009kv} is 
\begin{equation} \label{eq:sigma2}
 y_{\Sigma'}  = y_{\Sigma}, ~~~~~~~~~~~~~ Q^2_{\Sigma'}=Q^2_{\Sigma}\,,~~~~~~~~~~~~ 
x_{\Sigma'} = \frac{Q^2_{\Sigma}} {2 E_p (E-P_z) y_{\Sigma}} = \frac{Q_{\Sigma}^2} {2 E_p \Sigma_h}.
\end{equation} 
This modification takes
radiative effects at the lepton vertex into account by replacing
 the electron beam energy  in the calculation of $x$, in a similar manner 
to $y_{\Sigma}$. 

The  ``double angle method'' \cite{standa,hoegerda} is used  to  
reconstruct  $Q^2$ and $x$ from the electron and hadronic scattering angles as
\begin{equation}\label{qxda}
y_{DA} = \frac{\tan{(\theta_h/2)}}{\tan{(\theta_e/2)} + \tan{(\theta_h/2)}}\,,
~~~~~~Q^2_{DA}= 4 E_e^2 \cdot  
\frac{\cot{(\theta_e/2)}}{\tan{(\theta_e/2)} + \tan{(\theta_h/2)}}\,,~~~~~~~ x_{DA} = \frac{Q^2_{DA}}{s y_{DA}}\,.
\end{equation}
The method is largely insensitive to hadronization and, to first order, is independent of the 
detector energy scales. However, the hadronic angle is less well-determined
than the electron angle due to particle loss in the beampipe.
In the ``PT method'' of reconstruction 
\cite{Derrick:1996hn}  the well-measured 
electron kinematics is used to obtain a good
event-by-event estimate of the hadronic energy loss, 
by employing $\delta_{PT}=P_{T,h}/P_{T,e}$. This  improves
both the resolution and uncertainties of the reconstructed $y$ and $Q^2$.
The PT method uses all measured variables to optimise the resolution over the entire 
kinematic range measured, namely,
\begin{equation} \label{eq:ptmeth}
 \tan{\frac{\theta_{PT}}{2}} = \frac{\Sigma_{PT}}{P_{T,e}},{\rm~~~where}~~~~~~
 \Sigma_{PT} = 2E_e\frac{{C(\theta_h,P_{T,h},\delta_{PT})}\cdot\Sigma_h}
                       {\Sigma_e+{C(\theta_h,P_{T,h},\delta_{PT})}\cdot\Sigma_h}.
\end{equation}
The variable $\theta_{PT}$ is then substituted for $\theta_h$ in the formulae
for the double angle method to determine $x$, $y$ and $Q^2$. The 
detector-specific function
$C$ is calculated  using Monte Carlo simulations as $\Sigma_{true,h}/\Sigma_{h}$, depending on $\theta_h$, $P_{T,h}$ 
and $\delta_{PT}$.

The methods of kinematic reconstruction used
   by H1 and ZEUS are given in \Tab~\ref{tab:data} as
   part of the specification of the input data sets.

\subsection{Detectors}
\label{dets}
The H1 \cite{h1det,h1det2,spacalc} and ZEUS \cite{ZEUSDETECTOR} detectors had nearly
$4\pi$ hermetic coverage. They were built following similar physics considerations
but opted for different technical solutions, both for the calorimetric
and the tracking measurements.
 
The main component of the H1 detector was  the finely segmented Liquid Argon calorimeter (LAr).
It had an inner electromagnetic part and an outer hadronic part.
The resolutions, $\sigma_E$, as measured with test beams, are  about $0.11\sqrt{E/{\rm GeV}}$  
and $0.50\sqrt{E/{\rm GeV}}$ for electromagnetic and hadronic particles, respectively.  The LAr was surrounded by
a superconducting coil providing a solenoidal magnetic field of
$1.16$\,T and an instrumented iron structure acting as a shower
tail catcher and a muon detector. In $1996$ a major upgrade was performed
during which the backward detectors were replaced by a 
Backward Drift Chamber (BDC) attached to a new lead fibre calorimeter
(SpaCal) with a high resolution ($0.07\sqrt{E/\rm{GeV}}$)  electromagnetic section, which comprised
$27.5$ radiation lengths,
followed by a hadronic section. The SpaCal calorimeter had a total depth of two hadronic interaction lengths. 
In the upgraded configuration,
the innermost central tracking detector of H1 contained two silicon detectors,
the central and backward trackers,  CST and BST. These were
surrounded by the Central Inner Proportional chamber,
CIP, and the Central Jet Chamber, CJC, which was the main
tracking device of H1, with wires strung parallel to the beam axis.
The CJC had two concentric parts.
The  CJC was complemented by drift chambers to measure the
$z$ coordinate, and a further proportional chamber between
the CJC chambers was used for triggering.
Tracks in the forward direction were measured in the Forward Tracking
Detector, FTD, a set of  drift chamber modules of different orientation.

The main component of the ZEUS detector was the uranium-scintillator calorimeter  (CAL) consisting
of three parts: forward (FCAL), barrel (BCAL) and rear (RCAL). Each part was segmented
into one electromagnetic section (EMC) and either one (in RCAL) or two (in BCAL and FCAL)
hadronic sections (HAC). Under test-beam conditions, the energy resolutions were $0.18\sqrt{E/{\rm GeV}}$ and
$0.35\sqrt{E/\rm{GeV}}$ for the EMC
 and HAC sections, respectively. The timing resolution of the
CAL was $\sim$\,$1$\,ns for energy deposits greater than $4.5\,$GeV. Scintillator-tile presampler
detectors were mounted in front of the CAL. The RCAL was instrumented at a depth of $3.3$
radiation lengths with a silicon-pad hadron-electron separator (HES).
Charged particles were tracked in the central tracking detector (CTD) which operated in a magnetic field
of $1.4\,$T provided by a thin superconducting solenoid, positioned inside the BCAL and presampler.
The CTD was organised in nine superlayers
with five axial layers and four stereo layers. Planar drift chambers
provided additional tracking in the forward and rear directions.
The small angle rear tracking detector (SRTD), consisting of two orthogonal
planes of scintillator strips,  was used
to measure electrons at large $\theta_e$.
The angular coverage in the electron beam direction was extended with
the addition of a small Tungsten-scintillator calorimeter (BPC), located
behind the RCAL at $z=-294$\,cm close to the beam axis, and a
silicon microstrip tracking device (BPT) installed in front of the BPC.

Both H1 and ZEUS  were also equipped with photon taggers,
positioned at $\simeq$\,$100$\,m down the $e$ beam line,
for a determination of the luminosity
from Bethe-Heitler scattering, $ep \rightarrow ep \gamma$.
The measurement accuracy of the luminosity
was about $1 - 2$\% for each of the experiments.

\subsection{Data Samples}
A summary of the data used in this analysis is given in Table\,\ref{tab:data}.
In the first years until 1997, the proton beam energy
$E_p$ was set to $820$\,GeV. In 1998 it was increased to $920$\,GeV.
The NC data cover a wide range in $x$ and $Q^2$.
The lowest $Q^2 \ge 0.045$~GeV$^2$ data come from the measurements of ZEUS using
the BPC and BPT~\cite{Breitweg:1997hz,Breitweg:2000yn}. The $Q^2$ range from $0.2$~GeV$^2$ to $1.5$~GeV$^2$
is covered using special HERA runs, in which the interaction vertex
position was shifted forward
allowing for larger angles 
of the backward scattered electron
to be accepted~\cite{Adloff:1997mf,Breitweg:1998dz,Collaboration:2009bp}.
The lowest $Q^2$ for the shifted vertex data 
was reached using events, in which the effective electron
beam energy was reduced by initial state radiation~\cite{Collaboration:2009bp}. 
Values of  $Q^2\ge 1.5$~GeV$^2$ were measured using  the nominal vertex settings.
For $Q^2 \le 10$~GeV$^2$, the cross section is very high 
and the data were  collected using dedicated
trigger setups~\cite{h1alphas,Chekanov:2001qu,Collaboration:2009bp}. 
The highest accuracy of the cross-section measurement  is achieved
for  $10 \le Q^2 \le 100$~GeV$^2$  \cite{h1alphas,Chekanov:2001qu,Collaboration:2009kv}.
For $Q^2\ge 100$~GeV$^2$, the statistical uncertainty of the data becomes
relatively large.
The high $Q^2$ data included here were collected with  
positron~\cite{Adloff:1999ah,Chekanov:2001qu,Adloff:2003uh,Chekanov:2003yv} and with 
electron~\cite{Adloff:2000qj,Chekanov:2002ej} beams.  
The CC data for $e^+p$ and $e^-p$ scattering cover the range $300\le Q^2\le 30000$~GeV$^2$~\cite{Adloff:1999ah,zeuscc97,Chekanov:2002zs,Adloff:2003uh,Chekanov:2003vw}.

\begin{table}
\begin{center}
\begin{scriptsize}
\begin{tabular}{|lr|lr|lr|c|c|c|c|c|}
\hline
\multicolumn{2}{|c|}{Data Set} &
\multicolumn{2}{|c|}{$x$ Range} &
\multicolumn{2}{|c|}{$Q^2$ Range} &
${\cal L}$ & $e^+/e^-$ & $\sqrt{s}$ & $x$,$Q^2$ Reconstruction & Reference \\
\multicolumn{2}{|c|}{ } &
\multicolumn{2}{|c|}{ } &
\multicolumn{2}{|c|}{GeV$^2$} &
pb$^{-1}$ &  & GeV &  Method \EEq & \\
\hline
H1~svx-mb  & $95$-$00$ & $5\times 10^{-6}$  & $0.02$  & $0.2$ & $12$  & $2.1$ &$e^+p$   & $301$-$319$    & \ref{eq:emeth},\ref{eq:yh},\ref{eq:sigma2}    &\cite{Collaboration:2009bp} \\
H1~low~$Q^2$      & $96$-$00$ & $2\times 10^{-4}$  & $0.1$   & $12$  & $150$ & $22$ &$e^+p$   & $301$-$319$  &  \ref{eq:emeth},\ref{eq:yh},\ref{eq:sigma2}      &\cite{Collaboration:2009kv}\\
H1~NC             & $94$-$97$ & $0.0032$  &$0.65$   &$150$  &$30000$   & $35.6$  & $e^+p$ & $301$           &   \ref{eq:esigma}       &  \cite{Adloff:1999ah}\\ 
H1~CC             & $94$-$97$ & $0.013$   &$0.40$   &$300$  &$15000$   & $35.6$  & $e^+p$ & $301$            &   \ref{yjb}     &   \cite{Adloff:1999ah}\\ 
H1~NC             & $98$-$99$ & $0.0032$  &$0.65$   &$150$  &$30000$   & $16.4$  & $e^-p$ & $319$            &    \ref{eq:esigma}     & \cite{Adloff:2000qj}\\
H1~CC             & $98$-$99$ & $0.013$   &$0.40$   &$300$  &$15000$   & $16.4$  & $e^-p$ & $319$             &  \ref{yjb}       & \cite{Adloff:2000qj}\\
H1~NC HY          & $98$-$99$ & $0.0013$  &$0.01$   &$100$  &$800$   & $16.4$  & $e^-p$ & $319$               &  \ref{eq:emeth}     & \cite{Adloff:2003uh}\\
H1~NC             & $99$-$00$ & $0.0013$ &$0.65$   &$100$  &$30000$   & $65.2$  & $e^+p$ & $319$              &  \ref{eq:esigma}  & \cite{Adloff:2003uh}\\
H1~CC             & $99$-$00$ & $0.013$   &$0.40$   &$300$  &$15000$   & $65.2$  & $e^+p$ & $319$              &   \ref{yjb}    & \cite{Adloff:2003uh} \\
\hline
ZEUS~BPC            & $95$ & $2\times 10^{-6}$    & $6\times 10^{-5}$  &$0.11$  & $0.65$   & $1.65$  &  $e^+p$  & $301$ & \ref{eq:emeth} &      \cite{Breitweg:1997hz} \\
ZEUS~BPT            & $97$ & $6\times 10^{-7}$    & $0.001$   & $0.045$ & $0.65$  & $3.9$  & $e^+p$   &  $301$  & \ref{eq:emeth},  \ref{eq:esigma} &     \cite{Breitweg:2000yn}\\
ZEUS~SVX            & $95$ & $1.2\times 10^{-5}$  & $0.0019$  & $0.6$   & $17$  & $0.2$  & $e^+p$   &  $301$    &  \ref{eq:emeth} &     \cite{Breitweg:1998dz}\\
ZEUS~NC             & $96$-$97$ & $6\times10^{-5}$  &$0.65$& $2.7$ & $30000$  & $30.0$  & $e^+p$    & $301$     & \ref{eq:ptmeth}  &     \cite{Chekanov:2001qu}\\   
ZEUS~CC             & $94$-$97$ & $0.015$  & $0.42$  & $280$  & $17000$   &$47.7$   &  $e^+p$   & $301$         &  \ref{yjb}   &     \cite{zeuscc97}\\
ZEUS~NC             & $98$-$99$ & $0.005$  & $0.65$  & $200$  & $30000$  &$15.9$   & $e^-p$    & $319$          &  \ref{qxda} &     \cite{Chekanov:2002ej} \\   
ZEUS~CC             & $98$-$99$ & $0.015$  & $0.42$  & $280$  & $30000$  &$16.4$   & $e^-p$    & $319$          &  \ref{yjb}   &     \cite{Chekanov:2002zs} \\
ZEUS~NC             & $99$-$00$ & $0.005$  & $0.65$  & $200$  & $30000$  &$63.2$   & $e^+p$   & $319$           & \ref{qxda}   &     \cite{Chekanov:2003yv} \\   
ZEUS~CC             & $99$-$00$ & $0.008$  & $0.42$  & $280$  & $17000$  &$60.9$   &  $e^+p$   &$319$           &  \ref{yjb}  &      \cite{Chekanov:2003vw}\\
\hline
\end{tabular}
\end{scriptsize}
\end{center}
\caption{\label{tab:data}H1 and ZEUS data sets used for the combination. 
The H1~svx-mb~\cite{Collaboration:2009bp} 
and H1~low~$Q^2$~\cite{Collaboration:2009kv} data sets
comprise averages including data collected at 
$E_p=820$~GeV~\cite{Adloff:1997mf,h1alphas} and $E_p=920$~GeV. The formulae for $x,Q^2$ reconstruction are given in
\Sec~\ref{diskine}.}
\end{table} 
\section{Combination of the Measurements\label{sec:comb}}
\subsection{Combination Method\label{subsec:comb:method}}
The combination of the data sets uses the
$\chi^2$ minimisation method described in~\cite{Collaboration:2009bp}.
The $\chi^2$ function takes into account the correlated systematic uncertainties
for the H1 and ZEUS cross-section measurements.
For a single
data set the $\chi^2$ is defined as
\begin{equation}
 \chi^2_{\rm exp}\left(\boldsymbol{m},\boldsymbol{b}\right) = 
 \sum_i
 \frac{\left[m^i
- \sum_j \gamma^i_j m^i b_j  - {\mu^i} \right]^2}
{ \textstyle \delta^2_{i,{\rm stat}}\,{\mu^i}  \left(m^i -  \sum_j \gamma^i_j m^i b_j\right)+
\left(\delta_{i,{\rm uncor}}\,  m^i\right)^2}
 + \sum_j b^2_j\,.
\label{eq:ave}
\end{equation}
Here  ${\mu^i}$ is the  measured  value  at a point $i$ and
 $\gamma^i_j $, 
$\delta_{i,{\rm stat}} $ and 
$\delta_{i,{\rm uncor}}$ are relative
correlated systematic, relative statistical and relative uncorrelated systematic uncertainties,
respectively.
The function $\chi^2_{\rm exp}$ depends on the predictions $m^i$ 
for the measurements
(denoted as the vector $\boldsymbol{m}$) and 
 the shifts $b_j$ (denoted as the vector $\boldsymbol{b}$) of the correlated systematic error sources.
For the reduced cross-section  measurements  ${\mu^i} = \sigma_r^i$ and
$i$ denotes a $(x,Q^2)$ point. The summation over
$j$ extends over all correlated systematic sources. 
The predictions $m^i$ are 
given by the assumption that there is a single true value of the cross section 
corresponding to each data point $i$ and each process, neutral or 
charged current $e^+p$ or $e^-p$ scattering.

\EEq~\ref{eq:ave} takes into account that the quoted uncertainties  are based on measured cross sections, which are subject
to statistical fluctuations. 
Under the assumption that the statistical uncertainties are proportional
to the square root of the number of events and that the systematic
uncertainties are proportional to $\boldsymbol{m}$, the minimum of
 \Eq~\ref{eq:ave} provides an unbiased estimator of $\boldsymbol{m}$.
For the present cross section measurements, the leading systematic uncertainties
stem from the acceptance correction and luminosity determination. Therefore, 
the correlated and uncorrelated systematic uncertainties are of multiplicative nature, i.e. they increase
proportionally to the central values. In \Eq~\ref{eq:ave} the multiplicative nature of these uncertainties is taken into account by multiplying the
relative errors $\gamma^i_j$ and $\delta_{i,{\rm uncor}}$ by the expectation $m^i$. For the inclusive 
DIS cross-section measurements the 
background contribution is small and
the statistical uncertainties are defined by the square root of
the number of events used to determine $\sigma^i_r$. The expected number of events is determined by the expected cross section and it can be also
corrected for the biases due to the correlated systematic uncertainties.  This is taken into account by the 
$ \delta^2_{i,{\rm stat}}\,{\mu^i}  \left(m^i -  \sum_j \gamma^i_j m^i b_j\right)$ term.

Several data sets providing a number of measurements are represented
by a total $\chi^2$ function,
which is built from the sum of the $\chi^2_{\rm exp}$ functions for each data set $e$ 
\begin{equation}
\chi^2_{\rm tot} = \sum_e \chi^2_{{\rm exp},e}~. \label{eq:tot}
\end{equation}
The data averaging procedure allows the rearrangement of \Eq~\ref{eq:tot} such
that it takes a form similar to \Eq~\ref{eq:ave}
\begin{equation}
 \chi^2_{\rm tot}\left(\boldsymbol{m},\boldsymbol{b'}\right) = 
 \chi^2_{\rm min} + 
 \sum_{i=1}^{N_M}
 \frac{\left[m^i
- \sum_j \gamma^{i,{\rm ave}}_j m^i b'_j  - {\mu^{i,{\rm ave}}} \right]^2}
{ \textstyle \delta^2_{i,{\rm ave,stat}}\, \mu^{i,{\rm ave}}\left(m^i -  \sum_j \gamma^{i,{\rm ave}}_j m^i b'_j\right)+
\left(\delta_{i,{\rm ave,uncor}}\,  m^i\right)^2}
 + \sum_j (b'_j)^2.
\label{eq:avetot}
\end{equation}
Here $\mu^{i,{\rm ave}}$ is the average value at a point $i$  and
 $\gamma^{i,{\rm ave}}_j $, 
$\delta_{i,{\rm ave,stat}} $ and 
$\delta_{i,{\rm ave,uncor}}$ are its relative
correlated systematic, relative statistical and relative uncorrelated systematic uncertainties,
respectively. The value of $\chi^2_{\rm min}$ corresponds to the minimum of \Eq~\ref{eq:tot}.
The ratio $\chi^2_{\rm min}/n_{\rm dof}$ is a measure 
of the consistency of the data sets.
The number of degrees of
freedom, $n_{\rm dof}$, is calculated as the difference between the total 
number of measurements and 
the number of averaged points $N_M$. The systematic uncertainties $b'_j$
are obtained from the original ones, $b_j$, by an orthogonal 
transformation~\cite{Collaboration:2009bp}.
The summation of $j$ extends over all independent systematic error sources.

The original double differential cross-section measurements are published with their statistical
and  systematic uncertainties. The statistical uncertainties correspond to $\delta_{i,{\rm stat}}$ in \Eq~\ref{eq:ave}.
The systematic uncertainties are classified as either point-to-point correlated or point-to-point uncorrelated,
corresponding to $\gamma_i^j$ and $\delta_{i,{\rm uncor}}$, respectively. Several measurements~\cite{Breitweg:2000yn,Breitweg:1998dz,zeuscc97,Chekanov:2002ej,Chekanov:2002zs,Chekanov:2003yv,Chekanov:2003vw}
are reported
with asymmetric systematic uncertainties and they are symmetrised before performing the averaging. 
The resulting average is found to be insensitive to the details of the symmetrisation procedure.
Global normalisations of the data sets 
are split into an overall normalisation uncertainty of $0.5\%$, common to all data sets,
due to uncertainties of higher order corrections 
to the Bethe-Heitler process used for  the luminosity calculation,
and experimental uncertainties which are treated as correlated systematic sources $\gamma_i^j$.
Sources of point-to-point correlated uncertainties
may be in common for CC and NC data as well as for several data sets of the same experiment. 
The correlations of the systematic uncertainties across the data sets
are given in \Tab~\ref{tab:corel}. The systematic uncertainties are treated as independent between H1 and ZEUS 
apart from the $0.5\%$ overall normalisation
uncertainty. 

All the NC and CC cross-section data from H1 and ZEUS are combined in one simultaneous minimisation. 
Therefore resulting shifts of 
the correlated systematic uncertainties propagate coherently to both CC and NC data. 

\subsection{Corrections and Uncertainties\label{subsec:comb:corrs}}
\subsubsection{Extrapolation to Common $\boldsymbol{(x,Q^2)}$ Grid \label{subsubsec:extrapol}}
Prior to the combination, the H1 and ZEUS data are transformed to a common grid of $(x,Q^2)$ points.
The grid points are chosen such that the interpolation corrections are minimal taking advantage 
of the fact that the original $(x,Q^2)$ grids of the H1 and ZEUS experiments are similar.
Furthermore, the chosen grid ensures that no two separate measurements of the same data set interpolate to a common grid point\footnote{An exception
is made for the ZEUS SVX~\cite{Breitweg:1998dz} data for which five pairs of points are first averaged using statistical uncertainties and then added to the
combination.}.
For $Q^2\ge 0.2$~GeV$^2$, for the 
majority of the grid points both H1 and ZEUS measurements enter the combination.
For some of the grid points there is no nearby counterpart from the other experiment, giving points in
the combined cross section which originate from either H1 or ZEUS only. 
Note that through the systematic error correlation, such
data points may be nevertheless shifted with respect to the original measurement in the averaging procedure.

The transformation of a measurement from the given $(x,Q^2)$ to the 
nearest $(x_{\rm grid},Q^2_{\rm grid})$ point on the grid
is performed by multiplying the measured cross section 
by a ratio of theoretically calculated double differential cross sections at  
$(x_{\rm grid},Q^2_{\rm grid})$ 
and $(x,Q^2)$.
This interpolation is repeated iteratively. For the first iteration, 
the H1 PDF 2000 parametrisation~\cite{Adloff:2003uh}
is used for $Q^2 \ge 3$~GeV$^2$, where it is applicable, and the fractal model fit~\cite{Collaboration:2009bp} for
 $Q^2 < 3$~GeV$^2$. For the second iteration, a QCD fit to the first iteration of the averaged 
data (see \Sec~\ref{pdffit}) is used for $Q^2\ge 3$~GeV$^2$ and the fractal model fit for $Q^2<3$~GeV$^2$. 
The difference between cross-section measurements
obtained after the first and second iterations is smaller than a few per mille for the NC  
and  $2\%$ for the CC data. 
The QCD fit obtained using the data from the first iteration is to per mille precision identical 
to the fit obtained using the 
data from the second iteration. Therefore, no further iterations are performed. 
This procedure is checked using the ZEUS-JETS parametrisation~\cite{Chekanov:2005nn}  for the first iteration. 
The resulting cross-section difference is negligible compared to the experimental precision.

\subsubsection{Centre-of-Mass Energy Correction}
The data sets considered for the combination contain sub-samples taken with a proton 
beam energy of $E_p=820$~GeV and $E_p=920$~GeV. Both the 
CC and NC scattering reduced cross sections depend weakly on the
energy via terms containing the inelasticity $y$. 
For the CC data for all values of $y$ and for the NC data for $y<0.35$, 
the uncertainty on the theoretically estimated difference between cross sections 
for $E_p = 820$~GeV and $E_p = 920$~GeV is negligible compared to the experimental precision. 
Therefore the
data are corrected to a common centre-of-mass energy corresponding to $E_p=920$~GeV 
and then averaged. 
The NC data for $y\ge 0.35$ are kept separate for the two proton beam energies. 

The correction is calculated multiplicatively for the CC data which is given in 
full double differential form
\begin{equation} \label{eq:cccme}
\ccddne = \ccddre \left[  \ccddneth \left/ \ccddreth \right. \right]
\end{equation}
and additively for the NC data which is quoted in reduced double differential
 form
\begin{equation}
\ncredne = \ncredre + \Delta \ncredth (x, Q^2,y_{820},y_{920})\,.
\end{equation}
Here $y_{820}$ and $y_{920}$ are the inelasticities for the two proton beam energies calculated as $y = Q^2/ 4 E_e E_p x$.
The theoretical cross sections follow the same prescription used for the extrapolation 
correction in 
\Sec~\ref{subsubsec:extrapol}. For the neutral current case,
\begin{equation}
\Delta \ncredth  (x, Q^2,y_{820},y_{920}) = 
  \boldflth (x,Q^2)
\left[   \frac{\textstyle y_{820}^2}{\textstyle Y_{+\, 820}} - \frac{\textstyle y_{920}^2}{\textstyle Y_{+\, 920}} \right]
+
 \boldxftth (x,Q^2)
\left[   \pm \frac{\textstyle Y_{-\,820}}{\textstyle Y_{+\,820}} 
    \mp \frac{\textstyle Y_{-\,920}}{\textstyle Y_{+\,920}} \right]\,, \label{eq:delta}
\end{equation}
where $Y_{\pm\,820,920} = 1 \pm (1-y_{820,920})^2$.
The largest uncertainty in \Eq~\ref{eq:delta} stems from the calculation of the structure function $\boldflth$.
To estimate the uncertainty on the combined data arising from the centre-of-mass energy correction another average
is performed assuming $\boldflth=0$. The difference between this and the nominal average is at 
most $0.1\%$ 
and thus the uncertainty on the averaged cross section resulting 
from the centre-of-mass energy correction is negligible.
\subsubsection{Procedural Uncertainties \label{subsubsec:proc_errors}}
The $\chi^2$ function given by \Eq~\ref{eq:ave} treats all systematic uncertainties as multiplicative, i.e.
proportional to the expected central values. While this generally holds
for the normalisation uncertainties, this may not be the case for the other uncertainties.
To study the sensitivity of the average result to this issue, an alternative 
averaging is performed, for which only normalisation uncertainties are taken as
multiplicative while all other uncertainties are treated as additive. The difference between this average and the 
nominal average result is used as a correlated procedural error $\delta_{\rm ave,rel}$. The typical
size of $\delta_{\rm ave,rel}$ is below $0.5\%$ for the low $Q^2$ data reaching a few percent for high $Q^2$.

The H1 and ZEUS collaborations use similar methods for detector calibration 
and employ similar Monte Carlo simulation models for radiative corrections, for the  hadronic 
final state simulation and for photoproduction background subtraction. Such similarities may lead to 
correlations between the H1 and ZEUS measurements.
To investigate the effect of possible correlations,  $12$ 
sources of similar systematic uncertainties       of the two experiments are identified.
These are related to the photoproduction
background, the  electromagnetic and hadronic energy scales
and the electron scattering angle.  Then $2^{12}$ different averages are calculated assuming each of the 
$12$ pairs to be correlated or uncorrelated, and these alternative averages are compared to the
nominal average for which all sources are assumed to be uncorrelated. By studying these averages 
it is found that the only
two systematic sources 
which result in significantly different  average cross sections are the photoproduction background 
and the
hadronic energy scale. The differences between the nominal average  and the averages in which 
systematic sources for the photoproduction background and hadronic energy scale are considered to be correlated
are taken as additional procedural uncertainties  $\delta_{\rm ave,\gamma p}$ and $\delta_{\rm ave,had}$.
Typical values of $\delta_{\rm ave,\gamma p}$ and $\delta_{\rm ave,had}$ are below $0.5\%$.
As expected, $\delta_{\rm ave,\gamma p}$
is larger at high $y>0.5$ while $\delta_{\rm ave,had}$ is significant for low $y$ only.

\subsection{Results\label{subsec:comb:results}}
\def\totalave{ 1402}
\def\uniqueave{  741}
\def\chiavedat{609.6}
\def\chiave{636.5}
\def\dofave{656}
\def\chiavesys{26.9}

\begin{table}
\begin{center}
\begin{tabular}{lrr}
\hline
Data set & \multicolumn{2}{c}{Shift}  \\
         &  \multicolumn{1}{c}{ $b$}    &   \multicolumn{1}{c}{$(b \gamma_{\rm norm})$} \\
\hline
H1 svx-mb~\cite{Collaboration:2009bp} & $ -0.24$ & $( -0.2\%)$ \\ 
H1 low $Q^2$~\cite{Collaboration:2009kv} & $ -0.45$ & $( -0.4\%)$ \\ 
H1 $94-97$~\cite{Adloff:1999ah} & $ -0.65$ & $( -0.9\%)$ \\ 
H1 $98-99$~\cite{Adloff:2000qj,Adloff:2003uh} & $ -0.05$ & $( -0.1\%)$ \\ 
H1 $99-00$~\cite{Adloff:2003uh} & $ -0.19$ & $( -0.3\%)$ \\ 
ZEUS BPC~\cite{Breitweg:1997hz} & $  0.23$ & $(  0.3\%)$ \\ 
ZEUS BPT~\cite{Breitweg:2000yn} & $ -0.03$ & $( -0.1\%)$ \\ 
ZEUS SVX~\cite{Breitweg:1998dz} & $  0.78$ & $(  1.2\%)$ \\ 
ZEUS $94-97$~\cite{Chekanov:2001qu,zeuscc97} & $  0.44$ & $(  0.8\%)$ \\ 
ZEUS $96-97$ low $Q^2$~\cite{Chekanov:2001qu} & $ -1.10$ & $( -1.1\%)$ \\ 
ZEUS $98-99$~\cite{Chekanov:2002ej,Chekanov:2002zs} & $  0.05$ & $(  0.1\%)$ \\ 
ZEUS $99-00$~\cite{Chekanov:2003vw,Chekanov:2003yv} & $ -0.18$ & $( -0.4\%)$ \\ 

\hline
\end{tabular}
\end{center}
\caption{\label{tab:normshifts}Shifts of global normalisation 
of the data sets in units of the normalisation
uncertainty (and in percent of the original cross sections).  }
\end{table}
The average NC and CC cross sections and the structure function $F_2$
together with statistical, uncorrelated systematic and procedural
uncertainties are given in \Tabs~\ref{tab615a1}-\ref{tab3615a1}. The full information about correlation
between cross-section measurements can be found elsewhere~\cite{fullcorr}.
The total integrated luminosity of the combined data set corresponds to about $200$~pb$^{-1}$ for $e^+p$ and
 $30$~pb$^{-1}$ for $e^-p$.
In total $\totalave$ data points are combined to $\uniqueave$ cross-section measurements. 
The data show good consistency, with $\chi^2/n_{\rm dof} = \chiave/\dofave$. 
For the measurement at the point $i$ of the data set $e$ which has measurements of other data sets
 averaging to the same $(x,Q^2)$ point, the pull $p^{i,e}$ can be defined
\begin{equation}
p^{i,e} = \frac{\mu^{i,e}  - \mu^{i,{\rm ave}}\left(1- \sum_j \gamma^{i,e}_j b_{j,{\rm ave}}\right)}{\sqrt{\Delta_{i,e}^2 - \Delta_{i,{\rm ave}}^2}},
\end{equation}
where $\Delta_{i,e}$ ($\Delta_{i,{\rm ave}}$) is the statistical and uncorrelated systematic uncertainty added in quadrature for the measurement
$e$ (average). The distribution of pulls  shows no tensions 
for all processes across the kinematic plane, see \Fig~\ref{fig:pulls}. 

\begin{figure}
\centerline{\epsfig{file=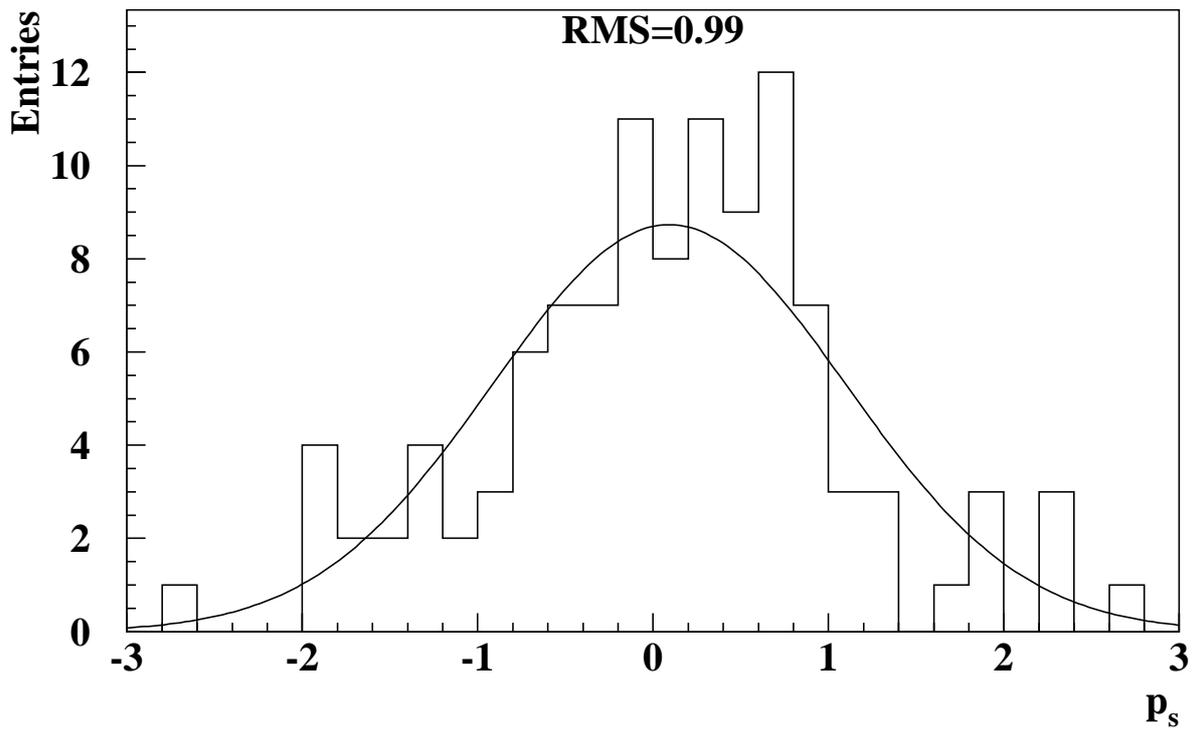 ,width=\linewidth}}
\caption{Distribution of pulls $p_s$ for correlated
systematic uncertainties including global normalisations. There are no entries outside the histogram range. 
RMS gives the root mean square of the distribution calculated as $\overline{p_s^2}$. 
The curve shows the result of a binned log-likelihood Gaussian
fit to the distribution.
\label{fig:syspulls}}
\end{figure}

There are in total $110$ sources of correlated systematic uncertainty, including global
normalisations, characterising 
the separate data sets. None of these systematic sources
shifts by more than $2\, \sigma$ of the nominal value in the averaging procedure. 
The distribution of pulls for the correlated systematic sources, defined as
$p_s = b_{j,{\rm ave}}/(1 - \Delta_{b_j,{\rm ave}}^2)^{1/2}$, where $\Delta_{b_j,{\rm ave}}$
is the uncertainty of the source $j$ after the averaging, is given in \Fig~\ref{fig:syspulls}.
The shifts of the global normalisation for each input data set 
are given in
\Tab~\ref{tab:normshifts}. 
The absolute normalisation of the combined data set
is to a large extent defined by the most precise measurements of
NC $e^+p$ cross section in the $10\le Q^2\le 100$~GeV$^2$ kinematic range. Here
the H1~\cite{Collaboration:2009kv} and ZEUS~\cite{Chekanov:2001qu} results move towards each other and 
the other data sets follow this adjustment.

The influence of several correlated systematic uncertainties is reduced
significantly for the averaged result. 
For example, the uncertainty due to the  H1 LAr calorimeter energy scale  is reduced by $55\%$  while the uncertainty
due to the
ZEUS photoproduction background is reduced by $65\%$. 
There are two main reasons for this significant reduction. 
Since H1 and ZEUS use different reconstruction methods, described in
 \Sec~\ref{diskine}, 
similar systematic sources influence 
the measured cross section differently as a function of $x$ and $Q^2$. 
Therefore, requiring the cross sections to agree at all $x$ and $Q^2$ constrains 
the systematics efficiently. In addition, for certain regions of the phase space, one of the two 
experiments has superior precision compared to the other.
For these regions, the less precise measurement is fitted to the more precise one, with a simultaneous reduction of the correlated systematic uncertainty.
This reduction propagates to the other average points, including those which are based solely on the measurement from one experiment.
  
For $Q^2\ge 100$~GeV$^2$ the precision of the H1 and
ZEUS measurements is about equal and thus the systematic uncertainties are reduced uniformly.
For $2.5\le Q^2 < 100$~GeV$^2$ and for $Q^2<1$~GeV$^2$ the precision is dominated by the H1~\cite{Collaboration:2009bp,Collaboration:2009kv}
and by the ZEUS~\cite{Breitweg:2000yn} measurements, respectively.
Therefore the overall reduction of the uncertainties is smaller, 
and it is essentially obtained from the reduction of the correlated systematic uncertainty.
For example, for $Q^2\sim 30$~GeV$^2$ the obtained precision of $1.1\%$ is dominated by the data of~\cite{Collaboration:2009kv}, 
where $1.3\%$ uncertainties
are reported.
The correlated error part of the full uncertainty is reduced from about 
75\% in~\cite{Collaboration:2009kv} to 55\% in the combined measurement. 
The total uncertainty is typically smaller than $2\%$ for $3 < Q^2 < 500$~GeV$^2$ and
reaches $1\%$ for $20 <  Q^2 < 100$~GeV$^2$. 
The uncertainties are larger for high inelasticity $y>0.6$ 
due to the photoproduction background.

In \Figs~\ref{fig:quality} and \ref{fig:quality2}, averaged data are compared to the 
input H1 and ZEUS data. 
In \Figs~\ref{fig:vsQ2l} and \ref{fig:vsQ2m}, 
the combined NC $e^+p$ data at very low $Q^2$ and at low $Q^2$ are shown.
In \Figs~\ref{fig:dataNCp} and \ref{fig:dataNCm}, the combined  NC $e^+p$ and $e^-p$ data at high $Q^2$ are
shown.
In \Fig~\ref{fig:scal} the $e^{+}p$ NC reduced cross section, for $Q^2 > 1$\,GeV$^2$, is shown
as a function of $Q^2$ for the 
HERA combined $e^+p$ data and for fixed-target data~\cite{bcdms,nmc} across 
the whole of the measured kinematic plane.
The combined NC $e^{\pm}p$ reduced cross sections are compared in the high-$Q^2$
region in \Fig~\ref{fig:ncepem}.
In \Figs~\ref{fig:dataCCp} and \ref{fig:dataCCm}
 the combined data set is shown for CC scattering at high $Q^2$.
The 
HERAPDF1.0 fit, described in the next section, 
is compared to the data
 in the kinematic region suitable
for the application of perturbative QCD. 
\section{QCD Analysis of the Combined Data \label{sec:fit}}
\label{pdffit}
\subsection{Definition of $\boldsymbol{\chi^2}$ and Treatment of Systematic Uncertainties}
The combined data set on NC and CC $e^+p$ and $e^-p$ inclusive cross sections 
is used as the sole input for a next-to-leading order (NLO) QCD PDF fit, called HERAPDF1.0. 
The form of the $\chi^2$ 
used for this fit follows \Eq~\ref{eq:ave}. 
The consistency of the present input data  
justifies the use of the conventional 
$\chi^2$ tolerance, $\Delta\chi^2=1$, when determining the experimental 
uncertainties on the HERAPDF1.0 fit.

The role of correlated systematic uncertainties should be considered 
carefully when defining the $\chi^2$ for PDF fits. Global fits have used the 
Hessian method~\cite{Pumplin:2002vw,Alekhin:2002fv}, the Offset 
method~\cite{Botje:1999dj,Chekanov:2002pv} and 
quadratic combination~\cite{Martin:2001es,Martin:2009iq}. In fits to HERA 
data alone, H1 have used 
the Hessian method~\cite{Adloff:2003uh,Collaboration:2009kv} and ZEUS have 
used the Offset method~\cite{Chekanov:2005nn}. 
These different methods can lead to 
quite different estimates of the PDF central values and 
uncertainties~\cite{CooperSarkar:2005zd,Alekhin:2003ye}, 
with the Offset method giving the largest uncertainties 
 if the correlated systematic uncertainties 
are larger than the statistical uncertainties of the data sets.
 
In the present analysis, the combination of the 
H1 and ZEUS data sets is done by a Hessian procedure 
under the assumption that there is only one true cross-section value for each 
process (NC or CC, $e^+p$ or $e^-p$ scattering) at each $(x,Q^2)$ point.
This results in a data set with correlated systematic 
uncertainties which are smaller or comparable to the statistical and uncorrelated 
uncertainties. 
Thus the central values and experimental uncertainties on the
PDFs which are extracted from the combined data are not much dependent 
on the method of treatment of correlated systematic uncertainties in the 
fitting procedure. 
The most conservative estimate of uncertainties still results from the use of the
Offset method for all $113$ systematic errors.
However, this procedure is very cumbersome and in practice a very similar uncertainty 
estimate is obtained from using the Offset method only for the largest correlated 
systematic errors, namely the three procedural errors 
(see \Sec~\ref{subsubsec:proc_errors}). Thus for the central fit, 
the $110$ systematic uncertainties 
which result from the ZEUS and H1 data sets are combined 
in quadrature, and the three sources of uncertainty which result from 
the combination procedure are treated by the offset method. 
 The $\chi^2$ per degree of freedom for this fit is $574/582$;
for a fit combining all $113$ uncertainties in quadrature the $\chi^2$ is $532$ 
and for a fit treating all $113$ by the Hessian method the $\chi^2$ is $579$.
The resulting experimental uncertainties on the  
PDFs are small. Therefore, a thorough consideration of further 
uncertainties due to model assumptions and parametrisation dependence is necessary. 

\subsection{Theoretical Formalism and Assumptions}
The QCD predictions for the structure functions 
are obtained by solving the DGLAP evolution 
equations~\cite{Gribov:1972ri,Gribov:1972rt,Lipatov:1974qm,Dokshitzer:1977sg,Altarelli:1977zs}
at NLO in the \msbar scheme with the
renormalisation and factorisation scales chosen to be $Q^2$.
The programme QCDNUM~\cite{QCDNUM} is used and checked against the 
programme QCDfit~\cite{Pascaud:1995qs}.
The DGLAP equations yield the PDFs
 at all values of $Q^2$ if they are provided
as functions of $x$ at some input scale $Q^2_0$. This scale is 
chosen to be $Q^2_0 = 1.9~$GeV$^2$ such that the starting scale is below the 
charm mass threshold, $Q_0^2 < m_c^2$. 
The light quark coefficient functions are calculated in QCDNUM.
The heavy quark 
coefficient functions are calculated in the general-mass
variable-flavour-number scheme of \cite{Thorne:1997ga}, with recent 
modifications~\cite{Thorne:2006qt,ThornePrivComm}.
The heavy quark masses  $m_c=1.4~$GeV and $m_b=4.75~$GeV are chosen following~\cite{Martin:2009iq}.
The strong coupling constant is fixed to 
$\asmz =  0.1176$~\cite{PDG}.

The HERA data have a minimum invariant mass of the hadronic system, $W$, 
of $15$\,GeV and a maximum $x$ of $0.65$, 
such that they are in a kinematic region where there is no
sensitivity to target mass and large-$x$ higher-twist contributions. 
A minimum $Q^2$ cut of $Q^2_{min} = 3.5$~GeV$^2$ is imposed 
to remain in the kinematic region where
perturbative QCD should be applicable.

PDFs are parametrised at the input scale by the generic form 
\begin{equation}
 xf(x) = A x^{B} (1-x)^{C} (1 + \epsilon\surd x + D x + E x^2).
\label{eqn:pdf}
\end{equation}
The parametrised PDFs  are the gluon distribution $xg$, the valence quark distributions 
$xu_v$, $xd_v$, and 
the $u$-type and $d$-type anti-quark distributions
$x\bar{U}$, $x\bar{D}$. Here $x\bar{U} = x\bar{u}$, 
$x\bar{D} = x\bar{d} +x\bar{s}$ at the chosen starting scale.
The central fit is found by first setting the $\epsilon$, $D$ and $E$ 
parameters to zero 
(this leaves $9$ parameters free) and then introducing them in the fit procedure, one at a time, 
to determine the best fit. The best $10$ parameter fit has $E_{u_v}\not= 0$.
The other $\epsilon$, $D$ and $E$ parameters are then added, one at a time, 
to determine the best $11$ parameter fit. The $11$ parameter fits do not 
represent a significant improvement in fit quality compared to the best $10$ 
parameter fit\footnote{The largest decrease in $\chi^2$ is $\Delta\chi^2 = -5$, 
for a fit which has $xd_{v} < x\bar{d}$ at large $x$.}. 
The $10$ parameter fit, selected as the central fit, 
has a good $\chi^2$ per degree of freedom, $574/582$, 
and satisfies the criteria that all the PDFs are positive 
and they obey the valence quark 
approximation that $xd_{v} > x\bar{d}$ at large $x$.  
The resulting parametrisations are
\begin{eqnarray}
\label{eq:xgpar}
xg(x) &=   & A_g x^{B_g} (1-x)^{C_g}  ,  \\
xu_v(x) &=  & A_{u_v} x^{B_{u_v}}  (1-x)^{C_{u_v}}\left(1+E_{u_v}x^2 \right) , \\
\label{eq:xuvpar}
xd_v(x) &=  & A_{d_v} x^{B_{d_v}}  (1-x)^{C_{d_v}} , \\
\label{eq:xdvpar}
x\bar{U}(x) &=  & A_{\bar{U}} x^{B_{\bar{U}}} (1-x)^{C_{\bar{U}}} , \\
\label{eq:xubarpar}
x\bar{D}(x) &= & A_{\bar{D}} x^{B_{\bar{D}}} (1-x)^{C_{\bar{D}}} .
\label{eq:xdbarpar}
\end{eqnarray}
The normalisation parameters, $A_g, A_{u_v}, A_{d_v}$, are constrained 
by the quark number sum rules and momentum sum rule. 
The $B$ parameters  $B_{\bar{U}}$ and $B_{\bar{D}}$ are set equal,
 $B_{\bar{U}}=B_{\bar{D}}$, such that 
there is a single $B$ parameter for the sea distributions. 
The strange quark distribution is expressed 
as $x$-independent fraction, $f_s$, of the $d$-type sea, 
$x\bar{s}= f_s x\bar{D}$ at $Q^2_0$.  The central value $f_s=0.31$ 
is chosen to be consistent with determinations 
of this fraction using neutrino-induced di-muon 
production~\cite{Martin:2009iq,Nadolsky:2008zw}. The further constraint 
$A_{\bar{U}}=A_{\bar{D}} (1-f_s)$, together with the requirement  
$B_{\bar{U}}=B_{\bar{D}}$,  ensures that 
$x\bar{u} \rightarrow x\bar{d}$ as $x \rightarrow 0$.
For the central fit, the valence $B$ parameters, 
$B_{u_v}$ and $B_{d_v}$  are also set equal, 
but this assumption is dropped when parametrisation variations are considered.
The central values of the parameters are given in Table~\ref{tab:param}.
\newcolumntype{d}{D{.}{.}{3}}
\newcolumntype{e}{D{.}{.}{1}}
\begin{table}[tbp]
\centerline{
\begin{tabular}{|l|ddee|}
\hline
\multicolumn{1}{|c}{ } &
\multicolumn{1}{c}{~~~$A$~~~} & 
\multicolumn{1}{c}{~~~$B$~~~} &
\multicolumn{1}{c}{~~~$C$~~~} &
\multicolumn{1}{c|}{~~~$E$~~~}  \\
\hline
 $xg$      & 6.8   &  0.22  & 9.0 &     \\
$xu_v$     & 3.7   &  0.67  & 4.7 & 9.7 \\
$xd_v$     & 2.2   &  0.67  & 4.3 &     \\
$x\bar{U}$ & 0.113 & -0.165 & 2.6 &     \\
$x\bar{D}$ & 0.163 & -0.165 & 2.4 &     \\
\hline
\end{tabular}}
\caption{Central values of the HERAPDF1.0 parameters.}
\label{tab:param}\end{table}

Model uncertainties and parametrisation uncertainties of the central fit solution
are evaluated by varying the input assumptions. The variation of numerical values chosen for 
the central fit is specified in  
Table~\ref{tab:model}. 
\begin{table}[tbp]
\centerline{
\begin{tabular}{|l|c|c|c|}
\hline
\multicolumn{1}{|c|}{Variation} &
\multicolumn{1}{c|}{Standard Value} &
\multicolumn{1}{c|}{Lower Limit} &
\multicolumn{1}{c|}{Upper Limit}  \\
\hline
$f_s$           & $0.31$              & $0.23\phantom{^{(a)}}$ & $0.38\phantom{^{(c,d)}}$   \\
$m_c$ [GeV]     & $1.4\phantom{0}$    & $1.35^{(a)}$           & $1.65\phantom{^{(c,d)}}$   \\
$m_b$ [GeV]     & $4.75$              & $4.3\phantom{0^{(a)}}$ & $5.0\phantom{0^{(c,d)}}$  \\
$Q^2_{min}$ [GeV$^2$] & $3.5\phantom{0}$  & $2.5\phantom{0^{(a)}}$              & $5.0\phantom{0^{(c,d)}}$   \\
$Q^2_0$ [GeV$^2$] & $1.9\phantom{0}$  & $1.5^{(b)}\phantom{0}$ & $2.5^{(c,d)}\phantom{0}$  \\
\hline
\multicolumn{2}{c}{ } &
\multicolumn{1}{l}{\footnotesize $^{(a)} Q^2_0=1.8$} &
\multicolumn{1}{l}{\footnotesize $^{(c)} m_c=1.6$} \\
\multicolumn{2}{c}{ } &
\multicolumn{1}{l}{\footnotesize $^{(b)} f_s=0.29$} &
\multicolumn{1}{l}{\footnotesize $^{(d)} f_s=0.34$} \\
\end{tabular}}
\caption{Standard values of input parameters  and the variations 
considered.  
}
\label{tab:model}
\end{table}
The variation of $f_s$ is chosen to span the ranges determined in~\cite{Martin:2009iq,Nadolsky:2008zw}. 
The variations of $Q^2_0$ and $f_s$ are not independent, since QCD evolution 
will ensure that the strangeness fraction increases as $Q^2_0$ increases. 
The value $f_s=0.29$ is used for $Q^2_0=1.5~$GeV$^2$ and the value $f_s=0.34$ 
is used for $Q^2_0=2.5~$GeV$^2$ in order 
to be consistent with the choice $f_s=0.31$ at $Q^2_0=1.9~$GeV$^2$.
The variations of $Q^2_0$ and $m_c$ are also not independent, 
since $Q_0 < m_c$ is required in the fit programme. Thus when $m_c = 1.35$\,GeV, 
the starting scale used is 
$Q^2_0=1.8$\,GeV$^2$. Similarly, when $Q^2_0 = 2.5~$GeV$^2$ the 
charm mass used is $m_c=1.6~$GeV. In practice, the variations of $f_s$, $m_c$,
$m_b$, mostly affect the model uncertainty of the $x\bar{s}$, $x\bar{c}$, 
$x\bar{b}$, quark distributions, respectively, and have little effect on other parton 
flavours. 
The difference between the central fit and the fits corresponding to model variations of $m_c$, 
$m_b$, $f_s$, $Q^2_{min}$ are 
added in quadrature, separately for positive and negative deviations, and 
represent the 
model uncertainty of the HERAPDF1.0 set. 

The variation in $Q^2_0$ is regarded as a parametrisation uncertainty, 
rather than a model uncertainty.
At the starting scale the gluon parametrisation is valence like.
For the downward variation of the starting scale, $Q^2_0 = 1.5~$GeV$^2$, 
a parametrisation 
which explicitly allows for a negative gluon contribution at low $x$ 
is considered:
a term of the form $A_g'x^{B_g'}(1-x)^{C_g'}$ is subtracted 
from the gluon of the standard parametrisation, where $C_g' = 25$ is 
fixed\footnote{The fit is not sensitive to this value provided it is high 
enough ($C_g' > 15$) that the term does not contribute at large $x$.}
and $A_g'$ and $B_g'$ are fitted. 
The variations of $Q^2_0$ mostly increase the PDF uncertainties of 
the sea and gluon  at small $x$. A further variation of parametrisation, 
allowing $B_{u_v}\not= B_{d_v}$,  increases the 
uncertainties on the valence quarks at low $x$. Finally, variation of 
the number of terms in
the polynomial $(1 + \epsilon\surd x + D x + E x^2)$ is considered for each 
fitted parton distribution. All the 11 parameter fits 
which have $E_{u_v} \not= 0$ and one more $\epsilon$, 
$D$ or $E$ parameter non-zero are 
considered as possible variants\footnote{The criteria that all PDFs should be 
positive or that $xd_v > x\bar{d}$ at high $x$ are not imposed on these 
variants.}. In practice only a small number of them 
have significantly different PDF shapes from the central fit, 
notably: $D_{u_v}\not= 0$, $D_{\bar{U}}\not= 0$ and $D_{\bar{D}}\not=0$. 
These variations mostly increase the PDF uncertainty at high $x$, 
but the valence PDFs at low $x$ are also affected 
because of the constraint of the quark number sum rules. 
The difference between all these parametrisation 
variations and the central fit is stored 
and an envelope representing the maximal deviation at each $x$ value
is constructed  
to represent the parametrisation uncertainty. 
This parametrisation uncertainty should be regarded as indicative
rather than exhaustive, for example the reduction on the
  high-$x$ gluon uncertainty at $x \approx 0.5$ is due to restrictions
  of the parameterisations rather than a strong data constraint.
The total PDF uncertainty is obtained by adding in quadrature experimental, model and parametrisation 
uncertainties.

In summary, the HERAPDF1.0 analysis has two notable features.
Firstly, 
it uses a consistent data set with small correlated 
systematic uncertainties and applies the conventional $\chi^2$ tolerance, 
$\Delta\chi^2=1$, when determining the experimental 
uncertainties on the PDFs. This data set includes four different processes,
NC and CC, $e^+p$ and $e^-p$ scattering, such that there is sufficient 
information to extract $x\bar{U}, x\bar{D}, xd_v, xu_v$ PDFs 
and the gluon PDF from the scaling violations. 
Secondly, this analysis makes an assessment of uncertainties introduced both 
by model assumptions and by assumptions about the form of the parametrisation.

\subsection{Fit Results}
\label{ssec:results} 

\begin{figure}[tbp]
\vspace{-0.3cm} 
\centerline{
\epsfig{figure=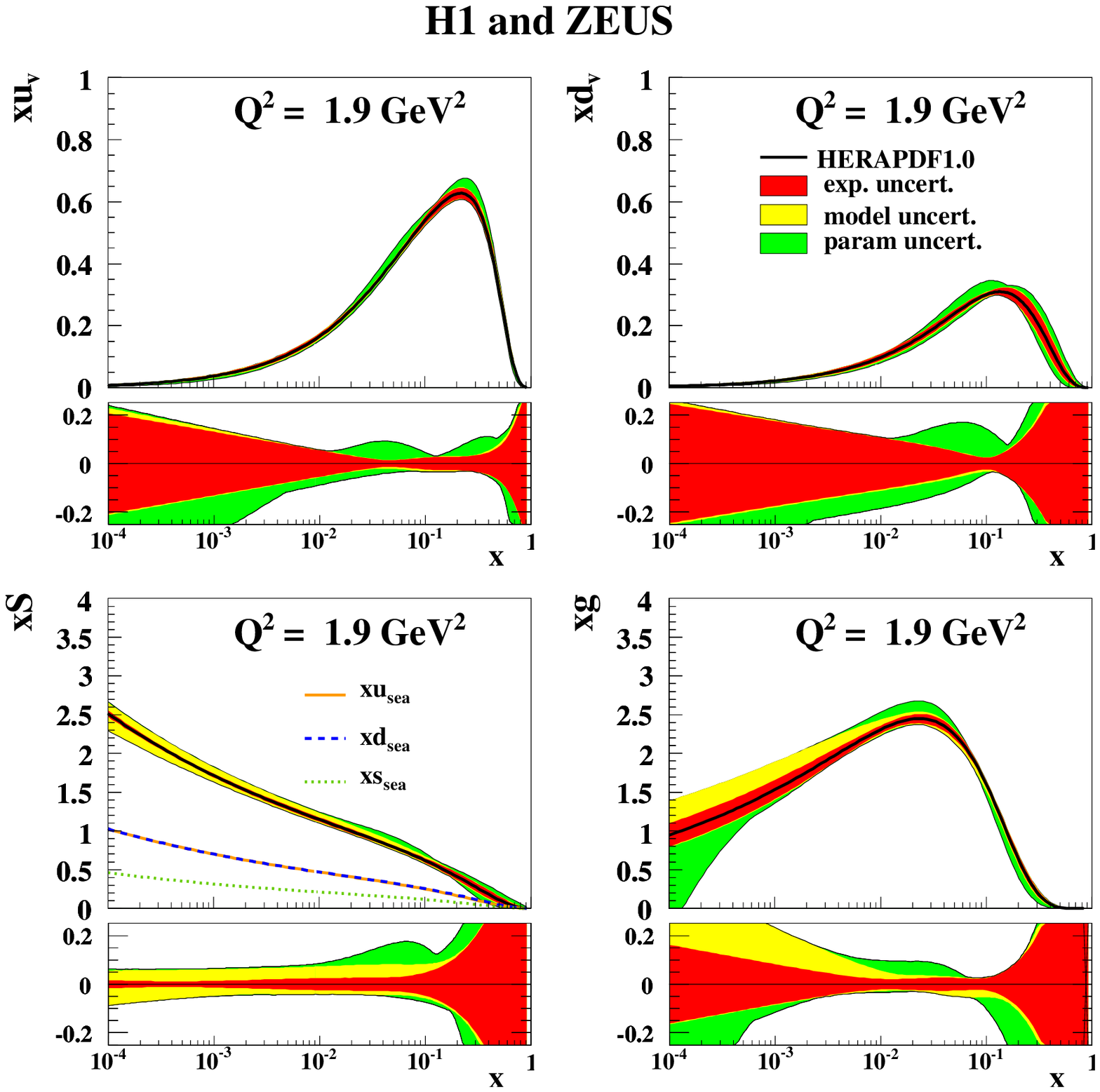  ,width=0.9\textwidth}}
\caption {
The parton distribution functions from 
HERAPDF1.0, $xu_v,xd_v,xS=2x(\bar{U}+\bar{D}),xg$, at 
$Q^2=1.9~$GeV$^2$. 
The break-up of the Sea PDF, $xS$, into the flavours,
$xu_{sea}=2x\bar{u}$, $xd_{sea}=2x\bar{d}$, $xs_{sea}=2x\bar{s}$, 
is illustrated.
Fractional uncertainty bands are shown 
below each PDF. 
The experimental, model and parametrisation 
uncertainties are shown 
separately. 
}
\label{fig:pdfs2}
\end{figure}

\begin{figure}[tbp]
\vspace{-0.3cm} 
\centerline{
\epsfig{figure=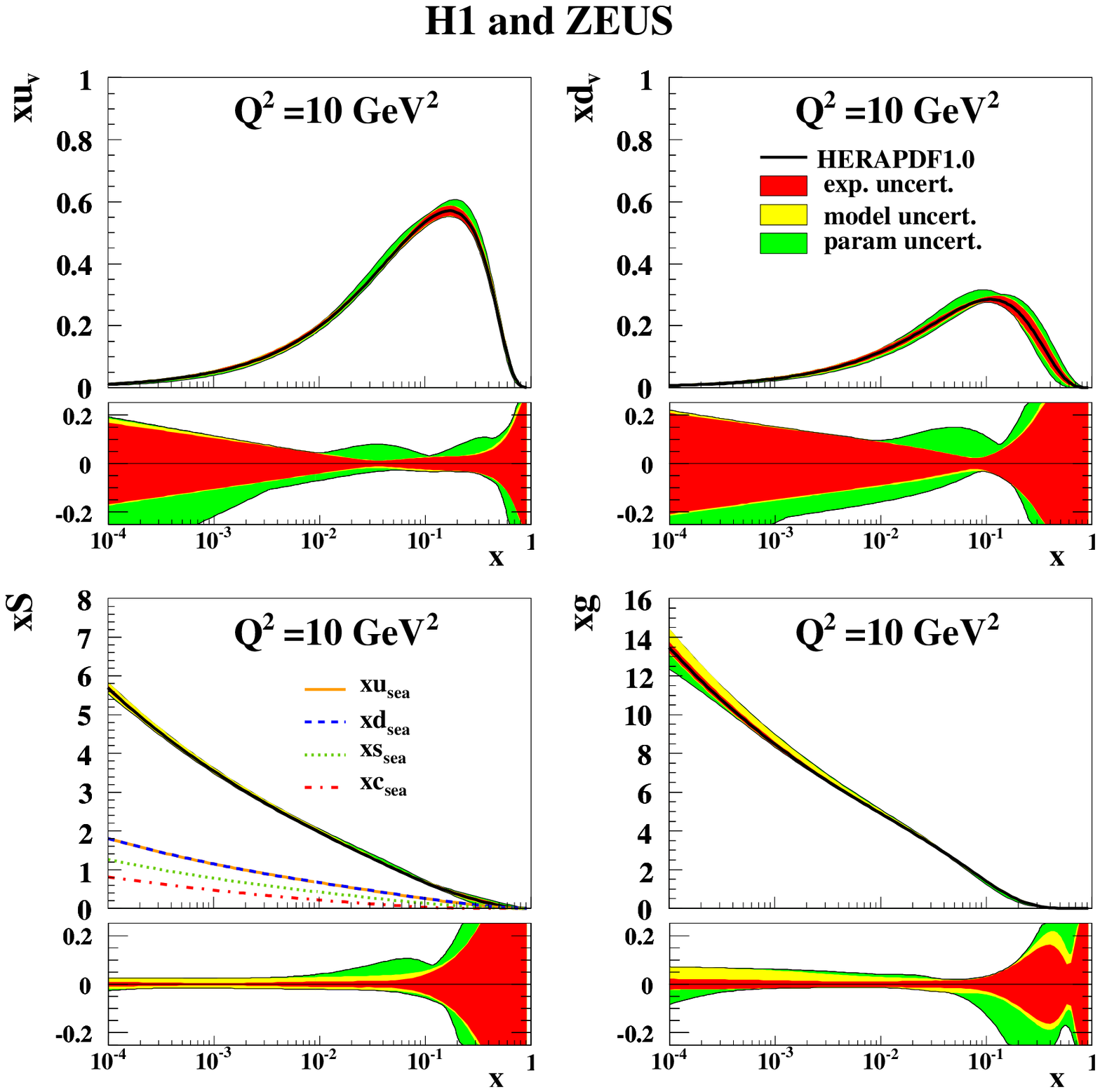  ,width=0.9\textwidth}}
\caption {
The parton distribution functions from 
HERAPDF1.0, $xu_v,xd_v,xS=2x(\bar{U}+\bar{D}),xg$, at 
$Q^2=10~$GeV$^2$. The break-up of the Sea PDF, $xS$, into the flavours,
$xu_{sea}=2x\bar{u}$, $xd_{sea}=2x\bar{d}$, $xs_{sea}=2x\bar{s}$, 
$xc_{sea}=2x\bar{c}$ is illustrated. Fractional uncertainty bands are shown 
below each PDF. 
The experimental, model and parametrisation 
uncertainties are shown 
separately. 
\label{fig:pdfs10}}
\end{figure}

\begin{figure}[tbp]
\vspace{-0.3cm} 
\centerline{
\epsfig{figure=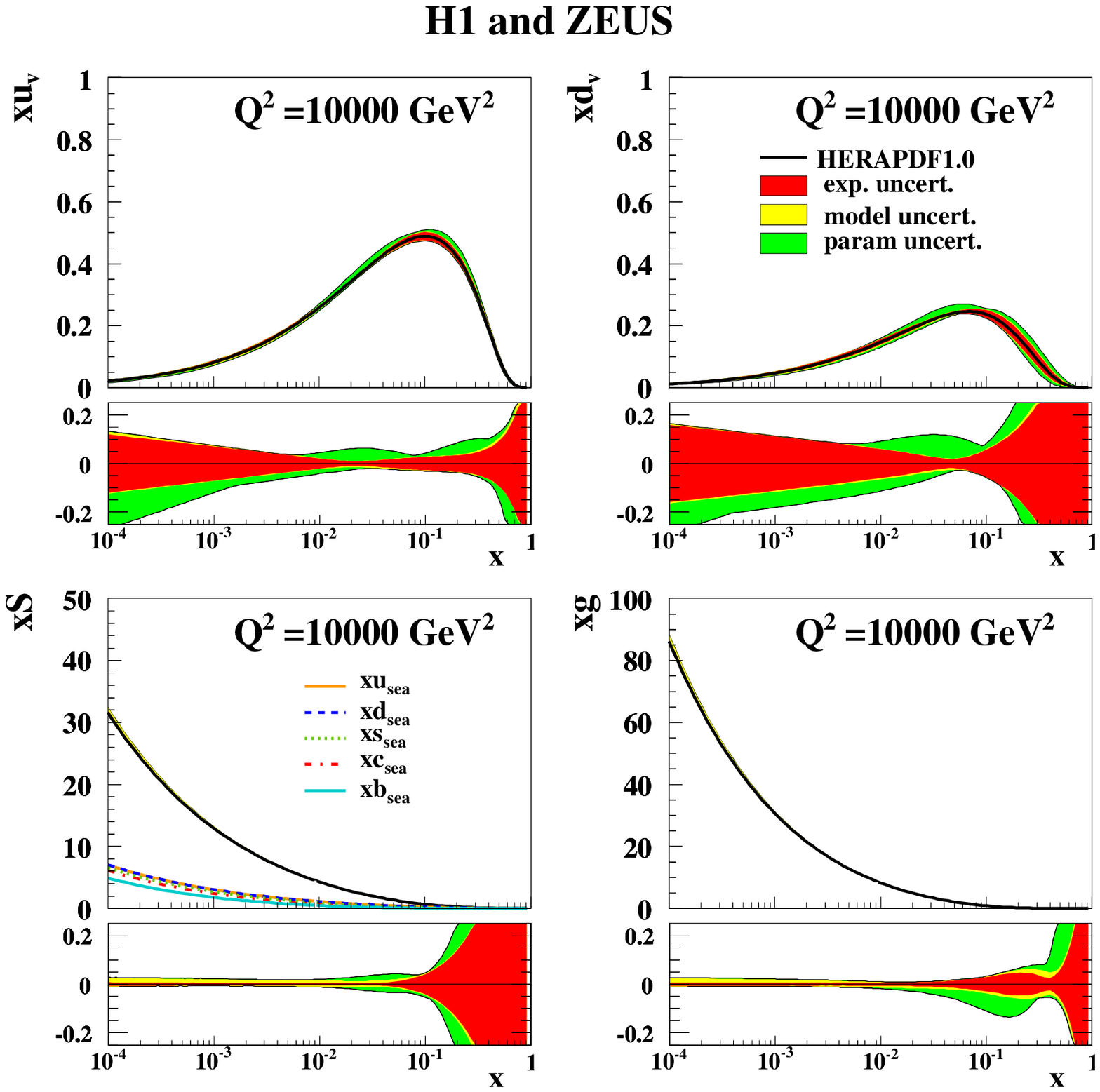  ,width=0.9\textwidth}}
\caption {
The parton distribution functions from 
HERAPDF1.0, $xu_v,xd_v,xS=2x(\bar{U}+\bar{D}),xg$, at 
$Q^2=10,000~$GeV$^2$. The break-up of the Sea PDF, $xS$, into the flavours,
$xu_{sea}=2x\bar{u}$, $xd_{sea}=2x\bar{d}$, $xs_{sea}=2x\bar{s}$, 
$xc_{sea}=2x\bar{c}$, $xb_{sea}=2x\bar{b}$ is illustrated. 
Fractional uncertainty bands are shown 
below each PDF. The experimental, model and parametrisation 
uncertainties are shown 
separately.}
\label{fig:pdfs10000}
\end{figure}

\FFigs~\ref{fig:pdfs2}-\ref{fig:pdfs10000} show the HERAPDF1.0 distributions, 
 $xu_v,xd_v,xS,xg$, as a function of $x$ at 
the starting scale $Q^2=1.9$\,GeV$^2$ and at $Q^2=10$ and $10000$~GeV$^2$,
where $xS =2x (\bar{U} + \bar{D})$ is the sea PDF. Note that for $Q^2 > m_c^2$,  
$x\bar{U} = x\bar{u} + x\bar{c}$, and for $Q^2 > m_b^2$, 
$x\bar{D} = x\bar{d} +x\bar{s} +x\bar{b}$, so that the heavy quarks are 
included in the sea distributions. The break-up of $xS$ into the flavours
$xu_{sea}=2x\bar{u}$, $xd_{sea}=2x\bar{d}$, $xs_{sea}=2x\bar{s}$, 
$xc_{sea}=2x\bar{c}$, $xb_{sea}=2x\bar{b}$ is illustrated so that the relative 
importance of each flavour at different $Q^2$ may be assessed. 
Fractional uncertainty bands are shown below each PDF. The experimental, 
model and parametrisation uncertainties are shown separately. 
The model and parametrisation uncertainties are asymmetric. For the sea 
and gluon distributions, the variations in parametrisation which have 
non-zero $\epsilon$, $D$ and $E$ affect the large-$x$ 
region, and the uncertainties arising from the 
variation of $Q^2_0$ and $Q^2_{min}$ affect the small-$x$ 
region. In particular the parametrisation variation which includes a negative
gluon term increases the uncertainty on the gluon at low $x$ and low $Q^2$. 
However the gluon distribution itself is not negative in the fitted 
kinematic region.  
For the valence distributions the non-zero $\epsilon$, $D$ and $E$ 
parametrisation uncertainty is important 
for all $x$, and is their dominant uncertainty. 
The total uncertainties at low $x$ decrease with increasing $Q^2$ due to QCD evolution and reach, for instance,
$2\%$ uncertainty on $xg$ at $Q^2=10000$~GeV$^2$ and $x<0.01$.

\begin{figure}[tbp]
\vspace{-0.3cm} 
\centerline{
\epsfig{figure=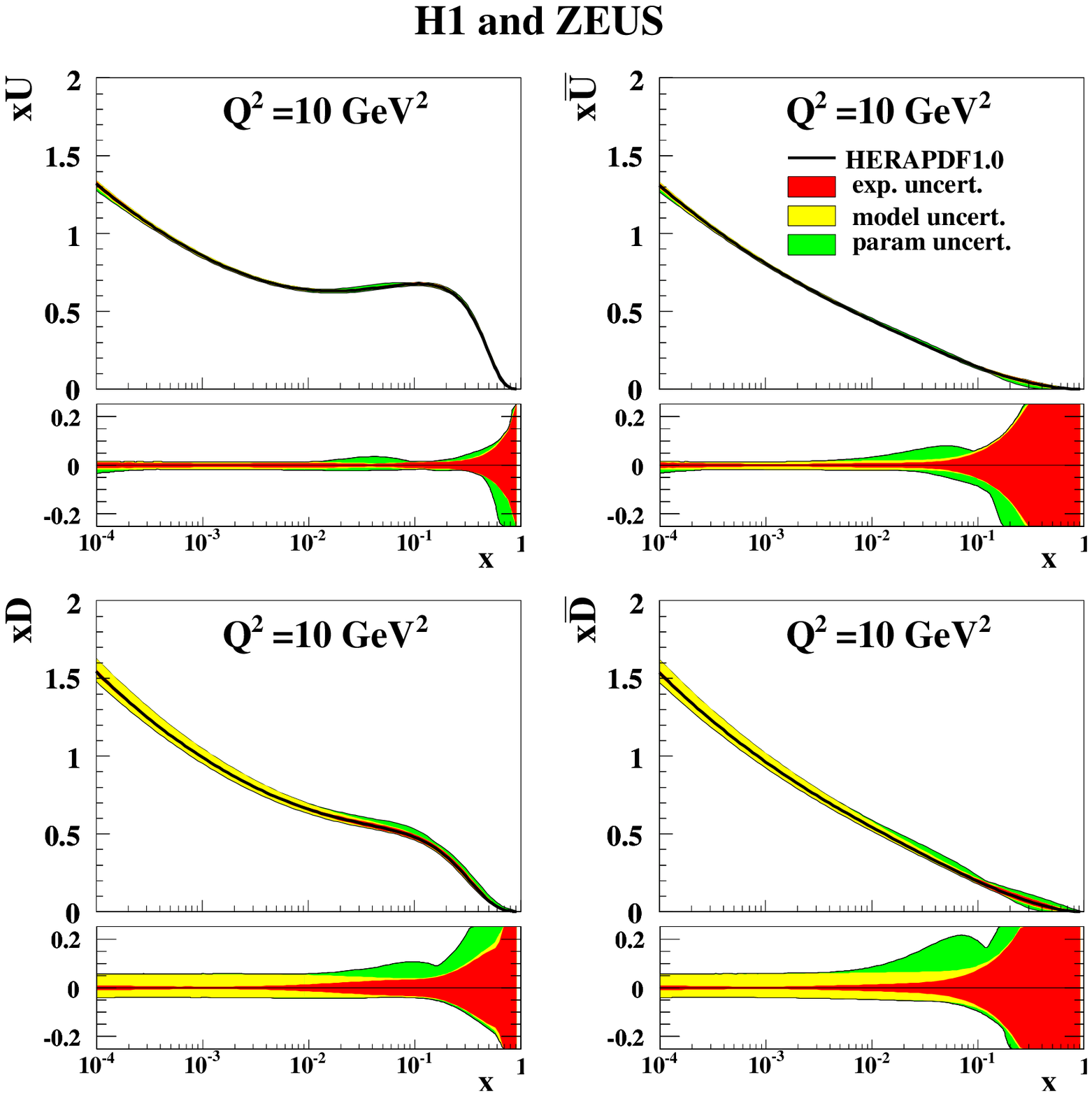  ,width=0.9\textwidth}}
\vspace{0.5cm}
\caption {
The parton distribution functions from 
HERAPDF1.0, $xU, xD, x\bar{U}, x\bar{D}$ at $Q^2=10~$GeV$^2$.
Fractional uncertainty bands are shown 
below each PDF. The experimental, model and parametrisation 
uncertainties are shown 
separately.}
\label{fig:pdfsUD}
\end{figure}

\begin{figure}[tbp]
\vspace{-0.3cm} 
\centerline{
\epsfig{figure=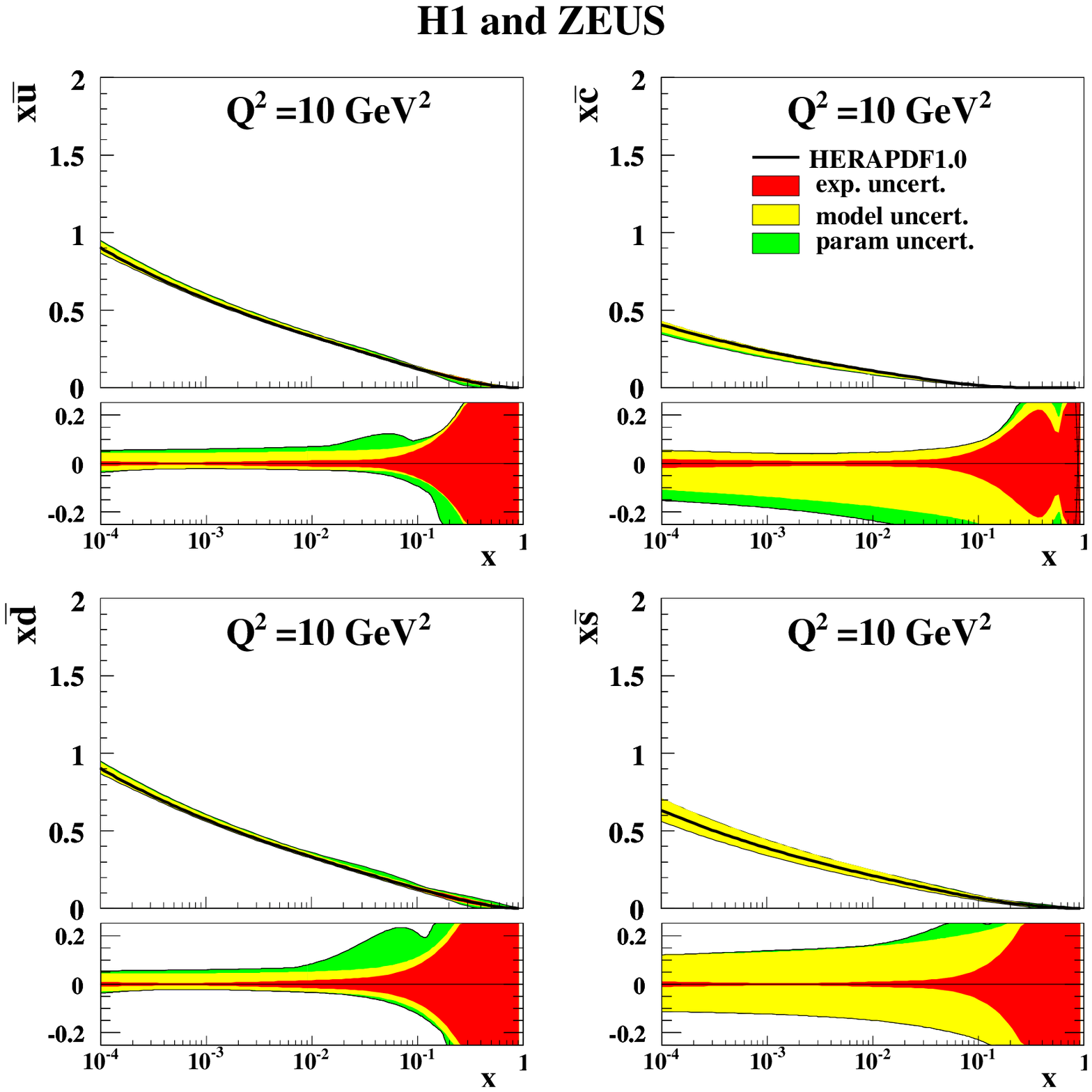  ,width=0.9\textwidth}}
\vspace{0.5cm}
\caption {
The parton distribution functions from 
HERAPDF1.0, $x\bar{u}, x\bar{d}, x\bar{c}, x\bar{s}$ at 
$Q^2=10~$GeV$^2$.
Fractional uncertainty bands are shown 
below each PDF. The experimental, model and parametrisation 
uncertainties are shown 
separately.}
\label{fig:pdfsudcs}
\end{figure}

The break-up of the PDFs into different flavours is further illustrated in 
\Figs~\ref{fig:pdfsUD} and \ref{fig:pdfsudcs}, where the quark distributions 
$xU, xD, x\bar{U}, x\bar{D}$ and 
$x\bar{u}, x\bar{d}, x\bar{c}, x\bar{s}$ are shown at $Q^2=10$~GeV$^2$. 
The PDFs $xU, xD, x\bar{U}, x\bar{D}$ are closely related to the measurements, 
see equations~\ref{ncfu},~\ref{ccupdo}, and are very well constrained at 
low $x$. The $U$ flavour is better constrained than the $D$ flavour because of 
the dominance of this flavour in all interactions except $e^+p$ CC scattering.
The quark distributions $x\bar{d}, x\bar{s}$ are derived from $x\bar{D}$ 
through the assumption on the value of $f_s$, and the uncertainty on $x\bar{s}$
directly reflects the uncertainty on this fraction. The uncertainty on 
$x\bar{d}$ follows closely that on $x\bar{D}$ and the uncertainty on 
$x\bar{u}$ follows closely that on $x\bar{U}$.  
The charm PDF, $x\bar{c}$, is shown at $Q^2=10$~GeV$^2$, so that the 
condition $Q^2 \gg m_c^2$ is fulfilled and it may be regarded as a 
fully active parton. However, it is strongly related to the 
gluon density such that it is affected by the same variations which 
affect the gluon PDF (variation of $Q^2_0$ and
 $Q^2_{min}$) as well as by the variation of $m_c$. 
The dominant uncertainty on the charm PDF comes from 
the variation $Q^2_0 = 2.5~$GeV$^2$.  The uncertainty on the
bottom PDF, $x\bar{b}$ (not shown), is dominated by the variation of $m_b$.

\begin{figure}[tbp]
\vspace{-0.5cm} 
\centerline{
\epsfig{figure=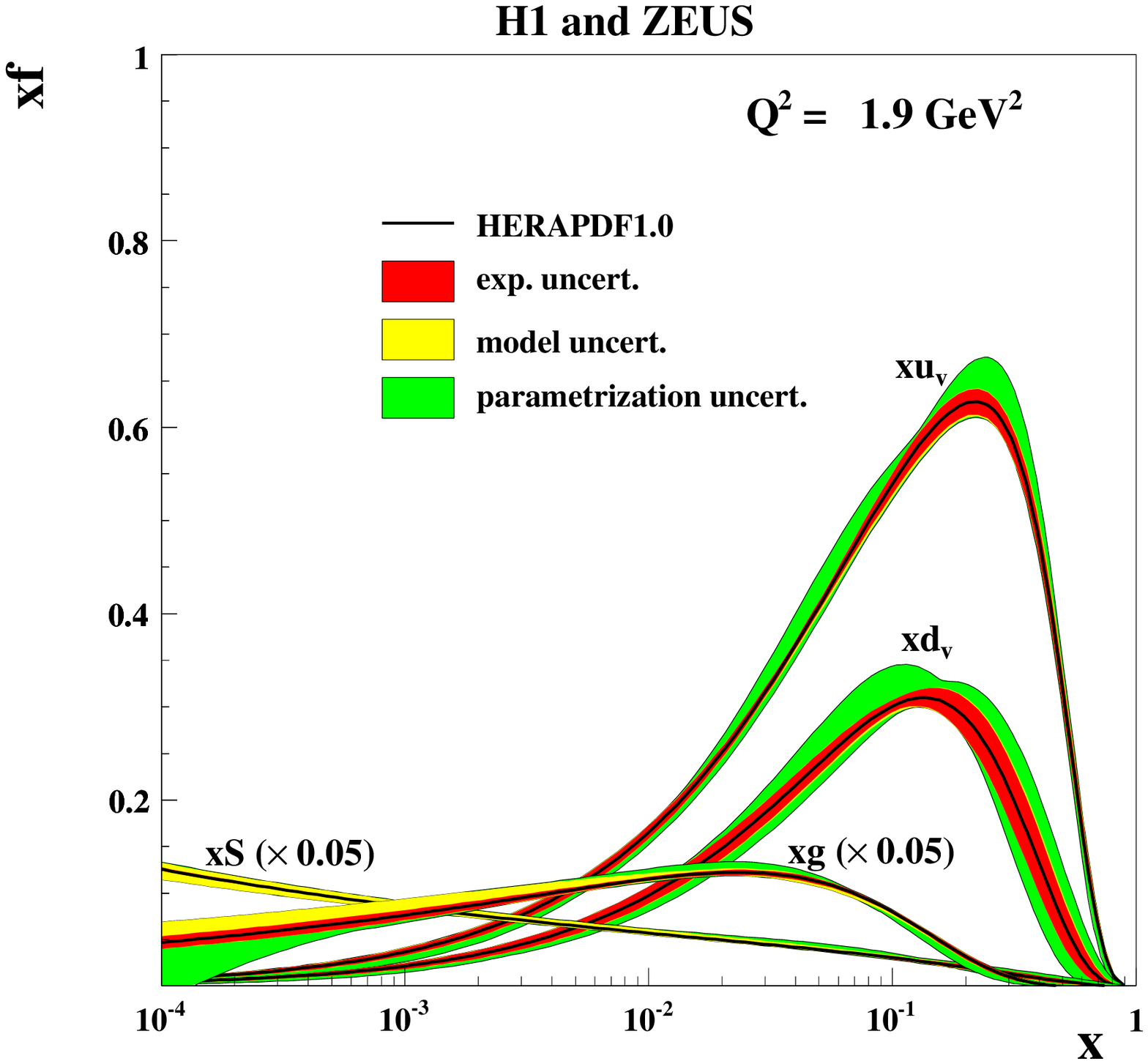 ,width=0.7\textwidth}}
\vspace*{-0.6cm}
\centerline{
\epsfig{figure=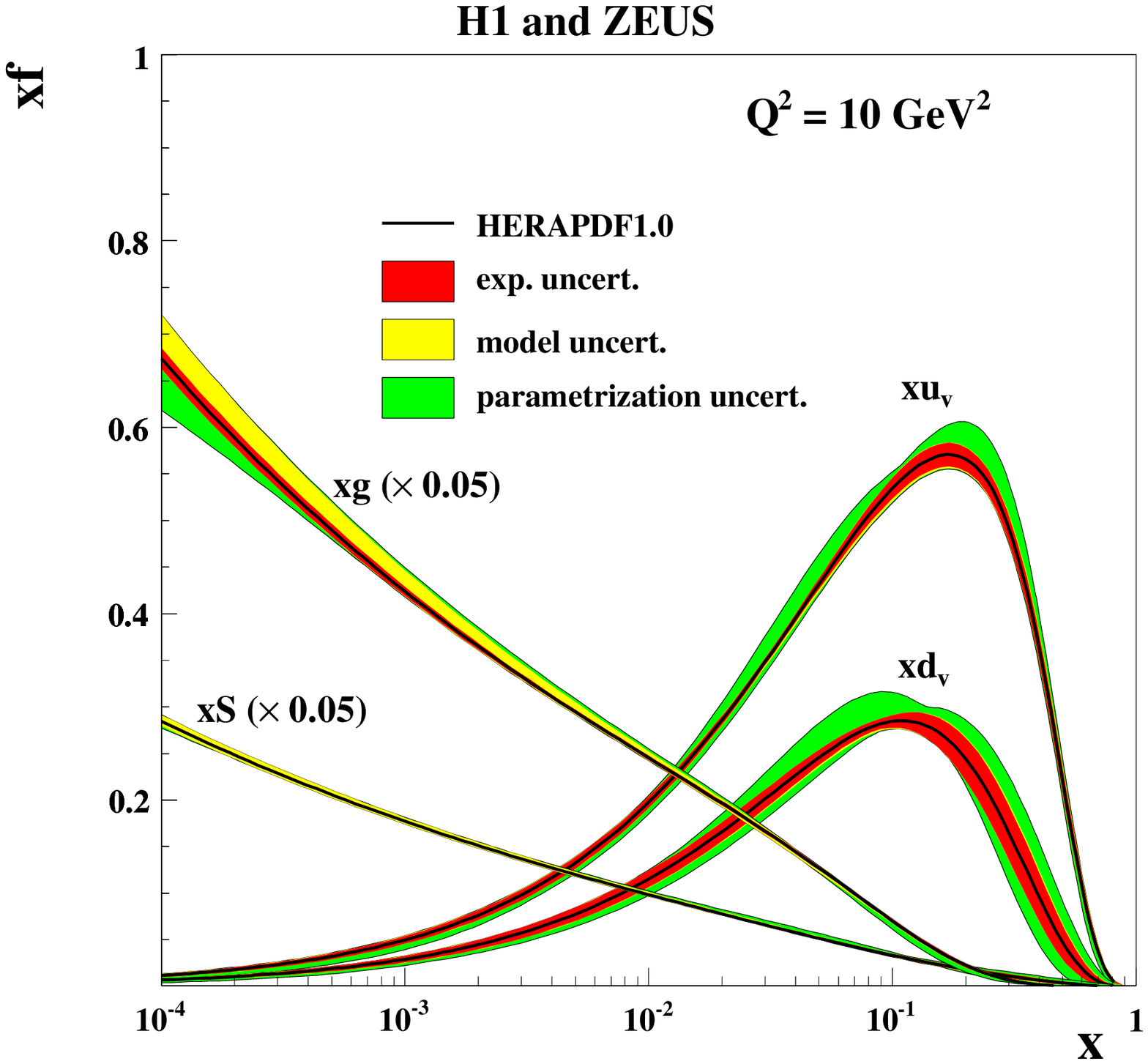 ,width=0.7\textwidth}}
\caption { 
The parton distribution functions from 
HERAPDF1.0, $xu_v,xd_v,xS=2x(\bar{U}+\bar{D}),xg$, at
$Q^2 = 1.9$~GeV$^2$ (top) and $Q^2 = 10$~GeV$^2$ (bottom).
The gluon and sea distributions are scaled down by a factor $20$.
The experimental, model and parametrisation 
uncertainties are shown 
separately.
}
\label{fig:summary}
\end{figure}

\begin{figure}[tbp]
\vspace{-0.5cm} 
\centerline{
\epsfig{figure=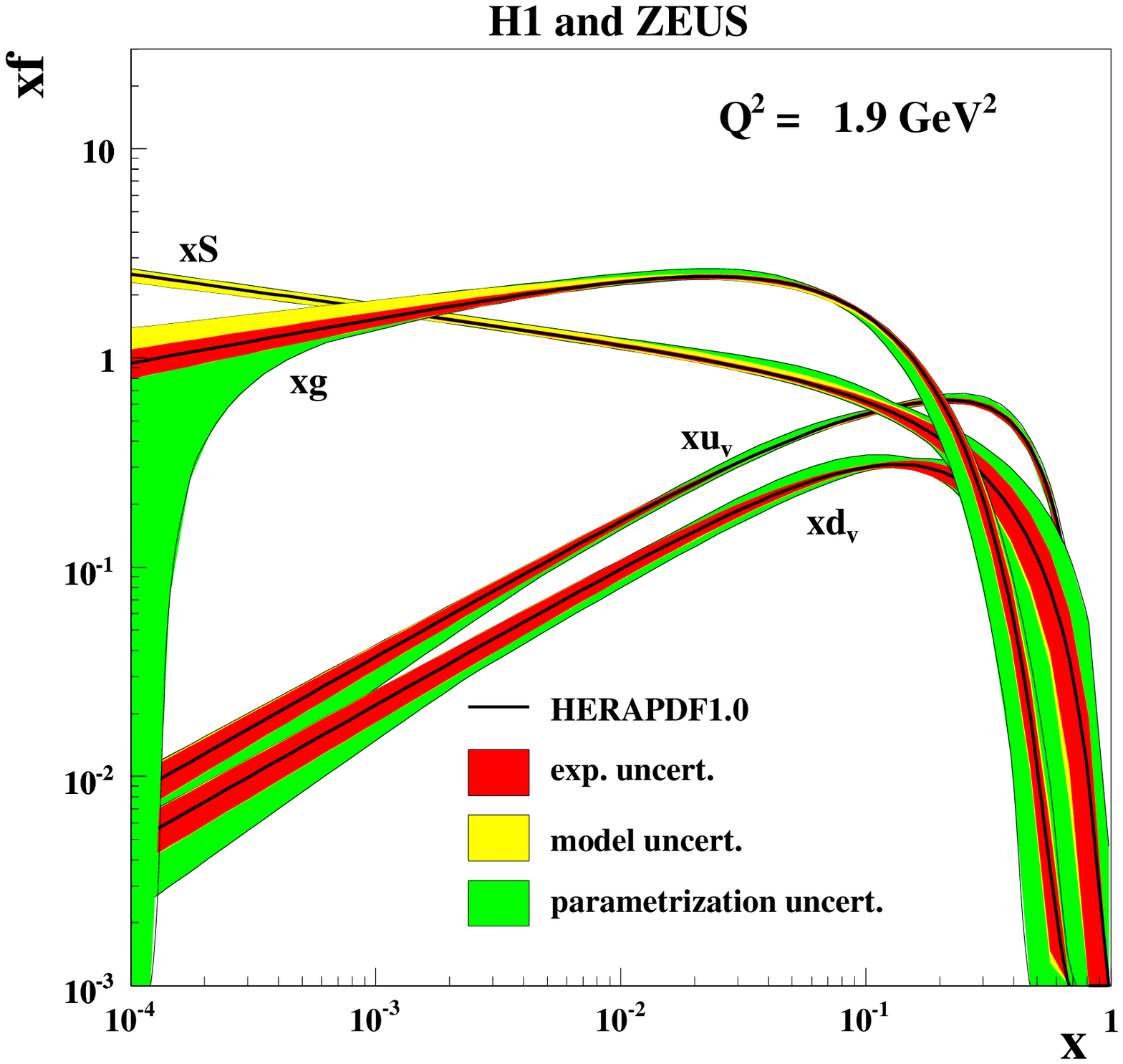 ,width=0.7\textwidth}}
\vspace*{-0.6cm}
\centerline{
\epsfig{figure=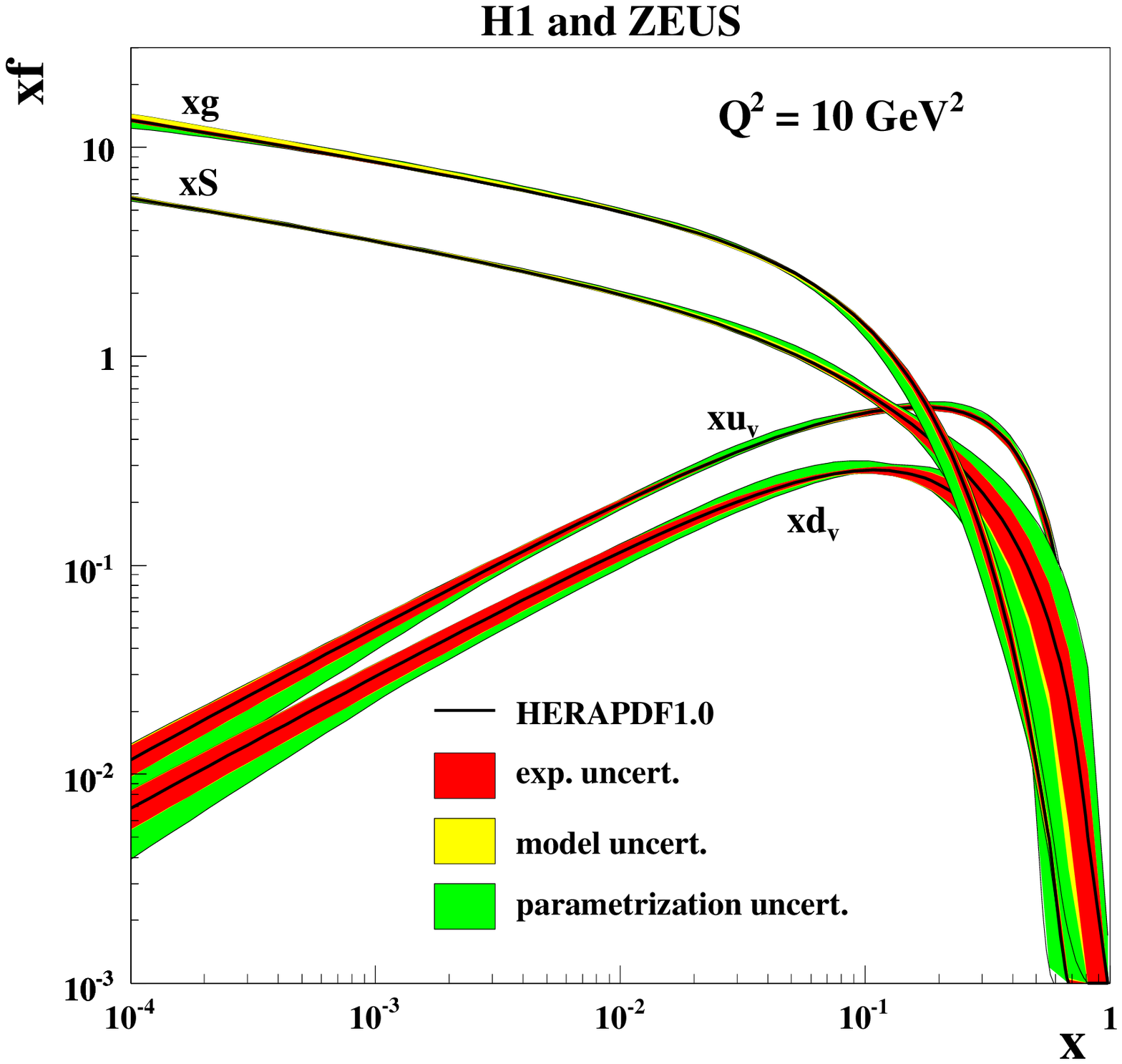 ,width=0.7\textwidth}}
\caption { 
The parton distribution functions from 
HERAPDF1.0, $xu_v,xd_v,xS=2x(\bar{U}+\bar{D}),xg$, at
$Q^2 = 1.9$~GeV$^2$ (top) and $Q^2 = 10$~GeV$^2$ (bottom).
The experimental, model and parametrisation 
uncertainties are shown 
separately.
}
\label{fig:summarylog}
\end{figure}

\FFig~\ref{fig:summary} shows summary plots of the HERAPDF1.0 distributions at the
starting scale $Q^2=1.9~$GeV$^2$ and at $Q^2=10~$GeV$^2$. 
\FFig~\ref{fig:summarylog} shows such plots on a logarithmic scale such that the 
behaviour and the uncertainties of the PDFs at high $x$ are emphasised.

\begin{figure}[tbp]
\vspace{-0.2cm} 
\centerline{
\epsfig{figure=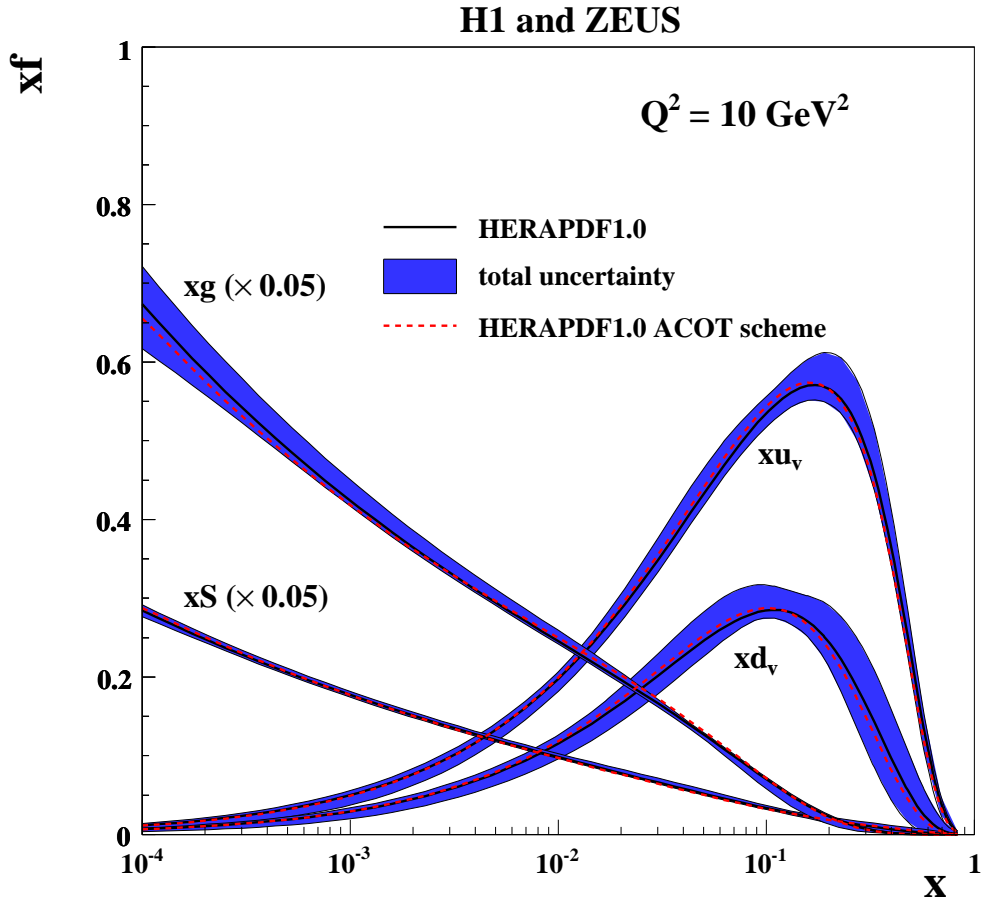  ,width=0.9\textwidth}}
\caption {
The parton distribution functions obtained using ACOT heavy-flavour scheme
compared to  
HERAPDF1.0 for $xu_v,xd_v,xS=2x(\bar{U}+\bar{D}),xg$, at $Q^2=10$~GeV$^2$. 
The bands show total uncertainties of the HERAPDF1.0 fit.
}
\label{fig:acot}
\end{figure}

A cross check of the sensitivity to the 
heavy-flavour treatment is made by performing a fit
 using an alternative general-mass variable-flavour-number scheme, the 
ACOT scheme~\cite{Kramer:2000hn}.
The result of this fit is shown for $Q^2=10$~GeV$^2$ in \Fig~\ref{fig:acot}. 
The PDFs determined using the
ACOT scheme are within the total uncertainty of the standard fit.

\begin{figure}[tbp]
\vspace{-0.2cm} 
\centerline{
\epsfig{figure=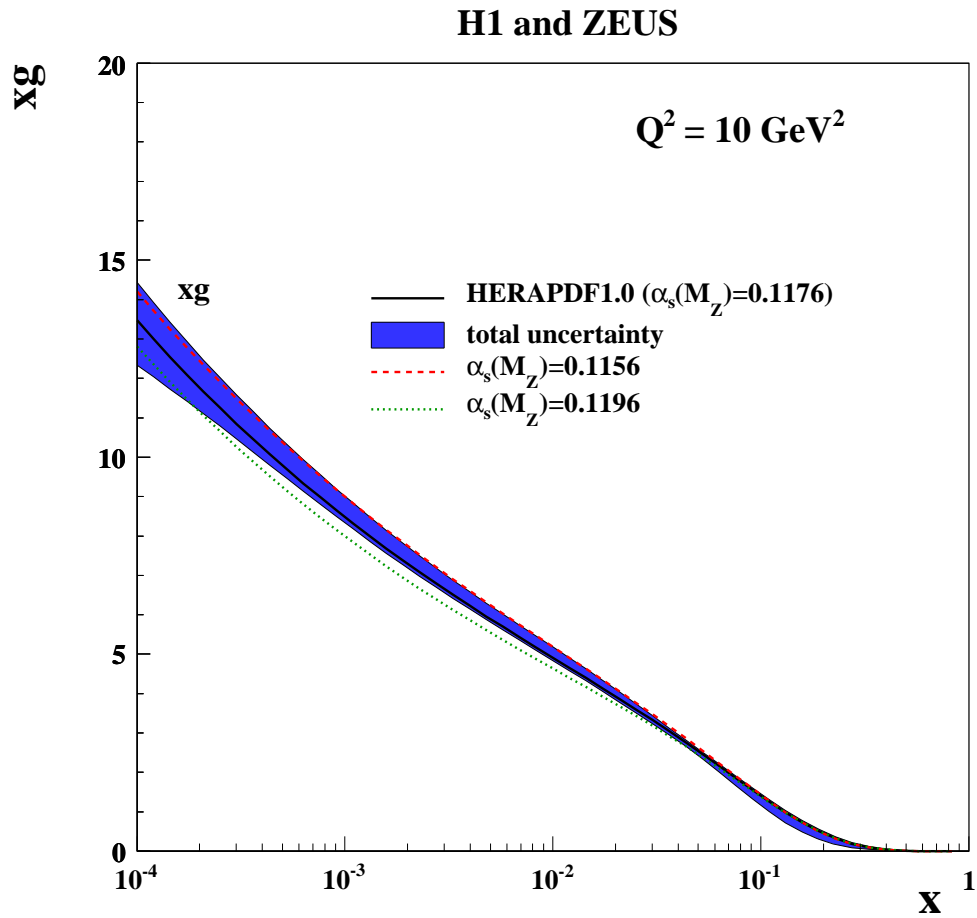  ,width=0.9\textwidth}}
\caption {
The gluon density $xg$ from HERAPDF1.0 compared for different values of $\alpha_S$.
The band shows total uncertainties of the HERAPDF1.0 fit.
}
\label{fig:alf}
\end{figure}

The sensitivity of the HERAPDF1.0 fit to the value of $\alpha_S$ is estimated by varying $\alpha_S$ by 
$\pm 0.002$~\cite{PDG}.
Only the gluon PDF is significantly affected.
The change  of the gluon PDF is 
outside the total uncertainty bands of the standard fit, as illustrated 
in \Fig~\ref{fig:alf}.

\section{Conclusions \label{sec:sum}}
Inclusive cross sections of neutral and charged current $e^{\pm}p$
scattering measured by the H1 and ZEUS Collaborations are combined.
The kinematic range of the NC data extends from $6 \cdot 10^{-7}$
to $0.65$ in Bjorken $x$ for $0.045 \leq Q^2 \leq 30000$\,GeV$^2$ and
inelasticity values $y$ between $0.005$ and $0.95$. The CC data cover
a range $ 0.013 \leq x \leq 0.4$ and $300 \leq Q^2 \leq 30000$\,GeV$^2$ for
$y$ between $0.037$ and $0.76$. The combination comprises all inclusive data
from the first period of the operation of HERA. 
The input data from H1 and ZEUS are consistent with each other at $\chi^2/n_{\rm dof} = 636.5/656$.
The total uncertainty of the combined data set reaches $1\%$ 
for NC scattering in the best measured region, $20 < Q^2 < 100$ GeV$^2$.

A NLO QCD analysis is performed based exclusively on 
these combined $e^{\pm}p$ scattering cross-section data.
A new set of parton distribution functions, HERAPDF1.0,
is obtained using a variable-flavour-number scheme.
The analysis yields small experimental uncertainties and includes estimates for model 
and parametrisation uncertainties. Due to the precision of the combined data set,
 the parametrisation HERAPDF1.0  has total uncertainties at the level of a few percent at low $x$.

\section*{Acknowledgements}
\refstepcounter{pdfadd} \pdfbookmark[0]{Acknowledgements}{s:acknowledge}

We are grateful to the HERA machine group whose outstanding
efforts have made these experiments possible.
We appreciate the contributions to the construction and maintenance of the H1 and ZEUS detectors of many people who are not listed as authors.
We thank our funding agencies for financial 
support, the DESY technical staff for continuous assistance and the 
DESY directorate for their support and for the hospitality they extended to the non-DESY members of the collaborations.

\clearpage
\bibliography{desy09-158}
  
\clearpage
\begin{table}
\begin{scriptsize}
\begin{center}
\begin{tabular}{ll}
      Source                           &  Data Samples  \\
\hline
 H1 $E'_e$ & $\delta_{1}$ H1 NC\cite{Adloff:2000qj} --- $\delta_{1}$ H1 NC~HY\cite{Adloff:2003uh} --- $\delta_{1}$ H1 NC\cite{Adloff:1999ah} --- $\delta_{1}$ H1 NC\cite{Adloff:2003uh}  \\ 
 H1 $E_h$ & $\delta_{2}$ H1 CC\cite{Adloff:1999ah} --- $\delta_{2}$ H1 CC\cite{Adloff:2003uh} --- $\delta_{2}$ H1 CC\cite{Adloff:2000qj} --- $\delta_{3}$ H1 NC\cite{Adloff:2000qj} --- $\delta_{3}$ H1 NC~HY\cite{Adloff:2003uh} --- $\delta_{3}$ H1 NC\cite{Adloff:1999ah} --- $\delta_{3}$ H1 NC\cite{Adloff:2003uh}  \\ 
 H1 $\gamma p$ asymmetry & $\delta_{6}$ H1 NC~HY\cite{Adloff:2003uh} --- $\delta_{6}$ H1 NC\cite{Adloff:2003uh}  \\ 
 H1 $\gamma p$ background & $\delta_{4}$ H1 CC\cite{Adloff:1999ah} --- $\delta_{4}$ H1 CC\cite{Adloff:2003uh} --- $\delta_{4}$ H1 CC\cite{Adloff:2000qj} --- $\delta_{5}$ H1 NC\cite{Adloff:2000qj} --- $\delta_{5}$ H1 NC~HY\cite{Adloff:2003uh} --- $\delta_{5}$ H1 NC\cite{Adloff:1999ah} --- $\delta_{5}$ H1 NC\cite{Adloff:2003uh}  \\ 
 H1 $\theta_e$ & $\delta_{2}$ H1 NC\cite{Adloff:2000qj} --- $\delta_{2}$ H1 NC~HY\cite{Adloff:2003uh} --- $\delta_{2}$ H1 NC\cite{Adloff:2003uh}  \\ 
 H1 CC cuts & $\delta_{1}$ H1 CC\cite{Adloff:2003uh} --- $\delta_{1}$ H1 CC\cite{Adloff:2000qj}  \\ 
 H1 LAr Noise & $\delta_{3}$ H1 CC\cite{Adloff:1999ah} --- $\delta_{3}$ H1 CC\cite{Adloff:2003uh} --- $\delta_{3}$ H1 CC\cite{Adloff:2000qj} --- $\delta_{4}$ H1 NC\cite{Adloff:2000qj} --- $\delta_{4}$ H1 NC~HY\cite{Adloff:2003uh} --- $\delta_{4}$ H1 NC\cite{Adloff:1999ah} --- $\delta_{4}$ H1 NC\cite{Adloff:2003uh}  \\ 
 H1 Lumi $94-97$ & $\delta_{5}$ H1 CC\cite{Adloff:1999ah} --- $\delta_{6}$ H1 NC\cite{Adloff:1999ah}  \\ 
 H1 Lumi $98-99$ & $\delta_{5}$ H1 CC\cite{Adloff:2000qj} --- $\delta_{6}$ H1 NC\cite{Adloff:2000qj} --- $\delta_{7}$ H1 NC~HY\cite{Adloff:2003uh}  \\ 
 H1 Lumi $99-00$ & $\delta_{5}$ H1 CC\cite{Adloff:2003uh} --- $\delta_{7}$ H1 NC\cite{Adloff:2003uh}  \\ 
 ZEUS $E'_e$ & $\delta_{1}$ ZEUS NC\cite{Chekanov:2002ej} --- $\delta_{1}$ ZEUS NC\cite{Chekanov:2003yv}  \\ 
 ZEUS $E_h$ a & $\delta_{1}$ ZEUS CC\cite{Chekanov:2002zs} --- $\delta_{1}$ ZEUS CC\cite{Chekanov:2003vw}  \\ 
 ZEUS $E_h$ b & $\delta_{2}$ ZEUS CC\cite{Chekanov:2002zs} --- $\delta_{2}$ ZEUS CC\cite{Chekanov:2003vw}  \\ 
 ZEUS $E_h$ in BCAL & $\delta_{2}$ ZEUS CC\cite{zeuscc97} --- $\delta_{6}$ ZEUS NC\cite{Chekanov:2001qu}  \\ 
 ZEUS $E_h$ in FCAL & $\delta_{1}$ ZEUS CC\cite{zeuscc97} --- $\delta_{5}$ ZEUS NC\cite{Chekanov:2001qu}  \\ 
 ZEUS $\delta$ cut & $\delta_{8}$ ZEUS BPC\cite{Breitweg:1997hz} --- $\delta_{1}$ ZEUS BPT\cite{Breitweg:2000yn}  \\ 
 ZEUS $\gamma p$ background & $\delta_{2}$ ZEUS NC\cite{Chekanov:2002ej} --- $\delta_{2}$ ZEUS NC\cite{Chekanov:2003yv}  \\ 
 ZEUS $\gamma p$ background & $\delta_{9}$ ZEUS BPC\cite{Breitweg:1997hz} --- $\delta_{14}$ ZEUS BPT\cite{Breitweg:2000yn} --- $\delta_{8}$ ZEUS SVX\cite{Breitweg:1998dz}  \\ 
 ZEUS $y_h$ cut & $\delta_{3}$ ZEUS BPC\cite{Breitweg:1997hz} --- $\delta_{2}$ ZEUS BPT\cite{Breitweg:2000yn}  \\ 
 ZEUS BPC linearity & $\delta_{5}$ ZEUS BPC\cite{Breitweg:1997hz} --- $\delta_{9}$ ZEUS BPT\cite{Breitweg:2000yn}  \\ 
 ZEUS BPC shower & $\delta_{4}$ ZEUS BPC\cite{Breitweg:1997hz} --- $\delta_{3}$ ZEUS BPT\cite{Breitweg:2000yn}  \\ 
 ZEUS CAL energy & $\delta_{2}$ ZEUS BPC\cite{Breitweg:1997hz} --- $\delta_{12}$ ZEUS BPT\cite{Breitweg:2000yn} --- $\delta_{9}$ ZEUS SVX\cite{Breitweg:1998dz}  \\ 
 ZEUS Cuts$_1$ & $\delta_{3}$ ZEUS NC\cite{Chekanov:2002ej} --- $\delta_{3}$ ZEUS NC\cite{Chekanov:2003yv}  \\ 
 ZEUS Cuts$_2$ & $\delta_{4}$ ZEUS NC\cite{Chekanov:2002ej} --- $\delta_{4}$ ZEUS NC\cite{Chekanov:2003yv}  \\ 
 ZEUS HFS model & $\delta_{3}$ ZEUS CC\cite{zeuscc97} --- $\delta_{3}$ ZEUS CC\cite{Chekanov:2002zs} --- $\delta_{6}$ ZEUS NC\cite{Chekanov:2002ej} --- $\delta_{6}$ ZEUS NC\cite{Chekanov:2003yv} --- $\delta_{3}$ ZEUS CC\cite{Chekanov:2003vw}  \\ 
 ZEUS Lumi $94-97$ & $\delta_{4}$ ZEUS CC\cite{zeuscc97} --- $\delta_{11}$ ZEUS NC\cite{Chekanov:2001qu}  \\ 
 ZEUS Lumi $98-99$ & $\delta_{4}$ ZEUS CC\cite{Chekanov:2002zs} --- $\delta_{7}$ ZEUS NC\cite{Chekanov:2002ej}  \\ 
 ZEUS Lumi $99-00$ & $\delta_{9}$ ZEUS NC\cite{Chekanov:2003yv} --- $\delta_{4}$ ZEUS CC\cite{Chekanov:2003vw}  \\ 

\hline
\end{tabular}
\end{center}
\end{scriptsize}
\caption{
\label{tab:corel}
List of systematic sources that are correlated across the data samples.
The type of the systematic uncertainty is given in the ``source'' column.
The labels $\delta_i$ denote 
the sources according to the sequential ordering in the list of the correlated systematic 
uncertainties of the  corresponding publication. An overall $0.5\%$ normalisation uncertainty, common to all
data sets, is not included in this list.}
\end{table}

\begin{table}
\begin{tiny}
\begin{center}
\begin{tabular}{l|cccccccccccccc}
\hline
\hline
Bin &  $Q^2$  & $x$ & $y$ &  $\sigma^{+~\rm ave}_{r}$ & $F_2^{\rm ave}$ 
 & $\delta_{\rm ave,stat}$ & $\delta_{\rm ave,uncor}$ & $\delta_{\rm ave,cor}$ & $\delta_{\rm ave,exp~tot}$
 & $\delta_{\rm ave,rel}$ & $\delta_{\rm ave,gp}$ & $\delta_{\rm ave,had}$ & $\delta_{\rm ave,tot}$ 
 & $\sqrt{s}$\\
$\#$     & GeV$^2$ &   &       &       &         &       \%           & \%    & \%  &    
\%              &   \% & \% & \% &\% & GeV \\
\hline

$    1 $ &$  0.045 $ &$ 0.621 \times 10^{-6}$ & $ 0.803 $ &$ 0.080 $ & ---  & $  5.05 $ &$  1.50 $ &$  7.29 $ &$ 8.99$ & $ 2.83$ & $ 2.28$ & $ 1.14$ & $ 9.77$ & $   301$ \\
$    2 $ &$  0.065 $ &$ 0.897 \times 10^{-6}$ & $ 0.803 $ &$ 0.111 $ & ---  & $  2.97 $ &$  1.50 $ &$  6.27 $ &$ 7.10$ & $ 2.78$ & $ 2.11$ & $ 1.21$ & $ 8.00$ & $   301$ \\
$    3 $ &$  0.065 $ &$ 0.102 \times 10^{-5}$ & $ 0.706 $ &$ 0.111 $ & ---  & $  4.69 $ &$  1.50 $ &$  5.98 $ &$ 7.75$ & $ 3.56$ & $ 1.16$ & $ 1.47$ & $ 8.73$ & $   301$ \\
$    4 $ &$  0.085 $ &$ 0.117 \times 10^{-5}$ & $ 0.805 $ &$ 0.136 $ & ---  & $  2.95 $ &$  1.50 $ &$  5.54 $ &$ 6.45$ & $ 2.57$ & $ 2.12$ & $ 1.37$ & $ 7.39$ & $   301$ \\
$    5 $ &$  0.085 $ &$ 0.134 \times 10^{-5}$ & $ 0.703 $ &$ 0.130 $ & ---  & $  2.27 $ &$  1.50 $ &$  4.01 $ &$ 4.85$ & $ 1.80$ & $ 1.57$ & $ 1.00$ & $ 5.49$ & $   301$ \\
$    6 $ &$  0.085 $ &$ 0.156 \times 10^{-5}$ & $ 0.604 $ &$ 0.126 $ & ---  & $  4.78 $ &$  1.50 $ &$  4.52 $ &$ 6.75$ & $ 1.03$ & $ 0.76$ & $ 0.24$ & $ 6.87$ & $   301$ \\
$    7 $ &$    0.1 $ &$ 0.151 \times 10^{-5}$ & $ 0.808 $ &$ 0.165 $ & ---  & $  3.48 $ &$  1.50 $ &$  5.49 $ &$ 6.67$ & $ 3.16$ & $ 1.97$ & $ 1.73$ & $ 7.83$ & $   301$ \\
$    8 $ &$    0.1 $ &$ 0.173 \times 10^{-5}$ & $ 0.705 $ &$ 0.166 $ & ---  & $  1.81 $ &$  0.99 $ &$  3.25 $ &$ 3.85$ & $ 2.28$ & $ 1.33$ & $ 0.99$ & $ 4.77$ & $   301$ \\
$    9 $ &$    0.1 $ &$ 0.202 \times 10^{-5}$ & $ 0.604 $ &$ 0.157 $ & ---  & $  1.79 $ &$  0.89 $ &$  2.43 $ &$ 3.15$ & $ 1.62$ & $ 0.67$ & $ 0.49$ & $ 3.64$ & $   301$ \\
$   10 $ &$    0.1 $ &$ 0.243 \times 10^{-5}$ & $ 0.502 $ &$ 0.149 $ &$ 0.154 $ &$  3.75 $ &$  1.50 $ &$  3.49 $ &$ 5.34$ & $ 1.11$ & $ 0.46$ & $ 0.20$ & $ 5.48$ & $   301$ \\
$   11 $ &$   0.15 $ &$ 0.207 \times 10^{-5}$ & $ 0.803 $ &$ 0.228 $ & ---  & $  4.35 $ &$  1.50 $ &$  4.58 $ &$ 6.49$ & $ 2.37$ & $ 1.94$ & $ 1.75$ & $ 7.39$ & $   301$ \\
$   12 $ &$   0.15 $ &$ 0.236 \times 10^{-5}$ & $ 0.705 $ &$ 0.199 $ & ---  & $  2.01 $ &$  1.02 $ &$  3.05 $ &$ 3.79$ & $ 1.71$ & $ 1.60$ & $ 0.74$ & $ 4.52$ & $   301$ \\
$   13 $ &$   0.15 $ &$ 0.276 \times 10^{-5}$ & $ 0.603 $ &$ 0.201 $ & ---  & $  1.62 $ &$  0.92 $ &$  2.24 $ &$ 2.91$ & $ 1.40$ & $ 0.87$ & $ 0.38$ & $ 3.37$ & $   301$ \\
$   14 $ &$   0.15 $ &$ 0.331 \times 10^{-5}$ & $ 0.502 $ &$ 0.201 $ &$ 0.207 $ &$  1.53 $ &$  0.78 $ &$  2.10 $ &$ 2.71$ & $ 1.31$ & $ 0.19$ & $ 0.34$ & $ 3.04$ & $   301$ \\
$   15 $ &$   0.15 $ &$ 0.414 \times 10^{-5}$ & $ 0.402 $ &$ 0.187 $ &$ 0.190 $ &$  1.97 $ &$  0.83 $ &$  2.32 $ &$ 3.15$ & $ 0.82$ & $ 0.11$ & $ 0.17$ & $ 3.27$ & $   301$ \\
$   16 $ &$   0.15 $ &$ 0.502 \times 10^{-5}$ & $ 0.295 $ &$ 0.185 $ &$ 0.187 $ &$  3.79 $ &$  1.50 $ &$  3.59 $ &$ 5.43$ & $ 1.33$ & $-0.24$ & $ 0.32$ & $ 5.60$ & $   319$ \\
$   17 $ &$    0.2 $ &$ 0.315 \times 10^{-5}$ & $ 0.704 $ &$ 0.250 $ & ---  & $  2.48 $ &$  0.94 $ &$  2.96 $ &$ 3.98$ & $ 1.89$ & $ 1.50$ & $ 0.99$ & $ 4.76$ & $   301$ \\
$   18 $ &$    0.2 $ &$ 0.368 \times 10^{-5}$ & $ 0.603 $ &$ 0.246 $ & ---  & $  1.98 $ &$  0.94 $ &$  2.10 $ &$ 3.04$ & $ 1.39$ & $ 0.72$ & $ 0.37$ & $ 3.44$ & $   301$ \\
$   19 $ &$    0.2 $ &$ 0.441 \times 10^{-5}$ & $ 0.503 $ &$ 0.240 $ &$ 0.247 $ &$  1.62 $ &$  0.74 $ &$  1.97 $ &$ 2.66$ & $ 1.15$ & $ 0.36$ & $ 0.17$ & $ 2.92$ & $   301$ \\
$   20 $ &$    0.2 $ &$ 0.552 \times 10^{-5}$ & $ 0.402 $ &$ 0.235 $ &$ 0.240 $ &$  1.48 $ &$  0.65 $ &$  1.84 $ &$ 2.45$ & $ 1.18$ & $ 0.04$ & $ 0.30$ & $ 2.74$ & $   301$ \\
$   21 $ &$    0.2 $ &$ 0.669 \times 10^{-5}$ & $ 0.295 $ &$ 0.227 $ &$ 0.229 $ &$  1.65 $ &$  0.78 $ &$  1.63 $ &$ 2.45$ & $ 0.86$ & $-0.24$ & $ 0.10$ & $ 2.61$ & $   319$ \\
$   22 $ &$    0.2 $ &$ 0.849 \times 10^{-5}$ & $ 0.233 $ &$ 0.224 $ &$ 0.225 $ &$  1.61 $ &$  0.61 $ &$  2.13 $ &$ 2.74$ & $ 1.06$ & $-0.19$ & $ 0.27$ & $ 2.96$ & $   319$ \\
$   23 $ &$    0.2 $ &$ 0.110 \times 10^{-4}$ & $ 0.180 $ &$ 0.208 $ &$ 0.209 $ &$  2.78 $ &$  1.50 $ &$  2.79 $ &$ 4.21$ & $ 0.95$ & $-0.21$ & $ 0.26$ & $ 4.33$ & $   319$ \\
$   24 $ &$    0.2 $ &$ 0.398 \times 10^{-4}$ & $ 0.050 $ &$ 0.211 $ &$ 0.211 $ &$ 14.92 $ &$ 11.94 $ &$  5.17 $ &$ 19.8$ & $-0.54$ & $-3.10$ & $-4.60$ & $20.57$ & $   319$ \\
$   25 $ &$    0.2 $ &$ 0.251 \times 10^{-3}$ & $ 0.008 $ &$ 0.181 $ &$ 0.181 $ &$ 13.45 $ &$  6.17 $ &$  2.96 $ &$ 15.1$ & $-0.03$ & $-1.23$ & $ 2.59$ & $15.36$ & $   319$ \\
$   26 $ &$   0.25 $ &$ 0.394 \times 10^{-5}$ & $ 0.703 $ &$ 0.280 $ & ---  & $  4.09 $ &$  1.50 $ &$  2.83 $ &$ 5.19$ & $ 1.78$ & $ 1.28$ & $ 0.89$ & $ 5.71$ & $   301$ \\
$   27 $ &$   0.25 $ &$ 0.460 \times 10^{-5}$ & $ 0.603 $ &$ 0.280 $ & ---  & $  2.09 $ &$  0.92 $ &$  2.58 $ &$ 3.44$ & $ 1.67$ & $ 1.27$ & $ 0.36$ & $ 4.05$ & $   301$ \\
$   28 $ &$   0.25 $ &$ 0.552 \times 10^{-5}$ & $ 0.502 $ &$ 0.282 $ &$ 0.291 $ &$  1.79 $ &$  0.92 $ &$  1.87 $ &$ 2.74$ & $ 1.27$ & $ 0.27$ & $ 0.28$ & $ 3.05$ & $   301$ \\
$   29 $ &$   0.25 $ &$ 0.690 \times 10^{-5}$ & $ 0.402 $ &$ 0.276 $ &$ 0.281 $ &$  1.56 $ &$  0.79 $ &$  1.84 $ &$ 2.54$ & $ 1.18$ & $ 0.02$ & $ 0.21$ & $ 2.81$ & $   301$ \\
$   30 $ &$   0.25 $ &$ 0.836 \times 10^{-5}$ & $ 0.295 $ &$ 0.266 $ &$ 0.269 $ &$  1.45 $ &$  0.73 $ &$  1.85 $ &$ 2.46$ & $ 1.17$ & $-0.23$ & $ 0.27$ & $ 2.75$ & $   319$ \\
$   31 $ &$   0.25 $ &$ 0.106 \times 10^{-4}$ & $ 0.233 $ &$ 0.261 $ &$ 0.262 $ &$  1.29 $ &$  0.66 $ &$  1.77 $ &$ 2.29$ & $ 1.16$ & $-0.32$ & $ 0.21$ & $ 2.59$ & $   319$ \\
$   32 $ &$   0.25 $ &$ 0.138 \times 10^{-4}$ & $ 0.179 $ &$ 0.250 $ &$ 0.251 $ &$  1.26 $ &$  0.71 $ &$  1.78 $ &$ 2.30$ & $ 1.22$ & $-0.28$ & $ 0.41$ & $ 2.65$ & $   319$ \\
$   33 $ &$   0.25 $ &$ 0.230 \times 10^{-4}$ & $ 0.107 $ &$ 0.245 $ &$ 0.245 $ &$  1.40 $ &$  1.50 $ &$  2.33 $ &$ 3.11$ & $ 1.92$ & $-0.18$ & $ 1.96$ & $ 4.15$ & $   319$ \\
$   34 $ &$   0.25 $ &$ 0.398 \times 10^{-4}$ & $ 0.062 $ &$ 0.237 $ &$ 0.237 $ &$  3.32 $ &$  1.54 $ &$  2.74 $ &$ 4.57$ & $ 0.79$ & $-0.67$ & $-0.96$ & $ 4.79$ & $   319$ \\
$   35 $ &$   0.25 $ &$ 0.110 \times 10^{-3}$ & $ 0.022 $ &$ 0.200 $ &$ 0.200 $ &$  3.95 $ &$  1.50 $ &$  2.45 $ &$ 4.88$ & $ 0.81$ & $-0.17$ & $-0.04$ & $ 4.95$ & $   319$ \\
$   36 $ &$   0.25 $ &$ 0.251 \times 10^{-3}$ & $ 0.010 $ &$ 0.196 $ &$ 0.196 $ &$  3.74 $ &$  1.44 $ &$  3.23 $ &$ 5.15$ & $-0.17$ & $-0.25$ & $-0.58$ & $ 5.19$ & $   319$ \\
$   37 $ &$   0.25 $ &$ 0.394 \times 10^{-3}$ & $ 0.006 $ &$ 0.195 $ &$ 0.195 $ &$  4.15 $ &$  1.50 $ &$  3.58 $ &$ 5.68$ & $ 1.49$ & $-0.22$ & $ 0.79$ & $ 5.93$ & $   319$ \\
$   38 $ &$   0.25 $ &$ 0.158 \times 10^{-2}$ & $ 0.002 $ &$ 0.199 $ &$ 0.199 $ &$ 10.97 $ &$  5.29 $ &$  2.36 $ &$ 12.4$ & $ 0.14$ & $-0.17$ & $ 2.67$ & $12.69$ & $   301$ \\
$   39 $ &$   0.35 $ &$ 0.512 \times 10^{-5}$ & $ 0.758 $ &$ 0.350 $ & ---  & $  3.42 $ &$  1.50 $ &$  3.33 $ &$ 5.00$ & $ 1.95$ & $ 0.85$ & $ 0.37$ & $ 5.45$ & $   301$ \\
$   40 $ &$   0.35 $ &$ 0.512 \times 10^{-5}$ & $ 0.675 $ &$ 0.437 $ & ---  & $ 22.06 $ &$ 12.78 $ &$  1.80 $ &$ 25.6$ & $-0.15$ & $-1.02$ & $ 0.68$ & $25.59$ & $   319$ \\
$   41 $ &$   0.35 $ &$ 0.610 \times 10^{-5}$ & $ 0.636 $ &$ 0.358 $ & ---  & $  5.74 $ &$ 11.03 $ &$  4.78 $ &$ 13.3$ & $ 0.64$ & $ 3.76$ & $ 1.03$ & $13.89$ & $   301$ \\
$   42 $ &$   0.35 $ &$ 0.662 \times 10^{-5}$ & $ 0.586 $ &$ 0.336 $ &$ 0.351 $ &$  2.07 $ &$  0.91 $ &$  1.88 $ &$ 2.94$ & $ 1.20$ & $ 0.41$ & $ 0.33$ & $ 3.22$ & $   301$ \\
$   43 $ &$   0.35 $ &$ 0.828 \times 10^{-5}$ & $ 0.469 $ &$ 0.335 $ &$ 0.344 $ &$  1.82 $ &$  0.93 $ &$  1.90 $ &$ 2.80$ & $ 1.25$ & $-0.07$ & $ 0.21$ & $ 3.07$ & $   301$ \\
$   44 $ &$   0.35 $ &$ 0.100 \times 10^{-4}$ & $ 0.346 $ &$ 0.348 $ &$ 0.353 $ &$  1.62 $ &$  0.79 $ &$  1.73 $ &$ 2.50$ & $ 1.20$ & $-0.28$ & $ 0.29$ & $ 2.80$ & $   319$ \\
$   45 $ &$   0.35 $ &$ 0.127 \times 10^{-4}$ & $ 0.272 $ &$ 0.326 $ &$ 0.329 $ &$  1.43 $ &$  0.79 $ &$  1.80 $ &$ 2.43$ & $ 1.16$ & $-0.29$ & $ 0.20$ & $ 2.72$ & $   319$ \\
$   46 $ &$   0.35 $ &$ 0.165 \times 10^{-4}$ & $ 0.210 $ &$ 0.316 $ &$ 0.317 $ &$  1.21 $ &$  0.63 $ &$  1.82 $ &$ 2.27$ & $ 1.20$ & $-0.28$ & $ 0.26$ & $ 2.59$ & $   319$ \\
$   47 $ &$   0.35 $ &$ 0.320 \times 10^{-4}$ & $ 0.108 $ &$ 0.299 $ &$ 0.299 $ &$  1.14 $ &$  0.72 $ &$  2.51 $ &$ 2.85$ & $ 2.25$ & $-0.39$ & $ 2.05$ & $ 4.19$ & $   319$ \\
$   48 $ &$   0.35 $ &$ 0.662 \times 10^{-4}$ & $ 0.052 $ &$ 0.283 $ &$ 0.284 $ &$  2.62 $ &$  1.50 $ &$  1.65 $ &$ 3.44$ & $ 0.45$ & $-0.24$ & $-0.18$ & $ 3.48$ & $   319$ \\
$   49 $ &$   0.35 $ &$ 0.130 \times 10^{-3}$ & $ 0.027 $ &$ 0.259 $ &$ 0.259 $ &$  2.52 $ &$  1.43 $ &$  1.46 $ &$ 3.24$ & $ 0.75$ & $-0.33$ & $ 0.06$ & $ 3.35$ & $   319$ \\
$   50 $ &$   0.35 $ &$ 0.220 \times 10^{-3}$ & $ 0.016 $ &$ 0.242 $ &$ 0.242 $ &$  2.67 $ &$  1.50 $ &$  1.95 $ &$ 3.63$ & $ 1.11$ & $-0.34$ & $ 0.16$ & $ 3.81$ & $   319$ \\
$   51 $ &$   0.35 $ &$ 0.500 \times 10^{-3}$ & $ 0.007 $ &$ 0.242 $ &$ 0.242 $ &$  2.51 $ &$  1.42 $ &$  1.75 $ &$ 3.37$ & $ 1.31$ & $-0.25$ & $ 1.12$ & $ 3.79$ & $   319$ \\
$   52 $ &$   0.35 $ &$ 0.251 \times 10^{-2}$ & $ 0.001 $ &$ 0.202 $ &$ 0.202 $ &$  9.97 $ &$  4.55 $ &$  1.47 $ &$ 11.1$ & $ 0.35$ & $ 0.16$ & $ 0.77$ & $11.09$ & $   319$ \\
$   53 $ &$    0.4 $ &$ 0.883 \times 10^{-5}$ & $ 0.502 $ &$ 0.354 $ &$ 0.365 $ &$  3.86 $ &$  1.50 $ &$  2.56 $ &$ 4.87$ & $ 1.29$ & $ 0.52$ & $ 0.16$ & $ 5.07$ & $   301$ \\
$   54 $ &$    0.4 $ &$ 0.110 \times 10^{-4}$ & $ 0.403 $ &$ 0.372 $ &$ 0.378 $ &$  2.18 $ &$  0.88 $ &$  2.20 $ &$ 3.22$ & $ 1.48$ & $ 0.04$ & $ 0.22$ & $ 3.55$ & $   301$ \\
$   55 $ &$    0.4 $ &$ 0.133 \times 10^{-4}$ & $ 0.297 $ &$ 0.357 $ &$ 0.360 $ &$  1.97 $ &$  0.88 $ &$  1.98 $ &$ 2.93$ & $ 1.34$ & $-0.27$ & $ 0.29$ & $ 3.25$ & $   319$ \\
$   56 $ &$    0.4 $ &$ 0.170 \times 10^{-4}$ & $ 0.233 $ &$ 0.355 $ &$ 0.357 $ &$  1.66 $ &$  0.83 $ &$  1.81 $ &$ 2.59$ & $ 1.18$ & $-0.31$ & $ 0.09$ & $ 2.86$ & $   319$ \\
$   57 $ &$    0.4 $ &$ 0.220 \times 10^{-4}$ & $ 0.180 $ &$ 0.336 $ &$ 0.337 $ &$  1.45 $ &$  0.78 $ &$  1.80 $ &$ 2.44$ & $ 1.25$ & $-0.32$ & $ 0.42$ & $ 2.79$ & $   319$ \\
$   58 $ &$    0.4 $ &$ 0.368 \times 10^{-4}$ & $ 0.107 $ &$ 0.332 $ &$ 0.333 $ &$  1.26 $ &$  0.77 $ &$  2.49 $ &$ 2.89$ & $ 2.27$ & $-0.33$ & $ 2.14$ & $ 4.27$ & $   319$ \\
$   59 $ &$    0.4 $ &$ 0.883 \times 10^{-4}$ & $ 0.045 $ &$ 0.321 $ &$ 0.321 $ &$  2.71 $ &$  1.50 $ &$  1.49 $ &$ 3.44$ & $ 0.98$ & $-0.26$ & $ 0.41$ & $ 3.61$ & $   319$ \\
$   60 $ &$    0.4 $ &$ 0.176 \times 10^{-3}$ & $ 0.022 $ &$ 0.288 $ &$ 0.288 $ &$  2.78 $ &$  1.50 $ &$  1.70 $ &$ 3.59$ & $ 0.72$ & $-0.32$ & $-0.22$ & $ 3.68$ & $   319$ \\
$   61 $ &$    0.4 $ &$ 0.294 \times 10^{-3}$ & $ 0.013 $ &$ 0.277 $ &$ 0.277 $ &$  2.74 $ &$  1.50 $ &$  1.59 $ &$ 3.51$ & $ 0.99$ & $-0.28$ & $ 0.33$ & $ 3.67$ & $   319$ \\
$   62 $ &$    0.4 $ &$ 0.631 \times 10^{-3}$ & $ 0.006 $ &$ 0.261 $ &$ 0.261 $ &$  2.74 $ &$  1.50 $ &$  2.00 $ &$ 3.71$ & $ 1.28$ & $-0.28$ & $ 0.38$ & $ 3.95$ & $   319$ \\

\hline
\hline
\end{tabular}
\end{center}
\end{tiny}
\tablecaption{\label{tab615a1}
HERA combined reduced cross section $\sigma^{+ {\rm~ave}}_{r}$  for NC $e^+p$  scattering. 
$F_2^{\rm ave}$ represents the structure function $F_2$ calculated
from $\sigma^{\rm ave}_{r}$ by using the HERAPDF1.0 parametrisation of $\tilde{F}_L, x\tilde{F}_3, F_2^{\gamma Z}$ for 
$Q^2 \ge 2$~GeV$^2$ and assuming $R\equiv F_L/(F_2-F_L)=0.25$ for $Q^2 < 2$~GeV$^2$. $F_2$ is given for $y<0.6$. 
$\delta_{\rm ave,stat}$, $\delta_{\rm ave,uncor}$, $\delta_{\rm ave,cor}$ and
$\delta_{\rm ave,exp~tot}$ represent the statistical, uncorrelated
systematic, correlated systematic and total experimental uncertainty, respectively. $\delta_{\rm ave,rel}$, $\delta_{\rm ave,gp}$ and $\delta_{\rm ave,had}$
are the three correlated sources of uncertainties arising from the combination
procedure, see \Sec~\ref{subsubsec:proc_errors}. $\delta_{\rm ave,tot}$ is the total uncertainty
calculated by adding $\delta_{\rm ave,exp~tot}$ and the procedural errors in quadrature. 
The uncertainties are quoted in percent relative to $\sigma^{+ \rm~ave}_{r}$. 
The overall normalisation uncertainty of $0.5\%$ is not included.
}
\end{table}

\begin{table}
\begin{tiny}
\begin{center}
\begin{tabular}{l|cccccccccccccc}
\hline
\hline
Bin &  $Q^2$  & $x$ & $y$ &  $\sigma^{+~\rm ave}_{r}$ & $F_2^{\rm ave}$ 
 & $\delta_{\rm ave,stat}$ & $\delta_{\rm ave,uncor}$ & $\delta_{\rm ave,cor}$ & $\delta_{\rm ave,exp~tot}$
 & $\delta_{\rm ave,rel}$ & $\delta_{\rm ave,gp}$ & $\delta_{\rm ave,had}$ & $\delta_{\rm ave,tot}$ 
 & $\sqrt{s}$\\
$\#$     & GeV$^2$ &   &       &       &         &       \%           & \%    & \%  &    
\%              &   \% & \% & \% &\% & GeV \\
\hline

\hline
$   63 $ &$    0.5 $ &$ 0.732 \times 10^{-5}$ & $ 0.675 $ &$ 0.428 $ & ---  & $  5.55 $ &$  5.73 $ &$  4.17 $ &$ 9.01$ & $-0.49$ & $-2.46$ & $-0.38$ & $ 9.36$ & $   319$ \\
$   64 $ &$    0.5 $ &$ 0.860 \times 10^{-5}$ & $ 0.645 $ &$ 0.443 $ & ---  & $  3.74 $ &$  9.17 $ &$  3.57 $ &$ 10.5$ & $ 0.58$ & $ 2.86$ & $ 0.70$ & $10.95$ & $   301$ \\
$   65 $ &$    0.5 $ &$ 0.158 \times 10^{-4}$ & $ 0.313 $ &$ 0.429 $ &$ 0.433 $ &$  3.53 $ &$  1.48 $ &$  2.33 $ &$ 4.48$ & $ 1.22$ & $-0.17$ & $ 0.14$ & $ 4.65$ & $   319$ \\
$   66 $ &$    0.5 $ &$ 0.212 \times 10^{-4}$ & $ 0.233 $ &$ 0.392 $ &$ 0.394 $ &$  2.06 $ &$  0.80 $ &$  1.95 $ &$ 2.94$ & $ 1.12$ & $-0.25$ & $-0.05$ & $ 3.16$ & $   319$ \\
$   67 $ &$    0.5 $ &$ 0.276 \times 10^{-4}$ & $ 0.179 $ &$ 0.379 $ &$ 0.380 $ &$  1.71 $ &$  0.76 $ &$  1.94 $ &$ 2.69$ & $ 1.35$ & $-0.30$ & $ 0.39$ & $ 3.05$ & $   319$ \\
$   68 $ &$    0.5 $ &$ 0.398 \times 10^{-4}$ & $ 0.124 $ &$ 0.367 $ &$ 0.368 $ &$  1.48 $ &$  0.81 $ &$  2.57 $ &$ 3.07$ & $ 2.32$ & $-0.52$ & $ 2.05$ & $ 4.39$ & $   319$ \\
\hline
\hline
\end{tabular}
\end{center}
\end{tiny}
\tablecaption{\label{tab615a2}Continuation of \Tab~\ref{tab615a1}.}
\end{table}

\begin{table}
\begin{tiny}
\begin{center}
\begin{tabular}{l|cccccccccccccc}
\hline
\hline
Bin &  $Q^2$  & $x$ & $y$ &  $\sigma^{+~\rm ave}_{r}$ & $F_2^{\rm ave}$ 
 & $\delta_{\rm ave,stat}$ & $\delta_{\rm ave,uncor}$ & $\delta_{\rm ave,cor}$ & $\delta_{\rm ave,exp~tot}$
 & $\delta_{\rm ave,rel}$ & $\delta_{\rm ave,gp}$ & $\delta_{\rm ave,had}$ & $\delta_{\rm ave,tot}$ 
 & $\sqrt{s}$\\
$\#$     & GeV$^2$ &   &       &       &         &       \%           & \%    & \%  &    
\%              &   \% & \% & \% &\% & GeV \\
\hline

$  134 $ &$    2.7 $ &$ 0.309 \times 10^{-4}$ & $ 0.864 $ &$ 0.855 $ & ---  & $  2.47 $ &$  2.67 $ &$  3.33 $ &$ 4.93$ & $ 0.06$ & $ 4.62$ & $ 0.25$ & $ 6.76$ & $   319$ \\
$  135 $ &$    2.7 $ &$ 0.366 \times 10^{-4}$ & $ 0.729 $ &$ 0.881 $ & ---  & $  2.29 $ &$  2.29 $ &$  1.72 $ &$ 3.67$ & $ 0.09$ & $ 2.19$ & $ 0.05$ & $ 4.27$ & $   319$ \\
$  136 $ &$    2.7 $ &$ 0.409 \times 10^{-4}$ & $ 0.732 $ &$ 0.975 $ & ---  & $  1.54 $ &$  6.21 $ &$  1.93 $ &$ 6.68$ & $ 0.57$ & $ 4.64$ & $ 0.61$ & $ 8.18$ & $   301$ \\
$  137 $ &$    2.7 $ &$ 0.500 \times 10^{-4}$ & $ 0.599 $ &$ 0.894 $ &$ 0.913 $ &$  1.06 $ &$  1.27 $ &$  1.30 $ &$ 2.11$ & $ 0.77$ & $ 2.43$ & $ 0.54$ & $ 3.35$ & $   301$ \\
$  138 $ &$    2.7 $ &$ 0.500 \times 10^{-4}$ & $ 0.534 $ &$ 0.882 $ &$ 0.901 $ &$  1.51 $ &$  1.65 $ &$  0.93 $ &$ 2.42$ & $ 0.17$ & $ 0.41$ & $ 0.10$ & $ 2.46$ & $   319$ \\
$  139 $ &$    2.7 $ &$ 0.800 \times 10^{-4}$ & $ 0.374 $ &$ 0.892 $ &$ 0.899 $ &$  3.28 $ &$  1.00 $ &$  1.81 $ &$ 3.87$ & $ 0.39$ & $ 0.65$ & $ 0.75$ & $ 4.02$ & $   301$ \\
\hline
\hline
\end{tabular}
\end{center}
\end{tiny}
\tablecaption{\label{tab615a3}Continuation of \Tab~\ref{tab615a1}.}
\end{table}
\begin{table}
\begin{tiny}
\begin{center}
\begin{tabular}{l|cccccccccccccc}
\hline
\hline
Bin &  $Q^2$  & $x$ & $y$ &  $\sigma^{+~\rm ave}_{r}$ & $F_2^{\rm ave}$ 
 & $\delta_{\rm ave,stat}$ & $\delta_{\rm ave,uncor}$ & $\delta_{\rm ave,cor}$ & $\delta_{\rm ave,exp~tot}$
 & $\delta_{\rm ave,rel}$ & $\delta_{\rm ave,gp}$ & $\delta_{\rm ave,had}$ & $\delta_{\rm ave,tot}$ 
 & $\sqrt{s}$\\
$\#$     & GeV$^2$ &   &       &       &         &       \%           & \%    & \%  &    
\%              &   \% & \% & \% &\% & GeV \\
\hline

$  205 $ &$    10. $ &$ 0.130 \times 10^{-3}$ & $ 0.853 $ &$ 1.141 $ & ---  & $  4.39 $ &$  1.41 $ &$  3.68 $ &$ 5.90$ & $ 1.21$ & $ 4.94$ & $ 0.71$ & $ 7.83$ & $   301$ \\
$  206 $ &$    10. $ &$ 0.200 \times 10^{-3}$ & $ 0.554 $ &$ 1.248 $ &$ 1.306 $ &$  2.10 $ &$  1.33 $ &$  1.34 $ &$ 2.83$ & $ 0.57$ & $ 1.54$ & $ 0.23$ & $ 3.28$ & $   301$ \\
$  207 $ &$    10. $ &$ 0.320 \times 10^{-3}$ & $ 0.309 $ &$ 1.153 $ &$ 1.169 $ &$  1.90 $ &$  1.12 $ &$  0.84 $ &$ 2.36$ & $ 0.37$ & $ 0.28$ & $-0.03$ & $ 2.41$ & $   319$ \\
$  208 $ &$    10. $ &$ 0.500 \times 10^{-3}$ & $ 0.198 $ &$ 1.069 $ &$ 1.074 $ &$  1.77 $ &$  1.19 $ &$  0.76 $ &$ 2.26$ & $ 0.17$ & $-0.06$ & $-0.22$ & $ 2.28$ & $   319$ \\
$  209 $ &$    10. $ &$ 0.800 \times 10^{-3}$ & $ 0.124 $ &$ 0.983 $ &$ 0.985 $ &$  1.83 $ &$  1.31 $ &$  0.73 $ &$ 2.37$ & $ 0.20$ & $-0.11$ & $-0.09$ & $ 2.38$ & $   319$ \\
$  210 $ &$    10. $ &$ 0.130 \times 10^{-2}$ & $ 0.076 $ &$ 0.898 $ &$ 0.899 $ &$  2.11 $ &$  1.21 $ &$  0.93 $ &$ 2.61$ & $-0.01$ & $ 0.01$ & $-0.03$ & $ 2.61$ & $   319$ \\
\hline
\hline
\end{tabular}
\end{center}
\end{tiny}
\tablecaption{\label{tab615a4}Continuation of \Tab~\ref{tab615a1}.}
\end{table}
\begin{table}
\begin{tiny}
\begin{center}
\begin{tabular}{l|cccccccccccccc}
\hline
\hline
Bin &  $Q^2$  & $x$ & $y$ &  $\sigma^{+~\rm ave}_{r}$ & $F_2^{\rm ave}$ 
 & $\delta_{\rm ave,stat}$ & $\delta_{\rm ave,uncor}$ & $\delta_{\rm ave,cor}$ & $\delta_{\rm ave,exp~tot}$
 & $\delta_{\rm ave,rel}$ & $\delta_{\rm ave,gp}$ & $\delta_{\rm ave,had}$ & $\delta_{\rm ave,tot}$ 
 & $\sqrt{s}$\\
$\#$     & GeV$^2$ &   &       &       &         &       \%           & \%    & \%  &    
\%              &   \% & \% & \% &\% & GeV \\
\hline

$  284 $ &$    45. $ &$ 0.800 \times 10^{-3}$ & $ 0.624 $ &$ 1.493 $ & ---  & $  1.80 $ &$  1.45 $ &$  1.41 $ &$ 2.71$ & $ 0.70$ & $ 1.23$ & $ 0.72$ & $ 3.14$ & $   301$ \\
$  285 $ &$    45. $ &$ 0.800 \times 10^{-3}$ & $ 0.556 $ &$ 1.468 $ &$ 1.571 $ &$  1.38 $ &$  0.89 $ &$  0.80 $ &$ 1.82$ & $ 0.25$ & $ 0.51$ & $ 0.17$ & $ 1.91$ & $   319$ \\
$  286 $ &$    45. $ &$ 0.130 \times 10^{-2}$ & $ 0.384 $ &$ 1.317 $ &$ 1.344 $ &$  0.95 $ &$  0.61 $ &$  0.78 $ &$ 1.37$ & $ 0.56$ & $ 0.12$ & $-0.10$ & $ 1.49$ & $   301$ \\
$  287 $ &$    45. $ &$ 0.130 \times 10^{-2}$ & $ 0.342 $ &$ 1.316 $ &$ 1.343 $ &$  0.87 $ &$  0.76 $ &$  0.65 $ &$ 1.32$ & $ 0.25$ & $ 0.14$ & $ 0.14$ & $ 1.36$ & $   319$ \\
$  288 $ &$    45. $ &$ 0.200 \times 10^{-2}$ & $ 0.222 $ &$ 1.163 $ &$ 1.172 $ &$  0.61 $ &$  0.57 $ &$  0.63 $ &$ 1.05$ & $ 0.28$ & $ 0.07$ & $ 0.08$ & $ 1.09$ & $   319$ \\
$  289 $ &$    45. $ &$ 0.320 \times 10^{-2}$ & $ 0.139 $ &$ 1.017 $ &$ 1.019 $ &$  0.62 $ &$  0.55 $ &$  0.65 $ &$ 1.05$ & $ 0.27$ & $ 0.05$ & $ 0.10$ & $ 1.09$ & $   319$ \\
\hline
\hline
\end{tabular}
\end{center}
\end{tiny}
\tablecaption{\label{tab615a5}Continuation of \Tab~\ref{tab615a1}.}
\end{table}
\begin{table}
\begin{tiny}
\begin{center}
\begin{tabular}{l|cccccccccccccc}
\hline
\hline
Bin &  $Q^2$  & $x$ & $y$ &  $\sigma^{+~\rm ave}_{r}$ & $F_2^{\rm ave}$ 
 & $\delta_{\rm ave,stat}$ & $\delta_{\rm ave,uncor}$ & $\delta_{\rm ave,cor}$ & $\delta_{\rm ave,exp~tot}$
 & $\delta_{\rm ave,rel}$ & $\delta_{\rm ave,gp}$ & $\delta_{\rm ave,had}$ & $\delta_{\rm ave,tot}$ 
 & $\sqrt{s}$\\
$\#$     & GeV$^2$ &   &       &       &         &       \%           & \%    & \%  &    
\%              &   \% & \% & \% &\% & GeV \\
\hline

$  367 $ &$   250. $ &$ 0.330 \times 10^{-2}$ & $ 0.749 $ &$ 1.247 $ & ---  & $  2.50 $ &$  2.70 $ &$  1.23 $ &$ 3.88$ & $ 0.08$ & $ 0.03$ & $-0.12$ & $ 3.89$ & $   319$ \\
$  368 $ &$   250. $ &$ 0.500 \times 10^{-2}$ & $ 0.554 $ &$ 1.079 $ &$ 1.123 $ &$  2.29 $ &$  1.61 $ &$  0.97 $ &$ 2.97$ & $ 1.02$ & $ 0.06$ & $ 0.18$ & $ 3.14$ & $   301$ \\
$  369 $ &$   250. $ &$ 0.500 \times 10^{-2}$ & $ 0.494 $ &$ 1.102 $ &$ 1.148 $ &$  1.61 $ &$  2.10 $ &$  0.90 $ &$ 2.80$ & $ 0.07$ & $ 0.02$ & $-0.12$ & $ 2.80$ & $   319$ \\
$  370 $ &$   250. $ &$ 0.800 \times 10^{-2}$ & $ 0.309 $ &$ 0.953 $ &$ 0.964 $ &$  1.04 $ &$  0.69 $ &$  0.72 $ &$ 1.44$ & $ 0.32$ & $ 0.03$ & $ 0.03$ & $ 1.48$ & $   319$ \\
$  371 $ &$   250. $ &$ 0.130 \times 10^{-1}$ & $ 0.190 $ &$ 0.825 $ &$ 0.828 $ &$  1.07 $ &$  0.64 $ &$  0.71 $ &$ 1.44$ & $ 0.11$ & $ 0.04$ & $ 0.21$ & $ 1.46$ & $   319$ \\
$  372 $ &$   250. $ &$ 0.200 \times 10^{-1}$ & $ 0.124 $ &$ 0.701 $ &$ 0.701 $ &$  1.18 $ &$  0.79 $ &$  0.70 $ &$ 1.59$ & $ 0.20$ & $ 0.03$ & $ 0.10$ & $ 1.60$ & $   319$ \\
\hline
\hline
\end{tabular}
\end{center}
\end{tiny}
\tablecaption{\label{tab615a6}Continuation of \Tab~\ref{tab615a1}.}
\end{table}

\begin{table}
\begin{tiny}
\begin{center}
\begin{tabular}{l|cccccccccccccc}
\hline
\hline
Bin &  $Q^2$  & $x$ & $y$ &  $\sigma^{+~\rm ave}_{r}$ & $F_2^{\rm ave}$ 
 & $\delta_{\rm ave,stat}$ & $\delta_{\rm ave,uncor}$ & $\delta_{\rm ave,cor}$ & $\delta_{\rm ave,exp~tot}$
 & $\delta_{\rm ave,rel}$ & $\delta_{\rm ave,gp}$ & $\delta_{\rm ave,had}$ & $\delta_{\rm ave,tot}$ 
 & $\sqrt{s}$\\
$\#$     & GeV$^2$ &   &       &       &         &       \%           & \%    & \%  &    
\%              &   \% & \% & \% &\% & GeV \\
\hline

$  446 $ &$  1000. $ &$ 0.130 \times 10^{-1}$ & $ 0.760 $ &$ 0.871 $ & ---  & $  5.04 $ &$  3.70 $ &$  1.31 $ &$ 6.39$ & $ 0.41$ & $-0.02$ & $-0.06$ & $ 6.40$ & $   319$ \\
$  447 $ &$  1000. $ &$ 0.200 \times 10^{-1}$ & $ 0.554 $ &$ 0.738 $ &$ 0.762 $ &$  5.46 $ &$  3.70 $ &$  1.00 $ &$ 6.67$ & $ 0.11$ & $ 0.01$ & $-0.07$ & $ 6.67$ & $   301$ \\
$  448 $ &$  1000. $ &$ 0.200 \times 10^{-1}$ & $ 0.494 $ &$ 0.732 $ &$ 0.756 $ &$  3.93 $ &$  3.00 $ &$  0.97 $ &$ 5.04$ & $ 0.28$ & $ 0.02$ & $-0.29$ & $ 5.06$ & $   319$ \\
$  449 $ &$  1000. $ &$ 0.320 \times 10^{-1}$ & $ 0.309 $ &$ 0.637 $ &$ 0.644 $ &$  3.29 $ &$  2.46 $ &$  1.00 $ &$ 4.23$ & $ 0.43$ & $ 0.03$ & $-0.09$ & $ 4.25$ & $   319$ \\
$  450 $ &$  1000. $ &$ 0.500 \times 10^{-1}$ & $ 0.198 $ &$ 0.556 $ &$ 0.557 $ &$  3.21 $ &$  2.35 $ &$  0.78 $ &$ 4.05$ & $ 0.27$ & $ 0.03$ & $-0.03$ & $ 4.06$ & $   319$ \\
$  451 $ &$  1000. $ &$ 0.800 \times 10^{-1}$ & $ 0.124 $ &$ 0.444 $ &$ 0.444 $ &$  3.35 $ &$  2.28 $ &$  0.81 $ &$ 4.13$ & $ 0.16$ & $ 0.03$ & $ 0.26$ & $ 4.15$ & $   319$ \\
\hline
\hline
\end{tabular}
\end{center}
\end{tiny}
\tablecaption{\label{tab615a7}Continuation of \Tab~\ref{tab615a1}.}
\end{table}

\begin{table}
\begin{tiny}
\begin{center}
\begin{tabular}{l|cccccccccccccc}
\hline
\hline
Bin &  $Q^2$  & $x$ & $y$ &  $\sigma^{+~\rm ave}_{r}$ & $F_2^{\rm ave}$ 
 & $\delta_{\rm ave,stat}$ & $\delta_{\rm ave,uncor}$ & $\delta_{\rm ave,cor}$ & $\delta_{\rm ave,exp~tot}$
 & $\delta_{\rm ave,rel}$ & $\delta_{\rm ave,gp}$ & $\delta_{\rm ave,had}$ & $\delta_{\rm ave,tot}$ 
 & $\sqrt{s}$\\
$\#$     & GeV$^2$ &   &       &       &         &       \%           & \%    & \%  &    
\%              &   \% & \% & \% &\% & GeV \\
\hline

$  522 $ &$ 20000. $ &$ 0.250 $ & $ 0.887 $ &$ 0.135 $ & ---  & $ 30.72 $ &$  3.58 $ &$  0.96 $ &$ 30.9$ & $ 0.85$ & $ 0.02$ & $ 0.36$ & $30.96$ & $   301$ \\
$  523 $ &$ 20000. $ &$ 0.250 $ & $ 0.791 $ &$ 0.096 $ & ---  & $ 22.47 $ &$  2.01 $ &$  2.50 $ &$ 22.7$ & $ 0.63$ & $ 0.05$ & $ 0.14$ & $22.71$ & $   319$ \\
$  524 $ &$ 20000. $ &$ 0.400 $ & $ 0.554 $ &$ 0.116 $ &$ 0.146 $ &$ 31.65 $ &$  7.66 $ &$  1.29 $ &$ 32.6$ & $ 1.60$ & $-0.01$ & $-0.51$ & $32.63$ & $   301$ \\
$  525 $ &$ 20000. $ &$ 0.400 $ & $ 0.494 $ &$ 0.047 $ &$ 0.059 $ &$ 35.20 $ &$  7.96 $ &$  2.49 $ &$ 36.2$ & $ 0.21$ & $ 0.01$ & $-0.03$ & $36.17$ & $   319$ \\
$  526 $ &$ 20000. $ &$ 0.650 $ & $ 0.304 $ &$ 0.014 $ &$ 0.014 $ &$ 50.08 $ &$ 28.03 $ &$  4.07 $ &$ 57.5$ & $ 7.64$ & $-0.01$ & $-0.23$ & $58.04$ & $   319$ \\
$  527 $ &$ 30000. $ &$ 0.400 $ & $ 0.831 $ &$ 0.110 $ & ---  & $ 66.01 $ &$  5.29 $ &$  1.46 $ &$ 66.2$ & $ 0.54$ & $ 0.00$ & $ 0.25$ & $66.24$ & $   301$ \\
$  528 $ &$ 30000. $ &$ 0.400 $ & $ 0.741 $ &$ 0.086 $ & ---  & $ 35.54 $ &$  4.75 $ &$ 13.06 $ &$ 38.2$ & $ 5.53$ & $-0.07$ & $ 0.09$ & $38.56$ & $   319$ \\

\hline
\hline
\end{tabular}
\end{center}
\end{tiny}
\tablecaption{\label{tab615a8}Continuation of \Tab~\ref{tab615a1}.}
\end{table}

\begin{table}
\begin{tiny}
\begin{center}
\begin{tabular}{l|cccccccccccccc}
\hline
\hline
Bin &  $Q^2$  & $x$ & $y$ &  $\sigma^{-~\rm ave}_{r}$ & $F_2^{\rm ave}$ 
 & $\delta_{\rm ave,stat}$ & $\delta_{\rm ave,uncor}$ & $\delta_{\rm ave,cor}$ & $\delta_{\rm ave,exp~tot}$
 & $\delta_{\rm ave,rel}$ & $\delta_{\rm ave,gp}$ & $\delta_{\rm ave,had}$ & $\delta_{\rm ave,tot}$ 
 & $\sqrt{s}$\\
 $\#$    & GeV$^2$ &   &       &       &         &       \%           & \%    & \%  &    
\%              &   \% & \% & \% &\% & GeV \\
\hline

$    1 $ &$    90. $ &$ 0.130 \times 10^{-2}$ & $ 0.684 $ &$ 1.372 $ & ---  & $  4.00 $ &$  5.50 $ &$  1.72 $ &$ 7.01$ & $-0.26$ & $ 0.01$ & $-0.10$ & $ 7.02$ & $   319$ \\
$    2 $ &$    90. $ &$ 0.200 \times 10^{-2}$ & $ 0.445 $ &$ 1.257 $ &$ 1.292 $ &$  3.43 $ &$  3.30 $ &$  1.87 $ &$ 5.11$ & $ 0.57$ & $-0.04$ & $-0.26$ & $ 5.15$ & $   319$ \\
$    3 $ &$   120. $ &$ 0.160 \times 10^{-2}$ & $ 0.741 $ &$ 1.343 $ & ---  & $  4.20 $ &$  4.80 $ &$  1.69 $ &$ 6.60$ & $-0.07$ & $ 0.01$ & $-0.16$ & $ 6.60$ & $   319$ \\
$    4 $ &$   120. $ &$ 0.200 \times 10^{-2}$ & $ 0.593 $ &$ 1.317 $ &$ 1.390 $ &$  3.31 $ &$  3.20 $ &$  1.95 $ &$ 5.00$ & $ 0.23$ & $-0.05$ & $ 0.26$ & $ 5.01$ & $   319$ \\
$    5 $ &$   120. $ &$ 0.320 \times 10^{-2}$ & $ 0.371 $ &$ 1.192 $ &$ 1.211 $ &$  3.01 $ &$  3.20 $ &$  1.59 $ &$ 4.67$ & $ 0.39$ & $-0.01$ & $-0.15$ & $ 4.69$ & $   319$ \\
$    6 $ &$   150. $ &$ 0.200 \times 10^{-2}$ & $ 0.741 $ &$ 1.331 $ & ---  & $  4.41 $ &$  4.80 $ &$  1.54 $ &$ 6.70$ & $ 0.24$ & $ 0.00$ & $-0.20$ & $ 6.71$ & $   319$ \\
\hline
\hline
\end{tabular}
\end{center}
\end{tiny}
\tablecaption{\label{tab515a1}
HERA combined reduced cross section $\sigma^{- {\rm~ave}}_{r}$  for NC $e^-p$  scattering. 
$F_2^{\rm ave}$ represents the structure function $F_2$ calculated
from $\sigma^{\rm ave}_{r}$ by using the HERAPDF1.0 parametrisation of $\tilde{F}_L, x\tilde{F}_3, F_2^{\gamma Z}$ for $y<0.6$. 
$\delta_{\rm ave,stat}$, $\delta_{\rm ave,uncor}$, $\delta_{\rm ave,cor}$ and
$\delta_{\rm ave,exp~tot}$ represent the statistical, uncorrelated
systematic, correlated systematic and total experimental uncertainty, respectively. $\delta_{\rm ave,rel}$, $\delta_{\rm ave,gp}$ and $\delta_{\rm ave,had}$
are the three correlated sources of uncertainties arising from the combination
procedure, see \Sec~\ref{subsubsec:proc_errors}. $\delta_{\rm ave,tot}$ is the total uncertainty
calculated by adding $\delta_{\rm ave,exp~tot}$ and the procedural errors in quadrature. 
The uncertainties are quoted in percent relative to $\sigma^{- \rm~ave}_{r}$. 
The overall normalisation uncertainty of $0.5\%$ is not included.
 }
\end{table}

\begin{table}
\begin{tiny}
\begin{center}
\begin{tabular}{l|cccccccccccccc}
\hline
\hline
Bin &  $Q^2$  & $x$ & $y$ &  $\sigma^{-~\rm ave}_{r}$ & $F_2^{\rm ave}$ 
 & $\delta_{\rm ave,stat}$ & $\delta_{\rm ave,uncor}$ & $\delta_{\rm ave,cor}$ & $\delta_{\rm ave,exp~tot}$
 & $\delta_{\rm ave,rel}$ & $\delta_{\rm ave,gp}$ & $\delta_{\rm ave,had}$ & $\delta_{\rm ave,tot}$ 
 & $\sqrt{s}$\\
 $\#$    & GeV$^2$ &   &       &       &         &       \%           & \%    & \%  &    
\%              &   \% & \% & \% &\% & GeV \\
\hline

$   66 $ &$   650. $ &$ 0.850 \times 10^{-2}$ & $ 0.756 $ &$ 0.972 $ & ---  & $  4.91 $ &$  2.20 $ &$  1.70 $ &$ 5.64$ & $ 0.03$ & $ 0.00$ & $ 0.01$ & $ 5.64$ & $   319$ \\
$   67 $ &$   650. $ &$ 0.130 \times 10^{-1}$ & $ 0.494 $ &$ 0.950 $ &$ 0.959 $ &$  3.64 $ &$  1.48 $ &$  1.56 $ &$ 4.23$ & $ 0.07$ & $ 0.00$ & $-0.17$ & $ 4.23$ & $   319$ \\
$   68 $ &$   650. $ &$ 0.200 \times 10^{-1}$ & $ 0.321 $ &$ 0.794 $ &$ 0.790 $ &$  4.59 $ &$  1.98 $ &$  1.35 $ &$ 5.18$ & $ 0.27$ & $ 0.00$ & $-0.23$ & $ 5.19$ & $   319$ \\
$   69 $ &$   650. $ &$ 0.320 \times 10^{-1}$ & $ 0.201 $ &$ 0.645 $ &$ 0.638 $ &$  5.01 $ &$  2.32 $ &$  1.32 $ &$ 5.68$ & $ 0.29$ & $ 0.00$ & $-0.13$ & $ 5.69$ & $   319$ \\
$   70 $ &$   650. $ &$ 0.500 \times 10^{-1}$ & $ 0.128 $ &$ 0.525 $ &$ 0.520 $ &$  5.14 $ &$  2.66 $ &$  1.41 $ &$ 5.96$ & $ 0.38$ & $ 0.00$ & $-0.04$ & $ 5.97$ & $   319$ \\
$   71 $ &$   650. $ &$ 0.800 \times 10^{-1}$ & $ 0.080 $ &$ 0.419 $ &$ 0.415 $ &$  5.48 $ &$  2.58 $ &$  1.39 $ &$ 6.21$ & $ 0.50$ & $ 0.00$ & $ 0.19$ & $ 6.23$ & $   319$ \\
$   72 $ &$   650. $ &$ 0.130 $ & $ 0.049 $ &$ 0.344 $ &$ 0.341 $ &$  5.96 $ &$  2.59 $ &$  1.41 $ &$ 6.65$ & $ 0.33$ & $ 0.00$ & $-0.14$ & $ 6.66$ & $   319$ \\
$   73 $ &$   650. $ &$ 0.180 $ & $ 0.036 $ &$ 0.336 $ &$ 0.333 $ &$  9.02 $ &$  6.80 $ &$  1.86 $ &$ 11.4$ & $ 0.71$ & $-0.01$ & $ 0.64$ & $11.49$ & $   319$ \\
$   74 $ &$   650. $ &$ 0.250 $ & $ 0.026 $ &$ 0.259 $ &$ 0.257 $ &$  6.68 $ &$  4.02 $ &$  1.49 $ &$ 7.94$ & $ 0.68$ & $-0.01$ & $ 0.12$ & $ 7.97$ & $   319$ \\
$   75 $ &$   650. $ &$ 0.400 $ & $ 0.016 $ &$ 0.088 $ &$ 0.087 $ &$ 23.18 $ &$  9.30 $ &$  1.90 $ &$ 25.1$ & $ 0.63$ & $ 0.00$ & $-0.64$ & $25.07$ & $   319$ \\
$   76 $ &$   800. $ &$ 0.105 \times 10^{-1}$ & $ 0.753 $ &$ 0.942 $ & ---  & $  9.18 $ &$  5.90 $ &$  1.46 $ &$ 11.0$ & $-0.19$ & $ 0.00$ & $-0.05$ & $11.01$ & $   319$ \\
$   77 $ &$   800. $ &$ 0.130 \times 10^{-1}$ & $ 0.608 $ &$ 0.885 $ & ---  & $  4.80 $ &$  1.63 $ &$  1.53 $ &$ 5.30$ & $ 0.20$ & $ 0.00$ & $ 0.06$ & $ 5.30$ & $   319$ \\
$   78 $ &$   800. $ &$ 0.200 \times 10^{-1}$ & $ 0.395 $ &$ 0.808 $ &$ 0.803 $ &$  4.98 $ &$  2.07 $ &$  1.39 $ &$ 5.57$ & $ 0.31$ & $-0.01$ & $-0.13$ & $ 5.58$ & $   319$ \\
$   79 $ &$   800. $ &$ 0.320 \times 10^{-1}$ & $ 0.247 $ &$ 0.662 $ &$ 0.653 $ &$  5.13 $ &$  1.94 $ &$  1.49 $ &$ 5.68$ & $ 0.55$ & $ 0.00$ & $-0.17$ & $ 5.71$ & $   319$ \\
$   80 $ &$   800. $ &$ 0.500 \times 10^{-1}$ & $ 0.158 $ &$ 0.590 $ &$ 0.581 $ &$  5.05 $ &$  1.86 $ &$  1.32 $ &$ 5.54$ & $ 0.05$ & $ 0.00$ & $ 0.14$ & $ 5.54$ & $   319$ \\
$   81 $ &$   800. $ &$ 0.800 \times 10^{-1}$ & $ 0.099 $ &$ 0.501 $ &$ 0.494 $ &$  5.58 $ &$  2.26 $ &$  1.43 $ &$ 6.19$ & $ 0.09$ & $ 0.00$ & $ 0.12$ & $ 6.19$ & $   319$ \\
$   82 $ &$   800. $ &$ 0.130 $ & $ 0.061 $ &$ 0.368 $ &$ 0.363 $ &$  6.89 $ &$  2.48 $ &$  1.92 $ &$ 7.57$ & $ 0.77$ & $ 0.01$ & $ 0.09$ & $ 7.61$ & $   319$ \\
$   83 $ &$   800. $ &$ 0.180 $ & $ 0.044 $ &$ 0.303 $ &$ 0.300 $ &$ 11.58 $ &$  6.20 $ &$  1.76 $ &$ 13.3$ & $ 0.43$ & $ 0.01$ & $ 0.44$ & $13.27$ & $   319$ \\
$   84 $ &$   800. $ &$ 0.250 $ & $ 0.032 $ &$ 0.241 $ &$ 0.239 $ &$  8.86 $ &$  3.38 $ &$  1.66 $ &$ 9.63$ & $ 0.62$ & $-0.01$ & $ 0.00$ & $ 9.65$ & $   319$ \\
$   85 $ &$   800. $ &$ 0.400 $ & $ 0.020 $ &$ 0.120 $ &$ 0.119 $ &$ 20.64 $ &$ 11.60 $ &$  2.31 $ &$ 23.8$ & $ 1.50$ & $-0.02$ & $ 0.60$ & $23.84$ & $   319$ \\
$   86 $ &$   800. $ &$ 0.650 $ & $ 0.012 $ &$ 0.015 $ &$ 0.015 $ &$ 21.50 $ &$ 12.50 $ &$  3.31 $ &$ 25.1$ & $ 2.87$ & $-0.04$ & $ 0.04$ & $25.25$ & $   319$ \\
$   87 $ &$  1000. $ &$ 0.130 \times 10^{-1}$ & $ 0.760 $ &$ 0.777 $ & ---  & $ 11.47 $ &$  6.50 $ &$  2.18 $ &$ 13.4$ & $ 0.07$ & $-0.06$ & $ 0.80$ & $13.39$ & $   319$ \\
$   88 $ &$  1000. $ &$ 0.200 \times 10^{-1}$ & $ 0.494 $ &$ 0.776 $ &$ 0.770 $ &$  7.96 $ &$  4.30 $ &$  1.60 $ &$ 9.18$ & $ 0.52$ & $-0.01$ & $-0.59$ & $ 9.22$ & $   319$ \\
$   89 $ &$  1000. $ &$ 0.320 \times 10^{-1}$ & $ 0.309 $ &$ 0.567 $ &$ 0.556 $ &$  9.04 $ &$  4.20 $ &$  1.45 $ &$ 10.1$ & $ 0.19$ & $ 0.00$ & $-0.36$ & $10.08$ & $   319$ \\
$   90 $ &$  1000. $ &$ 0.500 \times 10^{-1}$ & $ 0.198 $ &$ 0.575 $ &$ 0.563 $ &$  8.41 $ &$  4.30 $ &$  1.42 $ &$ 9.56$ & $-0.03$ & $ 0.01$ & $-0.12$ & $ 9.56$ & $   319$ \\
$   91 $ &$  1000. $ &$ 0.800 \times 10^{-1}$ & $ 0.124 $ &$ 0.453 $ &$ 0.444 $ &$  9.27 $ &$  4.90 $ &$  1.70 $ &$ 10.6$ & $ 0.75$ & $-0.01$ & $ 0.31$ & $10.65$ & $   319$ \\
$   92 $ &$  1000. $ &$ 0.130 $ & $ 0.076 $ &$ 0.494 $ &$ 0.486 $ &$ 10.27 $ &$  5.00 $ &$  1.65 $ &$ 11.5$ & $-0.44$ & $ 0.01$ & $ 0.00$ & $11.55$ & $   319$ \\
$   93 $ &$  1000. $ &$ 0.180 $ & $ 0.055 $ &$ 0.250 $ &$ 0.246 $ &$ 13.48 $ &$  5.40 $ &$  1.43 $ &$ 14.6$ & $ 0.43$ & $ 0.00$ & $ 0.05$ & $14.59$ & $   319$ \\
$   94 $ &$  1000. $ &$ 0.250 $ & $ 0.040 $ &$ 0.315 $ &$ 0.311 $ &$ 12.92 $ &$  8.20 $ &$  2.31 $ &$ 15.5$ & $ 0.64$ & $ 0.01$ & $ 0.94$ & $15.51$ & $   319$ \\
$   95 $ &$  1000. $ &$ 0.400 $ & $ 0.025 $ &$ 0.124 $ &$ 0.123 $ &$ 22.73 $ &$ 13.30 $ &$  2.13 $ &$ 26.4$ & $ 1.19$ & $-0.04$ & $ 0.29$ & $26.45$ & $   319$ \\
$   96 $ &$  1200. $ &$ 0.140 \times 10^{-1}$ & $ 0.847 $ &$ 0.902 $ & ---  & $  6.56 $ &$  1.95 $ &$  3.31 $ &$ 7.61$ & $ 0.03$ & $ 0.00$ & $-0.20$ & $ 7.61$ & $   319$ \\
$   97 $ &$  1200. $ &$ 0.200 \times 10^{-1}$ & $ 0.593 $ &$ 0.741 $ &$ 0.735 $ &$  5.63 $ &$  1.47 $ &$  1.36 $ &$ 5.98$ & $ 0.29$ & $-0.01$ & $-0.16$ & $ 5.98$ & $   319$ \\
$   98 $ &$  1200. $ &$ 0.320 \times 10^{-1}$ & $ 0.371 $ &$ 0.625 $ &$ 0.609 $ &$  5.60 $ &$  1.52 $ &$  1.38 $ &$ 5.97$ & $ 0.36$ & $ 0.00$ & $-0.16$ & $ 5.98$ & $   319$ \\
$   99 $ &$  1200. $ &$ 0.500 \times 10^{-1}$ & $ 0.237 $ &$ 0.552 $ &$ 0.536 $ &$  5.20 $ &$  1.34 $ &$  1.48 $ &$ 5.57$ & $ 0.23$ & $ 0.00$ & $-0.03$ & $ 5.57$ & $   319$ \\
$  100 $ &$  1200. $ &$ 0.800 \times 10^{-1}$ & $ 0.148 $ &$ 0.424 $ &$ 0.413 $ &$  5.54 $ &$  1.28 $ &$  1.40 $ &$ 5.86$ & $ 0.19$ & $ 0.00$ & $-0.03$ & $ 5.86$ & $   319$ \\
$  101 $ &$  1200. $ &$ 0.130 $ & $ 0.091 $ &$ 0.363 $ &$ 0.355 $ &$  6.00 $ &$  1.62 $ &$  1.42 $ &$ 6.38$ & $ 0.20$ & $ 0.00$ & $-0.02$ & $ 6.38$ & $   319$ \\
$  102 $ &$  1200. $ &$ 0.180 $ & $ 0.066 $ &$ 0.341 $ &$ 0.334 $ &$ 13.60 $ &$  4.70 $ &$  1.67 $ &$ 14.5$ & $ 0.41$ & $ 0.00$ & $ 0.40$ & $14.49$ & $   319$ \\
$  103 $ &$  1200. $ &$ 0.250 $ & $ 0.047 $ &$ 0.266 $ &$ 0.261 $ &$  6.82 $ &$  1.50 $ &$  1.41 $ &$ 7.13$ & $ 0.29$ & $ 0.00$ & $ 0.03$ & $ 7.13$ & $   319$ \\
$  104 $ &$  1200. $ &$ 0.400 $ & $ 0.030 $ &$ 0.104 $ &$ 0.103 $ &$ 12.36 $ &$  7.66 $ &$  1.89 $ &$ 14.7$ & $ 0.79$ & $-0.01$ & $ 0.32$ & $14.69$ & $   319$ \\
$  105 $ &$  1500. $ &$ 0.200 \times 10^{-1}$ & $ 0.741 $ &$ 0.838 $ & ---  & $  7.20 $ &$  2.18 $ &$  1.89 $ &$ 7.76$ & $ 0.00$ & $-0.01$ & $ 0.04$ & $ 7.76$ & $   319$ \\
$  106 $ &$  1500. $ &$ 0.320 \times 10^{-1}$ & $ 0.463 $ &$ 0.732 $ &$ 0.707 $ &$  6.45 $ &$  1.65 $ &$  1.53 $ &$ 6.83$ & $ 0.08$ & $ 0.00$ & $-0.20$ & $ 6.84$ & $   319$ \\
$  107 $ &$  1500. $ &$ 0.500 \times 10^{-1}$ & $ 0.296 $ &$ 0.578 $ &$ 0.555 $ &$  6.29 $ &$  1.26 $ &$  1.35 $ &$ 6.56$ & $ 0.27$ & $ 0.00$ & $-0.07$ & $ 6.57$ & $   319$ \\
$  108 $ &$  1500. $ &$ 0.800 \times 10^{-1}$ & $ 0.185 $ &$ 0.469 $ &$ 0.452 $ &$  6.53 $ &$  1.45 $ &$  1.43 $ &$ 6.84$ & $ 0.23$ & $ 0.00$ & $-0.09$ & $ 6.85$ & $   319$ \\
$  109 $ &$  1500. $ &$ 0.130 $ & $ 0.114 $ &$ 0.399 $ &$ 0.386 $ &$  7.83 $ &$  1.77 $ &$  1.47 $ &$ 8.16$ & $ 0.37$ & $ 0.00$ & $-0.06$ & $ 8.17$ & $   319$ \\
$  110 $ &$  1500. $ &$ 0.180 $ & $ 0.082 $ &$ 0.310 $ &$ 0.302 $ &$  8.69 $ &$  1.57 $ &$  1.39 $ &$ 8.94$ & $ 0.12$ & $ 0.00$ & $-0.04$ & $ 8.94$ & $   319$ \\
$  111 $ &$  1500. $ &$ 0.250 $ & $ 0.059 $ &$ 0.248 $ &$ 0.242 $ &$ 10.17 $ &$  2.27 $ &$  1.38 $ &$ 10.5$ & $ 0.66$ & $ 0.00$ & $ 0.17$ & $10.54$ & $   319$ \\
$  112 $ &$  1500. $ &$ 0.400 $ & $ 0.037 $ &$ 0.145 $ &$ 0.142 $ &$ 15.85 $ &$ 12.25 $ &$  2.05 $ &$ 20.1$ & $ 1.30$ & $-0.02$ & $ 0.48$ & $20.19$ & $   319$ \\
$  113 $ &$  1500. $ &$ 0.650 $ & $ 0.023 $ &$ 0.014 $ &$ 0.014 $ &$ 35.02 $ &$ 14.50 $ &$  2.93 $ &$ 38.0$ & $ 2.31$ & $-0.05$ & $ 0.46$ & $38.09$ & $   319$ \\
$  114 $ &$  2000. $ &$ 0.320 \times 10^{-1}$ & $ 0.618 $ &$ 0.751 $ & ---  & $  7.31 $ &$  1.99 $ &$  1.83 $ &$ 7.80$ & $ 0.26$ & $-0.01$ & $-0.21$ & $ 7.80$ & $   319$ \\
$  115 $ &$  2000. $ &$ 0.500 \times 10^{-1}$ & $ 0.395 $ &$ 0.558 $ &$ 0.524 $ &$  7.79 $ &$  1.65 $ &$  1.45 $ &$ 8.10$ & $ 0.46$ & $ 0.00$ & $-0.14$ & $ 8.11$ & $   319$ \\
$  116 $ &$  2000. $ &$ 0.800 \times 10^{-1}$ & $ 0.247 $ &$ 0.479 $ &$ 0.452 $ &$  7.63 $ &$  1.73 $ &$  1.67 $ &$ 8.00$ & $ 0.75$ & $ 0.00$ & $ 0.03$ & $ 8.04$ & $   319$ \\
$  117 $ &$  2000. $ &$ 0.130 $ & $ 0.152 $ &$ 0.326 $ &$ 0.310 $ &$  9.91 $ &$  2.49 $ &$  1.33 $ &$ 10.3$ & $ 0.86$ & $ 0.00$ & $ 0.11$ & $10.34$ & $   319$ \\
$  118 $ &$  2000. $ &$ 0.180 $ & $ 0.110 $ &$ 0.270 $ &$ 0.258 $ &$ 10.87 $ &$  1.98 $ &$  1.50 $ &$ 11.2$ & $ 0.33$ & $ 0.00$ & $-0.12$ & $11.15$ & $   319$ \\
$  119 $ &$  2000. $ &$ 0.250 $ & $ 0.079 $ &$ 0.258 $ &$ 0.248 $ &$ 11.64 $ &$  2.39 $ &$  1.51 $ &$ 12.0$ & $ 0.75$ & $ 0.00$ & $-0.02$ & $12.00$ & $   319$ \\
$  120 $ &$  2000. $ &$ 0.400 $ & $ 0.049 $ &$ 0.104 $ &$ 0.101 $ &$ 17.11 $ &$  5.25 $ &$  2.07 $ &$ 18.0$ & $ 0.81$ & $ 0.00$ & $ 0.18$ & $18.04$ & $   319$ \\
$  121 $ &$  3000. $ &$ 0.500 \times 10^{-1}$ & $ 0.593 $ &$ 0.626 $ &$ 0.558 $ &$  7.79 $ &$  2.90 $ &$  1.94 $ &$ 8.54$ & $ 0.62$ & $-0.03$ & $-0.44$ & $ 8.57$ & $   319$ \\
$  122 $ &$  3000. $ &$ 0.800 \times 10^{-1}$ & $ 0.371 $ &$ 0.498 $ &$ 0.446 $ &$  8.02 $ &$  2.06 $ &$  1.39 $ &$ 8.39$ & $ 0.31$ & $ 0.00$ & $-0.09$ & $ 8.40$ & $   319$ \\
$  123 $ &$  3000. $ &$ 0.130 $ & $ 0.228 $ &$ 0.425 $ &$ 0.387 $ &$  9.18 $ &$  3.08 $ &$  1.32 $ &$ 9.77$ & $-0.01$ & $ 0.00$ & $ 0.19$ & $ 9.77$ & $   319$ \\
$  124 $ &$  3000. $ &$ 0.180 $ & $ 0.165 $ &$ 0.320 $ &$ 0.294 $ &$ 10.64 $ &$  2.62 $ &$  1.36 $ &$ 11.1$ & $ 0.41$ & $ 0.00$ & $-0.09$ & $11.05$ & $   319$ \\
$  125 $ &$  3000. $ &$ 0.250 $ & $ 0.119 $ &$ 0.267 $ &$ 0.249 $ &$ 12.06 $ &$  3.13 $ &$  1.61 $ &$ 12.6$ & $ 0.19$ & $-0.01$ & $ 0.06$ & $12.56$ & $   319$ \\
$  126 $ &$  3000. $ &$ 0.400 $ & $ 0.074 $ &$ 0.145 $ &$ 0.137 $ &$ 16.10 $ &$  5.47 $ &$  1.82 $ &$ 17.1$ & $ 0.74$ & $-0.02$ & $ 0.39$ & $17.12$ & $   319$ \\
$  127 $ &$  3000. $ &$ 0.650 $ & $ 0.046 $ &$ 0.020 $ &$ 0.019 $ &$ 43.21 $ &$  5.95 $ &$  2.27 $ &$ 43.7$ & $ 0.17$ & $ 0.00$ & $ 0.06$ & $43.67$ & $   319$ \\
$  128 $ &$  5000. $ &$ 0.800 \times 10^{-1}$ & $ 0.618 $ &$ 0.637 $ & ---  & $  7.12 $ &$  1.92 $ &$  1.45 $ &$ 7.51$ & $ 0.39$ & $-0.01$ & $-0.28$ & $ 7.53$ & $   319$ \\
$  129 $ &$  5000. $ &$ 0.130 $ & $ 0.380 $ &$ 0.479 $ &$ 0.394 $ &$  9.55 $ &$  2.46 $ &$  1.34 $ &$ 9.95$ & $ 0.34$ & $ 0.00$ & $-0.20$ & $ 9.96$ & $   319$ \\
$  130 $ &$  5000. $ &$ 0.180 $ & $ 0.274 $ &$ 0.425 $ &$ 0.359 $ &$  9.68 $ &$  2.48 $ &$  1.38 $ &$ 10.1$ & $ 0.08$ & $ 0.00$ & $-0.01$ & $10.09$ & $   319$ \\
$  131 $ &$  5000. $ &$ 0.250 $ & $ 0.198 $ &$ 0.302 $ &$ 0.262 $ &$ 12.35 $ &$  3.53 $ &$  1.56 $ &$ 12.9$ & $ 1.16$ & $ 0.00$ & $ 0.20$ & $12.99$ & $   319$ \\
$  132 $ &$  5000. $ &$ 0.400 $ & $ 0.124 $ &$ 0.126 $ &$ 0.112 $ &$ 18.56 $ &$  5.69 $ &$  2.01 $ &$ 19.5$ & $ 1.58$ & $ 0.00$ & $-0.31$ & $19.58$ & $   319$ \\
$  133 $ &$  5000. $ &$ 0.650 $ & $ 0.076 $ &$ 0.010 $ &$ 0.009 $ &$ 40.99 $ &$ 17.70 $ &$  3.14 $ &$ 44.8$ & $ 1.73$ & $-0.03$ & $ 0.63$ & $44.79$ & $   319$ \\
$  134 $ &$  8000. $ &$ 0.130 $ & $ 0.608 $ &$ 0.586 $ & ---  & $ 10.57 $ &$  2.54 $ &$  2.58 $ &$ 11.2$ & $ 0.71$ & $ 0.00$ & $-0.25$ & $11.20$ & $   319$ \\
$  135 $ &$  8000. $ &$ 0.180 $ & $ 0.439 $ &$ 0.386 $ &$ 0.285 $ &$ 13.69 $ &$  2.83 $ &$  1.50 $ &$ 14.1$ & $ 0.04$ & $ 0.00$ & $-0.18$ & $14.06$ & $   319$ \\
$  136 $ &$  8000. $ &$ 0.250 $ & $ 0.316 $ &$ 0.301 $ &$ 0.233 $ &$ 15.36 $ &$  3.59 $ &$  1.55 $ &$ 15.8$ & $ 0.38$ & $-0.01$ & $ 0.05$ & $15.86$ & $   319$ \\
$  137 $ &$  8000. $ &$ 0.400 $ & $ 0.198 $ &$ 0.161 $ &$ 0.131 $ &$ 21.81 $ &$  7.34 $ &$  2.02 $ &$ 23.1$ & $ 0.99$ & $ 0.00$ & $-0.30$ & $23.13$ & $   319$ \\
$  138 $ &$  8000. $ &$ 0.650 $ & $ 0.122 $ &$ 0.020 $ &$ 0.017 $ &$ 64.03 $ &$  3.95 $ &$  1.56 $ &$ 64.2$ & $-0.06$ & $-0.01$ & $-0.02$ & $64.17$ & $   319$ \\

\hline
\hline
\end{tabular}
\end{center}
\end{tiny}
\tablecaption{\label{tab515a2}Continuation of \Tab~\ref{tab515a1}}
\end{table}

\begin{table}
\begin{tiny}
\begin{center}
\begin{tabular}{l|cccccccccccccc}
\hline
\hline
Bin &  $Q^2$  & $x$ & $y$ &  $\sigma^{-~\rm ave}_{r}$ & $F_2^{\rm ave}$ 
 & $\delta_{\rm ave,stat}$ & $\delta_{\rm ave,uncor}$ & $\delta_{\rm ave,cor}$ & $\delta_{\rm ave,exp~tot}$
 & $\delta_{\rm ave,rel}$ & $\delta_{\rm ave,gp}$ & $\delta_{\rm ave,had}$ & $\delta_{\rm ave,tot}$ 
 & $\sqrt{s}$\\
 $\#$    & GeV$^2$ &   &       &       &         &       \%           & \%    & \%  &    
\%              &   \% & \% & \% &\% & GeV \\
\hline

$  139 $ &$ 12000. $ &$ 0.180 $ & $ 0.659 $ &$ 0.453 $ & ---  & $ 16.43 $ &$  2.69 $ &$  3.02 $ &$ 16.9$ & $ 0.50$ & $ 0.00$ & $ 0.10$ & $16.92$ & $   319$ \\
$  140 $ &$ 12000. $ &$ 0.250 $ & $ 0.474 $ &$ 0.282 $ &$ 0.189 $ &$ 21.74 $ &$  4.83 $ &$  1.55 $ &$ 22.3$ & $ 0.05$ & $ 0.00$ & $-0.50$ & $22.33$ & $   319$ \\
$  141 $ &$ 12000. $ &$ 0.400 $ & $ 0.296 $ &$ 0.093 $ &$ 0.068 $ &$ 35.72 $ &$ 10.68 $ &$  2.55 $ &$ 37.4$ & $ 0.84$ & $-0.02$ & $-0.11$ & $37.38$ & $   319$ \\
$  142 $ &$ 20000. $ &$ 0.250 $ & $ 0.791 $ &$ 0.406 $ & ---  & $ 26.54 $ &$  3.29 $ &$  2.47 $ &$ 26.9$ & $ 0.38$ & $-0.01$ & $-0.07$ & $26.86$ & $   319$ \\
$  143 $ &$ 20000. $ &$ 0.400 $ & $ 0.494 $ &$ 0.194 $ &$ 0.118 $ &$ 33.37 $ &$  7.52 $ &$  1.55 $ &$ 34.2$ & $ 0.51$ & $ 0.00$ & $-0.25$ & $34.25$ & $   319$ \\
$  144 $ &$ 20000. $ &$ 0.650 $ & $ 0.304 $ &$ 0.013 $ &$ 0.009 $ &$ 73.47 $ &$ 34.51 $ &$  7.37 $ &$ 81.5$ & $ 3.18$ & $-0.04$ & $-0.40$ & $81.56$ & $   319$ \\
$  145 $ &$ 30000. $ &$ 0.400 $ & $ 0.741 $ &$ 0.247 $ & ---  & $ 47.25 $ &$  6.75 $ &$  5.97 $ &$ 48.1$ & $-0.63$ & $ 0.00$ & $ 0.52$ & $48.11$ & $   319$ \\

\hline
\hline
\end{tabular}
\end{center}
\end{tiny}
\tablecaption{\label{tab515a3}Continuation of \Tab~\ref{tab515a1}}
\end{table}

\begin{table}
\begin{tiny}
\begin{center}
\begin{tabular}{l|cccccccccccccc}
\hline
\hline
Bin &  $Q^2$  & $x$ & $y$ &  ${\rm d}^2 \sigma^{e^+p~{\rm ave}}_{CC}/{\rm d}x{\rm d}Q^2$ & $\sigma^{+~{\rm ave}}_{r,{\rm CC}} $
 & $\delta_{\rm ave,stat}$ & $\delta_{\rm ave,uncor}$ & $\delta_{\rm ave,cor}$ & $\delta_{\rm ave,exp~tot}$
 & $\delta_{\rm ave,rel}$ & $\delta_{\rm ave,gp}$ & $\delta_{\rm ave,had}$ & $\delta_{\rm ave,tot}$ 
 & $\sqrt{s}$\\
 $\#$    & GeV$^2$ &   &       &   pb/GeV$^2$            &       \%           & \%    & \%  &    
\%              &   \% & \% & \% &\% & GeV \\
\hline

$    1 $ &$     300. $ &$ 0.800 \times 10^{-2}$ & $ 0.371 $ &$ 0.136 $ & $ 1.417 $ &$ 19.12 $ &$  2.80 $ &$  3.97 $ &$ 19.7$ & $-1.25$ & $ 0.01$ & $ 0.09$ & $19.77$ & $   319$ \\
$    2 $ &$     300. $ &$ 0.130 \times 10^{-1}$ & $ 0.228 $ &$ 0.695 $ & $ 1.173 $ &$  8.83 $ &$  2.54 $ &$  3.59 $ &$ 9.86$ & $ 0.40$ & $-0.05$ & $-0.32$ & $ 9.87$ & $   319$ \\
$    3 $ &$     300. $ &$ 0.320 \times 10^{-1}$ & $ 0.093 $ &$ 0.210 $ & $ 0.875 $ &$  8.09 $ &$  1.86 $ &$  2.00 $ &$ 8.54$ & $ 1.20$ & $ 0.00$ & $ 0.15$ & $ 8.62$ & $   319$ \\
$    4 $ &$     300. $ &$ 0.800 \times 10^{-1}$ & $ 0.037 $ &$ 0.466 \times 10^{-1}$ & $ 0.484 $ &$ 10.74 $ &$  3.11 $ &$  1.68 $ &$ 11.3$ & $ 0.64$ & $-0.01$ & $ 0.36$ & $11.33$ & $   319$ \\
$    5 $ &$     500. $ &$ 0.130 \times 10^{-1}$ & $ 0.380 $ &$ 0.505 $ & $ 0.904 $ &$  8.09 $ &$  2.67 $ &$  2.65 $ &$ 8.92$ & $ 0.75$ & $-0.02$ & $-0.16$ & $ 8.95$ & $   319$ \\
$    6 $ &$     500. $ &$ 0.320 \times 10^{-1}$ & $ 0.154 $ &$ 0.168 $ & $ 0.740 $ &$  6.93 $ &$  1.84 $ &$  1.24 $ &$ 7.28$ & $ 0.59$ & $ 0.00$ & $ 0.14$ & $ 7.30$ & $   319$ \\
$    7 $ &$     500. $ &$ 0.800 \times 10^{-1}$ & $ 0.062 $ &$ 0.509 \times 10^{-1}$ & $ 0.561 $ &$  7.36 $ &$  1.74 $ &$  1.15 $ &$ 7.65$ & $-0.41$ & $ 0.03$ & $-0.04$ & $ 7.66$ & $   319$ \\
$    8 $ &$     500. $ &$ 0.130 $ & $ 0.038 $ &$ 0.242 \times 10^{-1}$ & $ 0.433 $ &$ 11.87 $ &$  2.63 $ &$  2.13 $ &$ 12.3$ & $ 0.16$ & $ 0.02$ & $ 0.22$ & $12.35$ & $   319$ \\
$    9 $ &$    1000. $ &$ 0.130 \times 10^{-1}$ & $ 0.760 $ &$ 0.323 $ & $ 0.664 $ &$ 10.67 $ &$  2.99 $ &$  2.93 $ &$ 11.5$ & $-0.80$ & $-0.03$ & $ 0.24$ & $11.49$ & $   319$ \\
$   10 $ &$    1000. $ &$ 0.320 \times 10^{-1}$ & $ 0.309 $ &$ 0.139 $ & $ 0.701 $ &$  5.11 $ &$  1.38 $ &$  1.16 $ &$ 5.42$ & $ 0.02$ & $-0.01$ & $ 0.04$ & $ 5.42$ & $   319$ \\
$   11 $ &$    1000. $ &$ 0.800 \times 10^{-1}$ & $ 0.124 $ &$ 0.434 \times 10^{-1}$ & $ 0.549 $ &$  5.73 $ &$  1.43 $ &$  0.77 $ &$ 5.95$ & $ 0.12$ & $ 0.02$ & $ 0.20$ & $ 5.96$ & $   319$ \\
$   12 $ &$    1000. $ &$ 0.130 $ & $ 0.076 $ &$ 0.213 \times 10^{-1}$ & $ 0.437 $ &$  8.25 $ &$  2.19 $ &$  0.96 $ &$ 8.59$ & $ 0.29$ & $ 0.01$ & $ 0.08$ & $ 8.59$ & $   319$ \\
$   13 $ &$    1000. $ &$ 0.250 $ & $ 0.040 $ &$ 0.702 \times 10^{-2}$ & $ 0.278 $ &$ 13.28 $ &$  4.01 $ &$  1.80 $ &$ 14.0$ & $ 1.01$ & $-0.01$ & $-0.38$ & $14.03$ & $   319$ \\
$   14 $ &$    1500. $ &$ 0.320 \times 10^{-1}$ & $ 0.463 $ &$ 0.961 \times 10^{-1}$ & $ 0.553 $ &$  7.91 $ &$  1.19 $ &$  1.27 $ &$ 8.10$ & $ 0.11$ & $ 0.02$ & $ 0.16$ & $ 8.10$ & $   319$ \\
$   15 $ &$    1500. $ &$ 0.800 \times 10^{-1}$ & $ 0.185 $ &$ 0.312 \times 10^{-1}$ & $ 0.449 $ &$  7.82 $ &$  0.57 $ &$  0.82 $ &$ 7.89$ & $-0.18$ & $ 0.00$ & $ 0.07$ & $ 7.89$ & $   319$ \\
$   16 $ &$    1500. $ &$ 0.130 $ & $ 0.114 $ &$ 0.156 \times 10^{-1}$ & $ 0.365 $ &$ 10.45 $ &$  0.74 $ &$  1.13 $ &$ 10.5$ & $-0.28$ & $ 0.02$ & $ 0.09$ & $10.54$ & $   319$ \\
$   17 $ &$    1500. $ &$ 0.250 $ & $ 0.059 $ &$ 0.555 \times 10^{-2}$ & $ 0.250 $ &$ 12.65 $ &$  1.40 $ &$  1.24 $ &$ 12.8$ & $-0.16$ & $-0.02$ & $ 0.57$ & $12.81$ & $   319$ \\
$   18 $ &$    2000. $ &$ 0.320 \times 10^{-1}$ & $ 0.618 $ &$ 0.806 \times 10^{-1}$ & $ 0.524 $ &$  7.67 $ &$  3.18 $ &$  1.84 $ &$ 8.50$ & $ 0.53$ & $ 0.03$ & $ 0.24$ & $ 8.52$ & $   319$ \\
$   19 $ &$    2000. $ &$ 0.800 \times 10^{-1}$ & $ 0.247 $ &$ 0.251 \times 10^{-1}$ & $ 0.408 $ &$  7.60 $ &$  2.88 $ &$  0.94 $ &$ 8.18$ & $ 0.28$ & $ 0.03$ & $ 0.30$ & $ 8.19$ & $   319$ \\
$   20 $ &$    2000. $ &$ 0.130 $ & $ 0.152 $ &$ 0.146 \times 10^{-1}$ & $ 0.386 $ &$  9.52 $ &$  3.79 $ &$  1.01 $ &$ 10.3$ & $ 1.11$ & $ 0.03$ & $ 0.61$ & $10.37$ & $   319$ \\
$   21 $ &$    2000. $ &$ 0.250 $ & $ 0.079 $ &$ 0.418 \times 10^{-2}$ & $ 0.212 $ &$ 13.93 $ &$  5.18 $ &$  1.13 $ &$ 14.9$ & $ 1.58$ & $ 0.01$ & $ 0.38$ & $14.99$ & $   319$ \\
$   22 $ &$    3000. $ &$ 0.800 \times 10^{-1}$ & $ 0.371 $ &$ 0.192 \times 10^{-1}$ & $ 0.390 $ &$  5.20 $ &$  1.61 $ &$  1.13 $ &$ 5.56$ & $ 0.29$ & $ 0.03$ & $-0.26$ & $ 5.58$ & $   319$ \\
$   23 $ &$    3000. $ &$ 0.130 $ & $ 0.228 $ &$ 0.939 \times 10^{-2}$ & $ 0.310 $ &$  6.65 $ &$  1.54 $ &$  1.08 $ &$ 6.91$ & $ 0.38$ & $ 0.01$ & $ 0.08$ & $ 6.92$ & $   319$ \\
$   24 $ &$    3000. $ &$ 0.250 $ & $ 0.119 $ &$ 0.284 \times 10^{-2}$ & $ 0.181 $ &$  8.98 $ &$  2.25 $ &$  1.18 $ &$ 9.33$ & $ 0.42$ & $ 0.00$ & $ 0.02$ & $ 9.34$ & $   319$ \\
$   25 $ &$    3000. $ &$ 0.400 $ & $ 0.074 $ &$ 0.731 \times 10^{-3}$ & $ 0.074 $ &$ 20.21 $ &$  5.06 $ &$  2.06 $ &$ 20.9$ & $ 1.08$ & $-0.06$ & $ 0.35$ & $20.96$ & $   319$ \\
$   26 $ &$    5000. $ &$ 0.800 \times 10^{-1}$ & $ 0.618 $ &$ 0.826 \times 10^{-2}$ & $ 0.246 $ &$  8.94 $ &$  2.30 $ &$  1.73 $ &$ 9.40$ & $ 0.66$ & $ 0.04$ & $-0.08$ & $ 9.42$ & $   319$ \\
$   27 $ &$    5000. $ &$ 0.130 $ & $ 0.380 $ &$ 0.482 \times 10^{-2}$ & $ 0.234 $ &$  7.32 $ &$  1.83 $ &$  1.18 $ &$ 7.64$ & $ 0.61$ & $ 0.02$ & $ 0.00$ & $ 7.66$ & $   319$ \\
$   28 $ &$    5000. $ &$ 0.250 $ & $ 0.198 $ &$ 0.175 \times 10^{-2}$ & $ 0.163 $ &$  8.63 $ &$  1.98 $ &$  1.48 $ &$ 8.98$ & $ 0.93$ & $ 0.00$ & $ 0.51$ & $ 9.04$ & $   319$ \\
$   29 $ &$    5000. $ &$ 0.400 $ & $ 0.124 $ &$ 0.462 \times 10^{-3}$ & $ 0.069 $ &$ 16.59 $ &$  3.98 $ &$  2.24 $ &$ 17.2$ & $ 2.48$ & $-0.03$ & $ 0.47$ & $17.40$ & $   319$ \\
$   30 $ &$    8000. $ &$ 0.130 $ & $ 0.608 $ &$ 0.175 \times 10^{-2}$ & $ 0.135 $ &$ 11.30 $ &$  3.60 $ &$  3.06 $ &$ 12.2$ & $ 0.90$ & $ 0.05$ & $-0.13$ & $12.29$ & $   319$ \\
$   31 $ &$    8000. $ &$ 0.250 $ & $ 0.316 $ &$ 0.808 \times 10^{-3}$ & $ 0.120 $ &$ 10.68 $ &$  3.25 $ &$  2.12 $ &$ 11.4$ & $ 1.51$ & $ 0.01$ & $-0.01$ & $11.46$ & $   319$ \\
$   32 $ &$    8000. $ &$ 0.400 $ & $ 0.198 $ &$ 0.222 \times 10^{-3}$ & $ 0.053 $ &$ 19.92 $ &$  6.77 $ &$  3.35 $ &$ 21.3$ & $ 1.86$ & $-0.05$ & $ 0.42$ & $21.39$ & $   319$ \\
$   33 $ &$   15000. $ &$ 0.250 $ & $ 0.593 $ &$ 0.120 \times 10^{-3}$ & $ 0.039 $ &$ 21.82 $ &$  5.09 $ &$  3.75 $ &$ 22.7$ & $ 1.75$ & $ 0.06$ & $-0.17$ & $22.79$ & $   319$ \\
$   34 $ &$   15000. $ &$ 0.400 $ & $ 0.371 $ &$ 0.113 \times 10^{-3}$ & $ 0.059 $ &$ 17.93 $ &$  7.02 $ &$  3.41 $ &$ 19.6$ & $ 2.16$ & $-0.03$ & $ 0.42$ & $19.68$ & $   319$ \\

\hline
\hline
\end{tabular}
\end{center}
\end{tiny}
\tablecaption{\label{tab3615a1}
HERA combined double differential  ${\rm d}^2 \sigma^{e^+p~{\rm ave}}_{CC}/{\rm d}x{\rm d}Q^2$ and reduced $\sigma^{+~{\rm ave}}_{r,{\rm CC}}$ cross section  for CC $e^+p$ scattering.
$\delta_{\rm ave,stat}$, $\delta_{\rm ave,uncor}$, $\delta_{\rm ave,cor}$ and
$\delta_{\rm ave,exp~tot}$ represent the statistical, uncorrelated
systematic, correlated systematic and total experimental uncertainty, respectively. 
$\delta_{\rm ave,rel}$, $\delta_{\rm ave,gp}$ and $\delta_{\rm ave,had}$
are the three correlated sources of uncertainties arising from the combination
procedure, see \Sec~\ref{subsubsec:proc_errors}. $\delta_{\rm ave,tot}$ is the total uncertainty
calculated by adding $\delta_{\rm ave,exp~tot}$ and the procedural errors in quadrature. 
The uncertainties are quoted in percent relative to  ${\rm d}^2 \sigma^{e^+p~{\rm ave}}_{CC}/{\rm d}x{\rm d}Q^2$.
The overall normalisation uncertainty of $0.5\%$ is not included.
}
\end{table}

\begin{table}
\begin{tiny}
\begin{center}
\begin{tabular}{l|cccccccccccccc}
\hline
\hline
Bin &  $Q^2$  & $x$ & $y$ &  ${\rm d}^2 \sigma^{e^-p~{\rm ave}}_{CC}/{\rm d}x{\rm d}Q^2$ & $\sigma^{-~{\rm ave}}_{r,{\rm CC}} $
 & $\delta_{\rm ave,stat}$ & $\delta_{\rm ave,uncor}$ & $\delta_{\rm ave,cor}$ & $\delta_{\rm ave,exp~tot}$
 & $\delta_{\rm ave,rel}$ & $\delta_{\rm ave,gp}$ & $\delta_{\rm ave,had}$ & $\delta_{\rm ave,tot}$ 
 & $\sqrt{s}$\\
 $\#$    & GeV$^2$ &   &       &   pb/GeV$^2$            &       \%           & \%    & \%  &    
\%              &   \% & \% & \% &\% & GeV \\
\hline

$    1 $ &$     300. $ &$ 0.130 \times 10^{-1}$ & $ 0.228 $ &$ 0.393 $ & $ 0.663 $ &$ 60.18 $ &$  7.49 $ &$ 11.91 $ &$ 61.8$ & $ 7.73$ & $-0.24$ & $ 0.43$ & $62.29$ & $   319$ \\
$    2 $ &$     300. $ &$ 0.320 \times 10^{-1}$ & $ 0.093 $ &$ 0.286 $ & $ 1.188 $ &$ 22.51 $ &$  6.15 $ &$  6.22 $ &$ 24.1$ & $ 6.87$ & $-0.06$ & $ 0.45$ & $25.11$ & $   319$ \\
$    3 $ &$     300. $ &$ 0.800 \times 10^{-1}$ & $ 0.037 $ &$ 0.632 \times 10^{-1}$ & $ 0.657 $ &$ 42.56 $ &$  8.72 $ &$  6.47 $ &$ 43.9$ & $ 3.32$ & $-0.20$ & $ 0.86$ & $44.05$ & $   319$ \\
$    4 $ &$     500. $ &$ 0.130 \times 10^{-1}$ & $ 0.380 $ &$ 0.477 $ & $ 0.854 $ &$ 27.75 $ &$  7.42 $ &$  6.96 $ &$ 29.6$ & $ 1.64$ & $-0.03$ & $ 0.18$ & $29.60$ & $   319$ \\
$    5 $ &$     500. $ &$ 0.320 \times 10^{-1}$ & $ 0.154 $ &$ 0.249 $ & $ 1.097 $ &$ 17.39 $ &$  4.45 $ &$  3.94 $ &$ 18.4$ & $ 2.27$ & $-0.12$ & $ 0.51$ & $18.52$ & $   319$ \\
$    6 $ &$     500. $ &$ 0.800 \times 10^{-1}$ & $ 0.062 $ &$ 0.764 \times 10^{-1}$ & $ 0.841 $ &$ 19.54 $ &$  4.50 $ &$  2.36 $ &$ 20.2$ & $-0.28$ & $ 0.00$ & $ 0.58$ & $20.20$ & $   319$ \\
$    7 $ &$     500. $ &$ 0.130 $ & $ 0.038 $ &$ 0.545 \times 10^{-1}$ & $ 0.976 $ &$ 29.14 $ &$  6.50 $ &$  1.99 $ &$ 29.9$ & $-0.40$ & $-0.01$ & $ 0.89$ & $29.94$ & $   319$ \\
$    8 $ &$    1000. $ &$ 0.130 \times 10^{-1}$ & $ 0.760 $ &$ 0.708 $ & $ 1.456 $ &$ 25.99 $ &$ 16.51 $ &$  5.59 $ &$ 31.3$ & $-1.81$ & $ 0.00$ & $ 0.07$ & $31.35$ & $   319$ \\
$    9 $ &$    1000. $ &$ 0.320 \times 10^{-1}$ & $ 0.309 $ &$ 0.169 $ & $ 0.855 $ &$ 14.43 $ &$  4.05 $ &$  2.32 $ &$ 15.2$ & $ 0.28$ & $-0.01$ & $ 0.35$ & $15.17$ & $   319$ \\
$   10 $ &$    1000. $ &$ 0.800 \times 10^{-1}$ & $ 0.124 $ &$ 0.628 \times 10^{-1}$ & $ 0.794 $ &$ 13.16 $ &$  4.18 $ &$  2.13 $ &$ 14.0$ & $-0.15$ & $ 0.00$ & $ 0.31$ & $13.97$ & $   319$ \\
$   11 $ &$    1000. $ &$ 0.130 $ & $ 0.076 $ &$ 0.344 \times 10^{-1}$ & $ 0.708 $ &$ 17.24 $ &$  3.71 $ &$  2.66 $ &$ 17.8$ & $-0.30$ & $ 0.02$ & $ 0.47$ & $17.84$ & $   319$ \\
$   12 $ &$    1000. $ &$ 0.250 $ & $ 0.040 $ &$ 0.140 \times 10^{-1}$ & $ 0.554 $ &$ 37.44 $ &$ 10.50 $ &$  1.47 $ &$ 38.9$ & $ 0.16$ & $-0.01$ & $-0.06$ & $38.91$ & $   319$ \\
$   13 $ &$    1500. $ &$ 0.320 \times 10^{-1}$ & $ 0.463 $ &$ 0.155 $ & $ 0.895 $ &$ 18.04 $ &$  8.80 $ &$  2.66 $ &$ 20.2$ & $ 0.21$ & $-0.01$ & $ 0.00$ & $20.24$ & $   319$ \\
$   14 $ &$    1500. $ &$ 0.800 \times 10^{-1}$ & $ 0.185 $ &$ 0.549 \times 10^{-1}$ & $ 0.791 $ &$ 14.02 $ &$  3.60 $ &$  1.86 $ &$ 14.6$ & $-0.01$ & $-0.01$ & $-0.03$ & $14.60$ & $   319$ \\
$   15 $ &$    1500. $ &$ 0.130 $ & $ 0.114 $ &$ 0.321 \times 10^{-1}$ & $ 0.752 $ &$ 18.04 $ &$  3.10 $ &$  2.24 $ &$ 18.4$ & $-0.15$ & $-0.01$ & $-0.13$ & $18.44$ & $   319$ \\
$   16 $ &$    1500. $ &$ 0.250 $ & $ 0.059 $ &$ 0.102 \times 10^{-1}$ & $ 0.458 $ &$ 24.02 $ &$  2.05 $ &$  2.56 $ &$ 24.2$ & $-0.37$ & $ 0.00$ & $-0.11$ & $24.24$ & $   319$ \\
$   17 $ &$    2000. $ &$ 0.320 \times 10^{-1}$ & $ 0.618 $ &$ 0.131 $ & $ 0.854 $ &$ 15.55 $ &$  4.30 $ &$  2.30 $ &$ 16.3$ & $ 0.35$ & $ 0.00$ & $ 0.47$ & $16.30$ & $   319$ \\
$   18 $ &$    2000. $ &$ 0.800 \times 10^{-1}$ & $ 0.247 $ &$ 0.567 \times 10^{-1}$ & $ 0.923 $ &$ 13.04 $ &$  3.80 $ &$  1.51 $ &$ 13.7$ & $-0.11$ & $ 0.00$ & $ 0.31$ & $13.67$ & $   319$ \\
$   19 $ &$    2000. $ &$ 0.130 $ & $ 0.152 $ &$ 0.197 \times 10^{-1}$ & $ 0.521 $ &$ 21.20 $ &$  4.40 $ &$  1.48 $ &$ 21.7$ & $-0.24$ & $ 0.00$ & $ 0.47$ & $21.71$ & $   319$ \\
$   20 $ &$    2000. $ &$ 0.250 $ & $ 0.079 $ &$ 0.861 \times 10^{-2}$ & $ 0.438 $ &$ 25.51 $ &$  6.50 $ &$  1.41 $ &$ 26.4$ & $-0.11$ & $ 0.00$ & $ 0.30$ & $26.36$ & $   319$ \\
$   21 $ &$    3000. $ &$ 0.800 \times 10^{-1}$ & $ 0.371 $ &$ 0.333 \times 10^{-1}$ & $ 0.677 $ &$  9.91 $ &$  2.60 $ &$  1.75 $ &$ 10.4$ & $-0.10$ & $ 0.00$ & $ 0.28$ & $10.40$ & $   319$ \\
$   22 $ &$    3000. $ &$ 0.130 $ & $ 0.228 $ &$ 0.211 \times 10^{-1}$ & $ 0.698 $ &$ 10.88 $ &$  3.49 $ &$  1.45 $ &$ 11.5$ & $ 0.84$ & $ 0.01$ & $-0.01$ & $11.55$ & $   319$ \\
$   23 $ &$    3000. $ &$ 0.250 $ & $ 0.119 $ &$ 0.746 \times 10^{-2}$ & $ 0.474 $ &$ 13.44 $ &$  3.33 $ &$  1.36 $ &$ 13.9$ & $-0.08$ & $ 0.00$ & $ 0.21$ & $13.91$ & $   319$ \\
$   24 $ &$    3000. $ &$ 0.400 $ & $ 0.074 $ &$ 0.246 \times 10^{-2}$ & $ 0.250 $ &$ 26.82 $ &$ 10.03 $ &$  2.99 $ &$ 28.8$ & $ 1.59$ & $-0.05$ & $-1.08$ & $28.86$ & $   319$ \\
$   25 $ &$    5000. $ &$ 0.800 \times 10^{-1}$ & $ 0.618 $ &$ 0.195 \times 10^{-1}$ & $ 0.581 $ &$ 12.66 $ &$  4.23 $ &$  3.05 $ &$ 13.7$ & $ 0.88$ & $ 0.02$ & $ 0.07$ & $13.72$ & $   319$ \\
$   26 $ &$    5000. $ &$ 0.130 $ & $ 0.380 $ &$ 0.129 \times 10^{-1}$ & $ 0.626 $ &$ 11.18 $ &$  3.42 $ &$  1.33 $ &$ 11.8$ & $ 0.37$ & $ 0.00$ & $-0.12$ & $11.77$ & $   319$ \\
$   27 $ &$    5000. $ &$ 0.250 $ & $ 0.198 $ &$ 0.429 \times 10^{-2}$ & $ 0.400 $ &$ 13.35 $ &$  2.74 $ &$  1.41 $ &$ 13.7$ & $ 0.59$ & $ 0.00$ & $ 0.10$ & $13.71$ & $   319$ \\
$   28 $ &$    5000. $ &$ 0.400 $ & $ 0.124 $ &$ 0.189 \times 10^{-2}$ & $ 0.283 $ &$ 18.80 $ &$  5.71 $ &$  2.53 $ &$ 19.8$ & $ 2.41$ & $-0.02$ & $-0.96$ & $19.97$ & $   319$ \\
$   29 $ &$    8000. $ &$ 0.130 $ & $ 0.608 $ &$ 0.802 \times 10^{-2}$ & $ 0.619 $ &$ 12.72 $ &$  4.37 $ &$  2.16 $ &$ 13.6$ & $ 0.44$ & $ 0.01$ & $-0.43$ & $13.64$ & $   319$ \\
$   30 $ &$    8000. $ &$ 0.250 $ & $ 0.316 $ &$ 0.380 \times 10^{-2}$ & $ 0.565 $ &$ 11.44 $ &$  3.23 $ &$  1.59 $ &$ 12.0$ & $ 0.54$ & $ 0.00$ & $-0.34$ & $12.01$ & $   319$ \\
$   31 $ &$    8000. $ &$ 0.400 $ & $ 0.198 $ &$ 0.105 \times 10^{-2}$ & $ 0.250 $ &$ 19.41 $ &$  4.61 $ &$  2.91 $ &$ 20.2$ & $ 0.99$ & $ 0.00$ & $-0.92$ & $20.20$ & $   319$ \\
$   32 $ &$   15000. $ &$ 0.250 $ & $ 0.593 $ &$ 0.133 \times 10^{-2}$ & $ 0.434 $ &$ 16.39 $ &$  6.52 $ &$  3.49 $ &$ 18.0$ & $ 0.27$ & $ 0.01$ & $-0.83$ & $18.00$ & $   319$ \\
$   33 $ &$   15000. $ &$ 0.400 $ & $ 0.371 $ &$ 0.372 \times 10^{-3}$ & $ 0.195 $ &$ 23.55 $ &$  5.91 $ &$  3.33 $ &$ 24.5$ & $ 1.54$ & $ 0.01$ & $-1.36$ & $24.59$ & $   319$ \\
$   34 $ &$   30000. $ &$ 0.400 $ & $ 0.741 $ &$ 0.106 \times 10^{-3}$ & $ 0.160 $ &$ 54.99 $ &$  3.60 $ &$  9.58 $ &$ 55.9$ & $ 1.38$ & $ 0.03$ & $-0.20$ & $55.95$ & $   319$ \\

\hline
\hline
\end{tabular}
\end{center}
\end{tiny}
\tablecaption{\label{tab3515a1}
HERA combined double differential  ${\rm d}^2 \sigma^{e^-p~{\rm ave}}_{CC}/{\rm d}x{\rm d}Q^2$ and reduced $\sigma^{-~{\rm ave}}_{r,{\rm CC}}$ cross section  for CC $e^-p$ scattering.
$\delta_{\rm ave,stat}$, $\delta_{\rm ave,uncor}$, $\delta_{\rm ave,cor}$ and
$\delta_{\rm ave,exp~tot}$ represent the statistical, uncorrelated
systematic, correlated systematic and total experimental uncertainty, respectively. 
$\delta_{\rm ave,rel}$, $\delta_{\rm ave,gp}$ and $\delta_{\rm ave,had}$
are the three correlated sources of uncertainties arising from the combination
procedure, see \Sec~\ref{subsubsec:proc_errors}. $\delta_{\rm ave,tot}$ is the total uncertainty
calculated by adding $\delta_{\rm ave,exp~tot}$ and the procedural errors in quadrature. 
The uncertainties are quoted in percent relative to  ${\rm d}^2 \sigma^{e^-p~{\rm ave}}_{CC}/{\rm d}x{\rm d}Q^2$.
The overall normalisation uncertainty of $0.5\%$ is not included.
}
\end{table}

\clearpage

\begin{figure}
\centerline{\epsfig{file=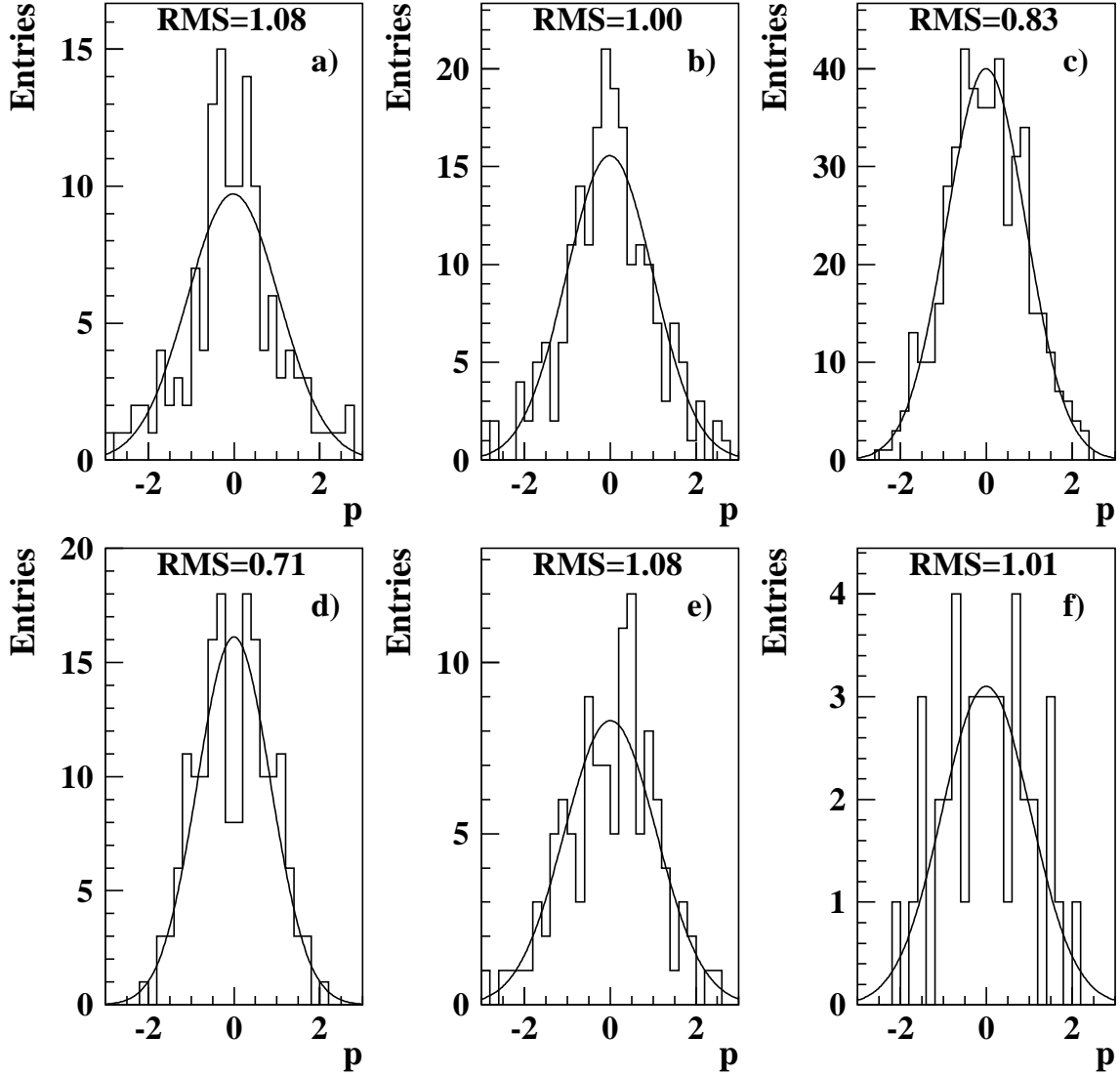 ,width=\linewidth}}
\caption{Distribution of pulls $p$ for the following samples:
a) NC $e^+p$ for $Q^2< 3.5$~GeV$^2$;
b) NC $e^+p$ for $3.5 \le Q^2 < 100$~GeV$^2$;
c) NC $e^+p$ for $Q^2\ge 100$~GeV$^2$;
d) NC $e^-p$;
e) CC $e^+p$  and 
f) CC $e^-p$.  There are no entries outside the histogram ranges. \label{fig:pulls}
RMS gives the root mean square of each distribution calculated as $\overline{p^2}$.
The curves show the results of binned log-likelihood Gaussian
fits to the distributions.
}
\end{figure}

\clearpage

\begin{figure}[tbp]
\vspace{-0.5cm} 
\centerline{
\epsfig{figure=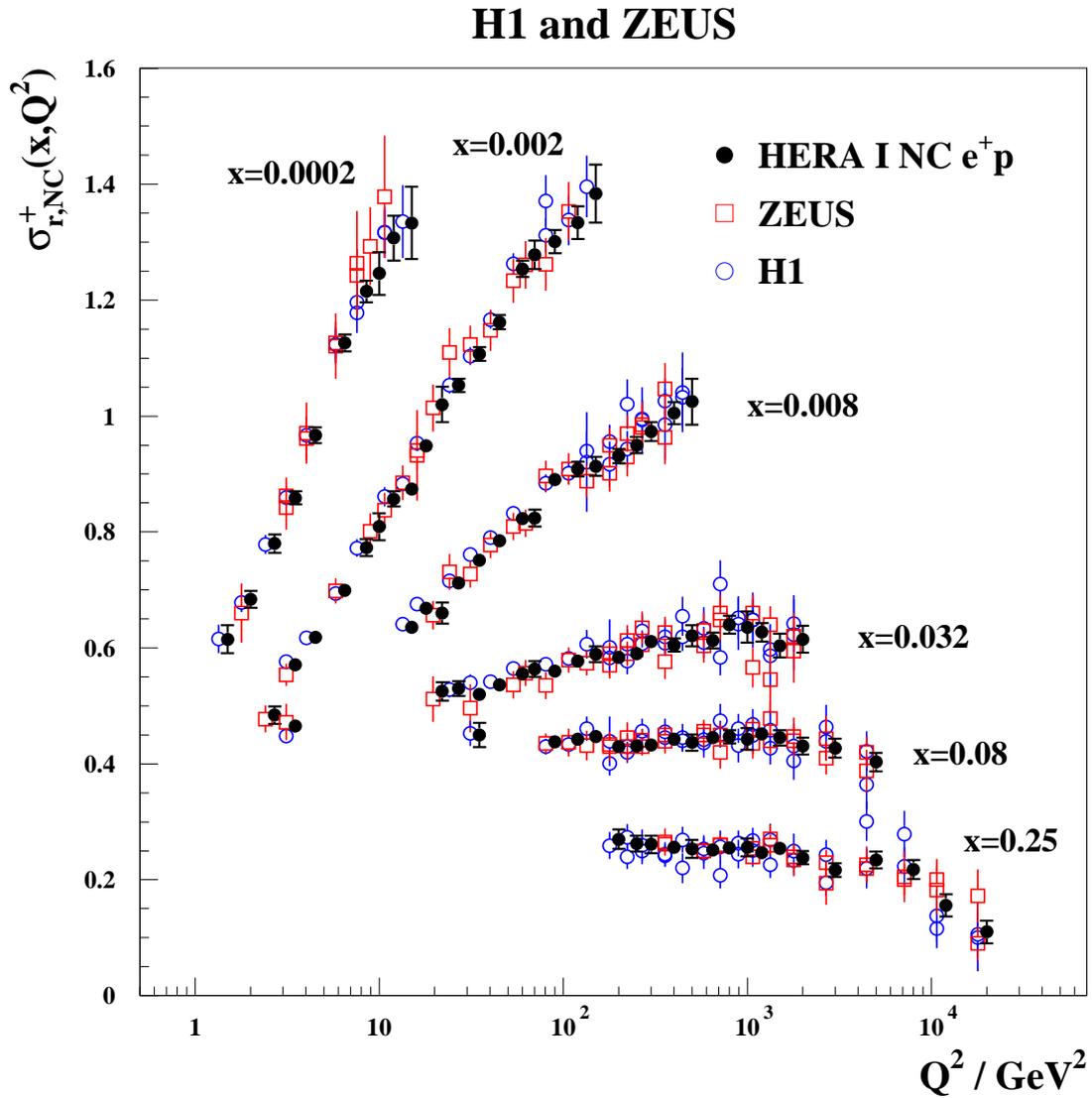 ,width=\linewidth}}
\caption {HERA combined NC $e^+p$ reduced 
cross section as a function of 
$Q^2$ for six $x$-bins compared to the separate 
H1 and ZEUS data input to the averaging procedure.
The error bars indicate the total experimental uncertainty.
The individual measurements are displaced horizontally for better visibility.
}
\label{fig:quality}
\end{figure}
\begin{figure}[tbp]
\centerline{
\epsfig{figure=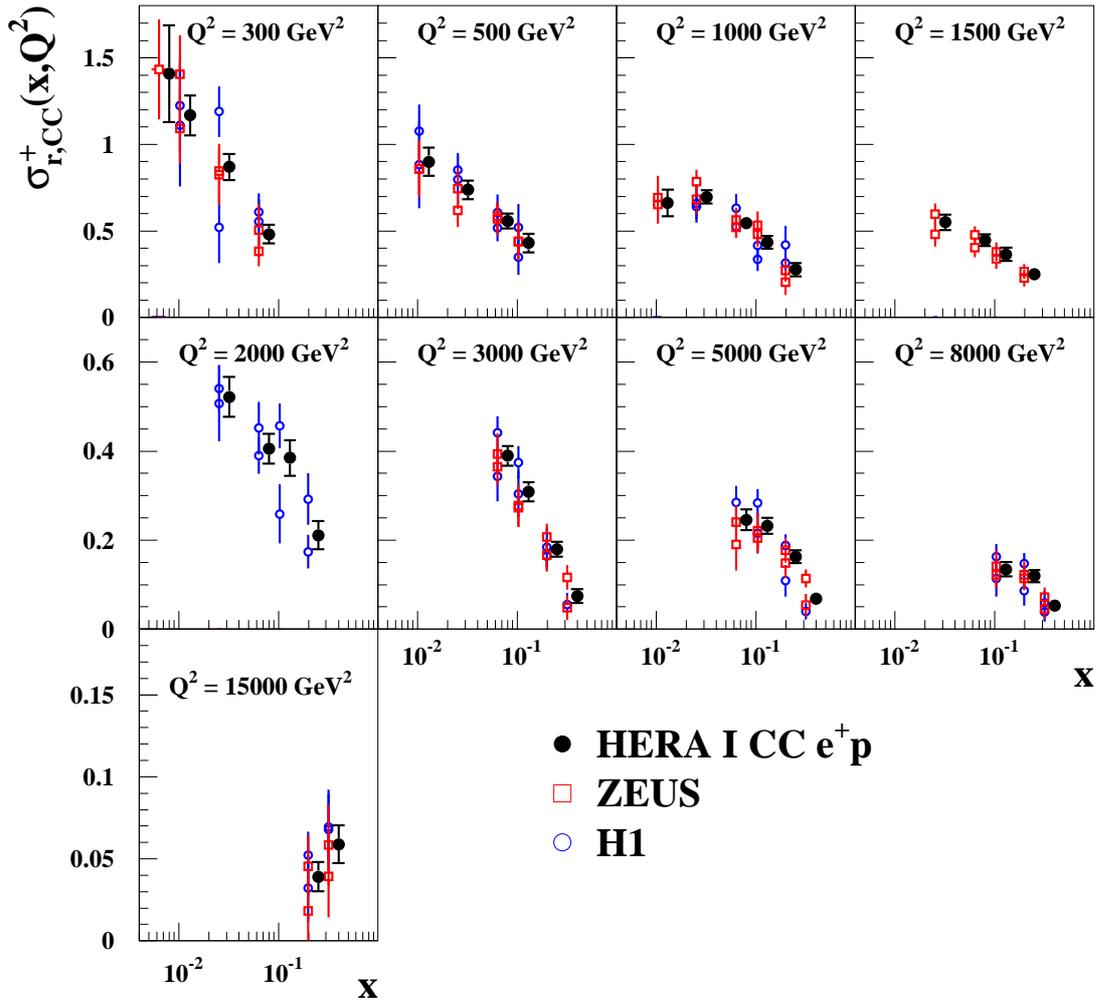 ,width=\linewidth}
}
\caption {HERA combined CC $e^+p$  reduced cross section
 compared to the separate 
H1 and ZEUS data input to the averaging procedure. 
The error bars indicate the total experimental uncertainty.
The individual measurements are displaced horizontally for better visibility.
 }
\label{fig:quality2}
\end{figure}


\begin{figure}[tbp]
\vspace{-0.5cm} 
\centerline{
\epsfig{figure=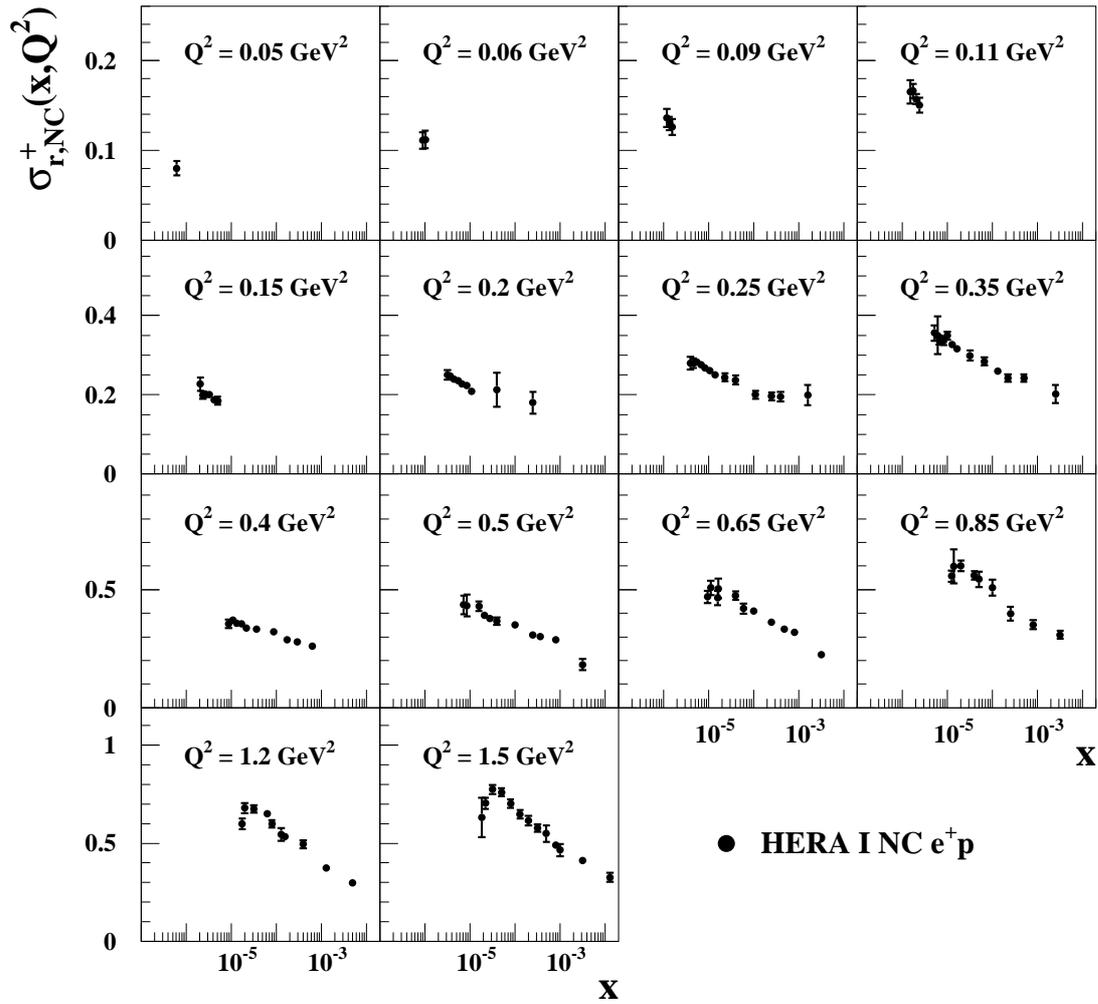 ,width=\linewidth}
}
\caption {HERA combined NC $e^+p$ reduced cross section at very low $Q^2$. 
The error bars indicate the total experimental uncertainty.
}
\label{fig:vsQ2l}
\end{figure}

\begin{figure}[tbp]
\vspace{-0.5cm} 
\centerline{
\epsfig{figure=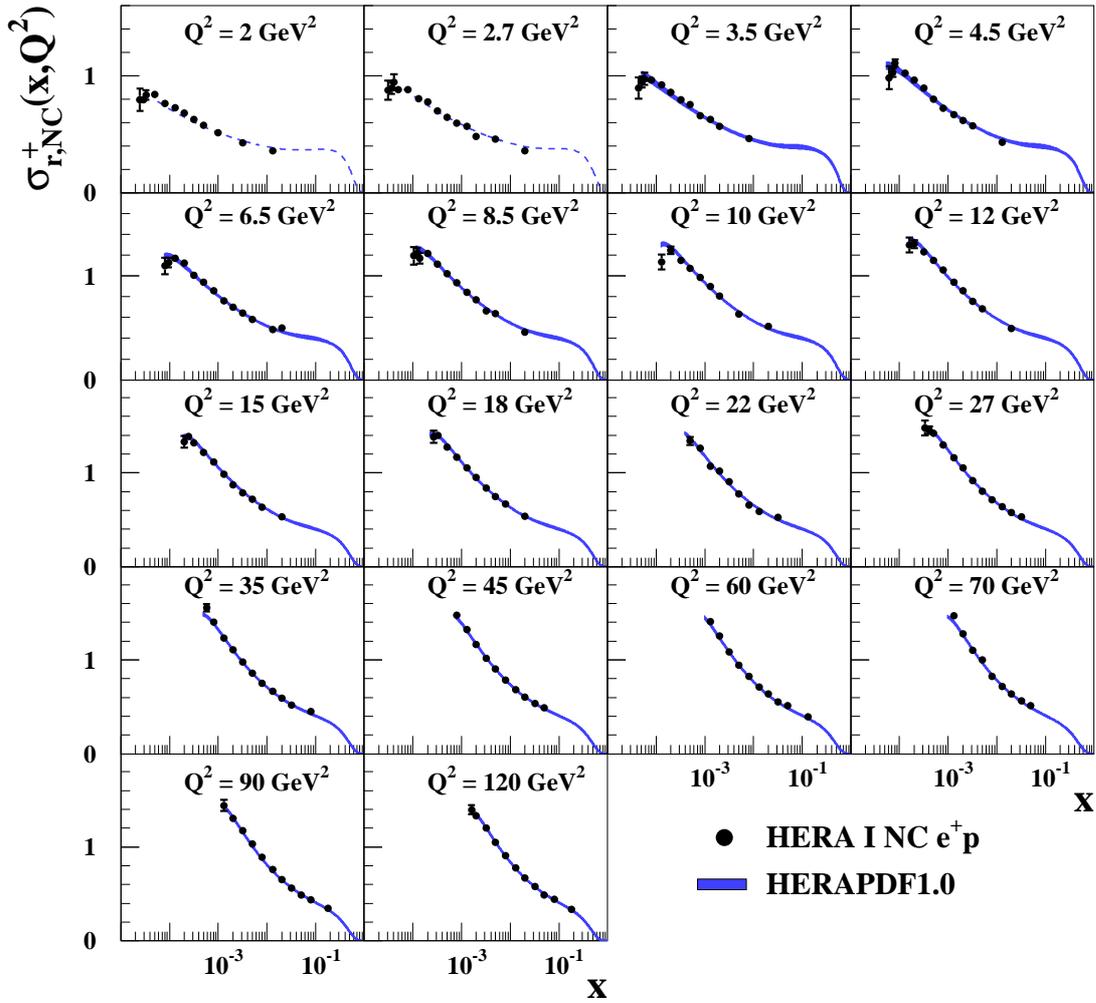 ,width=\linewidth}
}
\caption {HERA combined NC $e^+p$  reduced cross section at low $Q^2$.
The error bars indicate the total experimental uncertainty.
The  HERAPDF1.0 fit is superimposed.
 The  bands represent the total uncertainty of the fit.
 Dashed lines are shown for $Q^2$ bins not included in the QCD analysis. 
}
\label{fig:vsQ2m}
\end{figure}

\begin{figure}[tbp]
\centerline{
\epsfig{figure=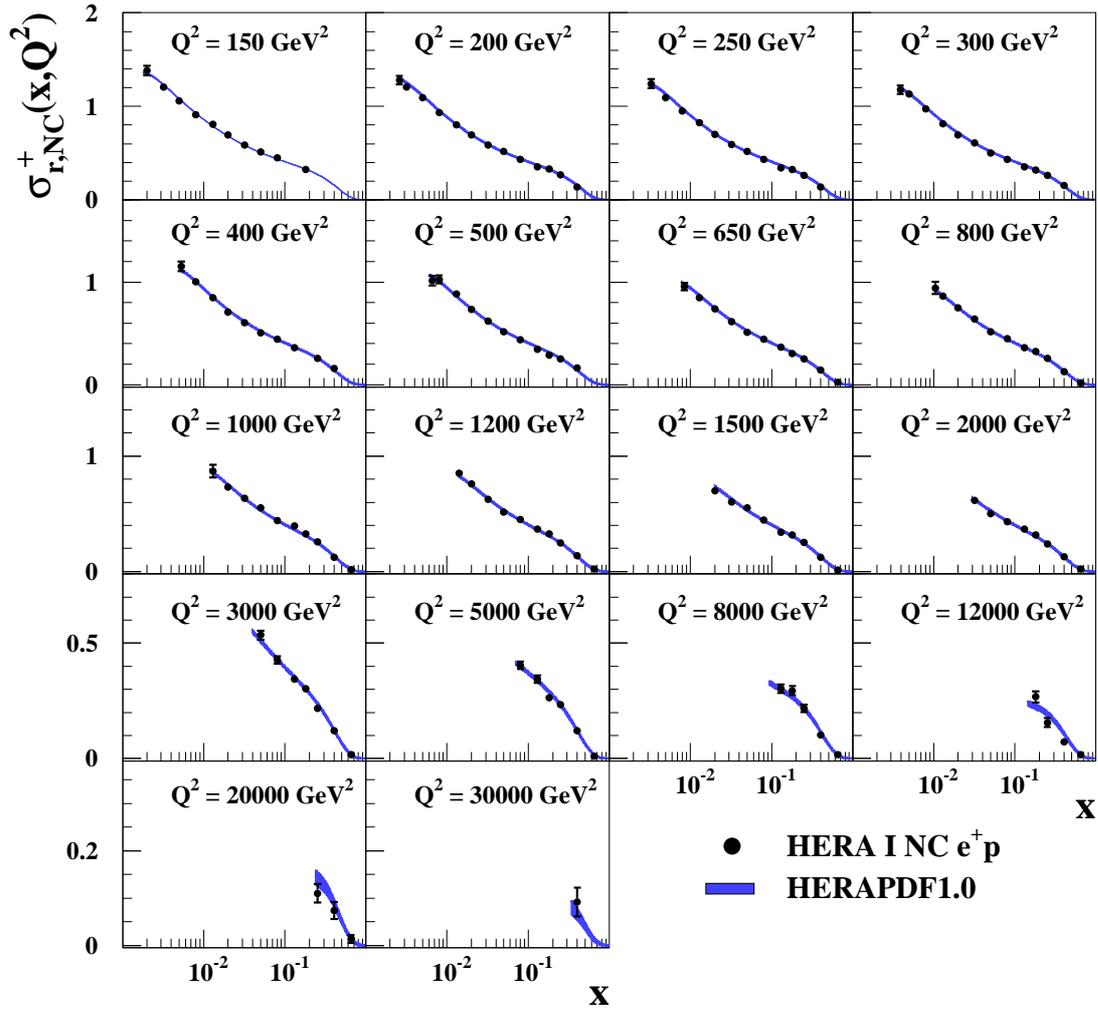 ,width=\linewidth}}
\caption {HERA combined NC $e^+p$ reduced cross section for high $Q^2$. 
The error bars indicate the total experimental uncertainty.
The predictions of the HERAPDF1.0 fit are superimposed.
 The  bands represent the total uncertainty of the fit.
}
\label{fig:dataNCp}
\end{figure}

\begin{figure}[tbp]
\centerline{
\epsfig{figure=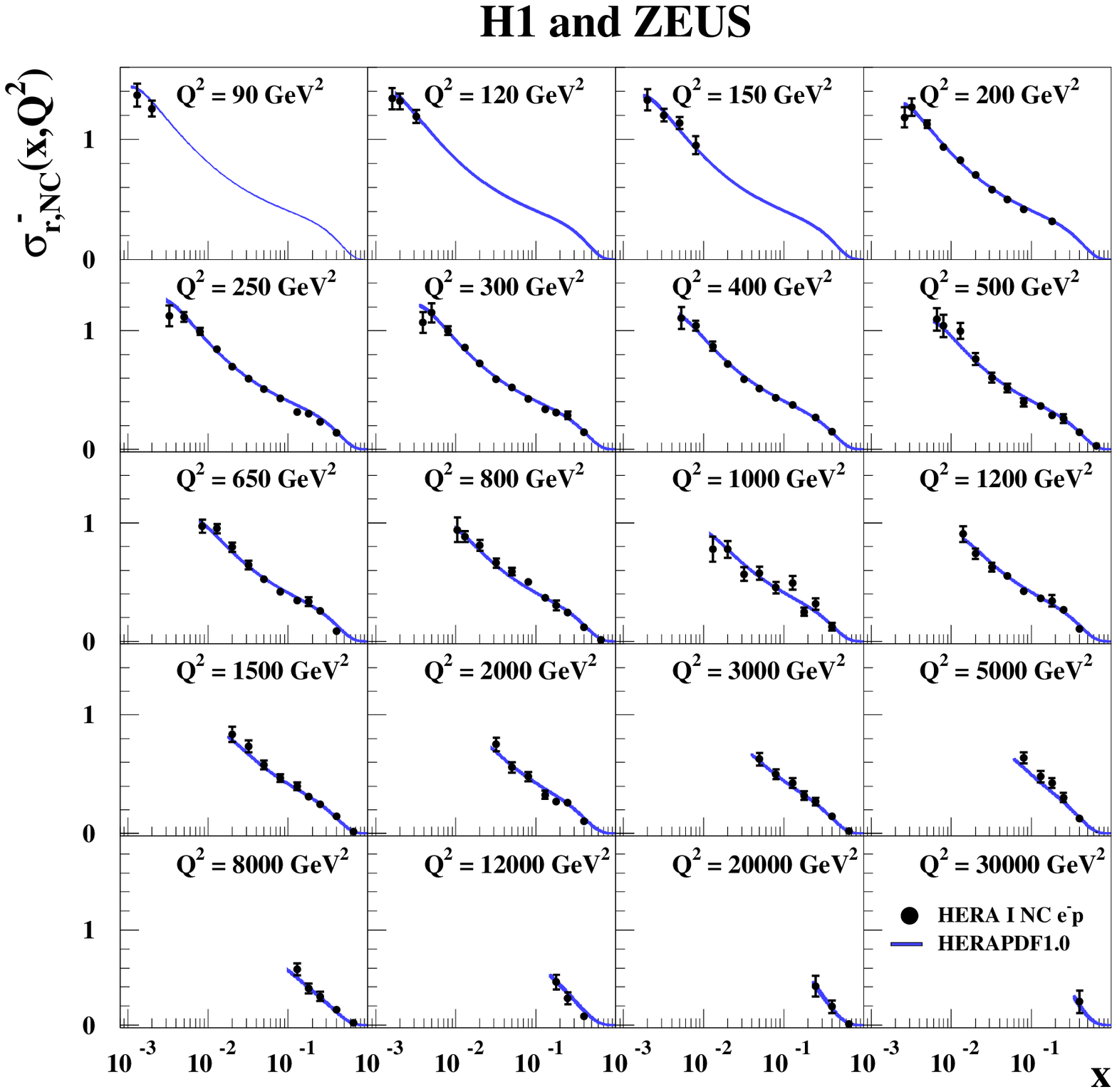 ,width=\linewidth}}
\caption {HERA combined  NC $e^-p$  reduced cross section.
The error bars indicate the total experimental uncertainty.
The HERAPDF1.0 fit is superimposed.
 The  bands represent the total uncertainty of the fit.
}
\label{fig:dataNCm}
\end{figure}

\begin{figure}[tbp]
\vspace{-0.5cm} 
\centerline{
\epsfig{figure=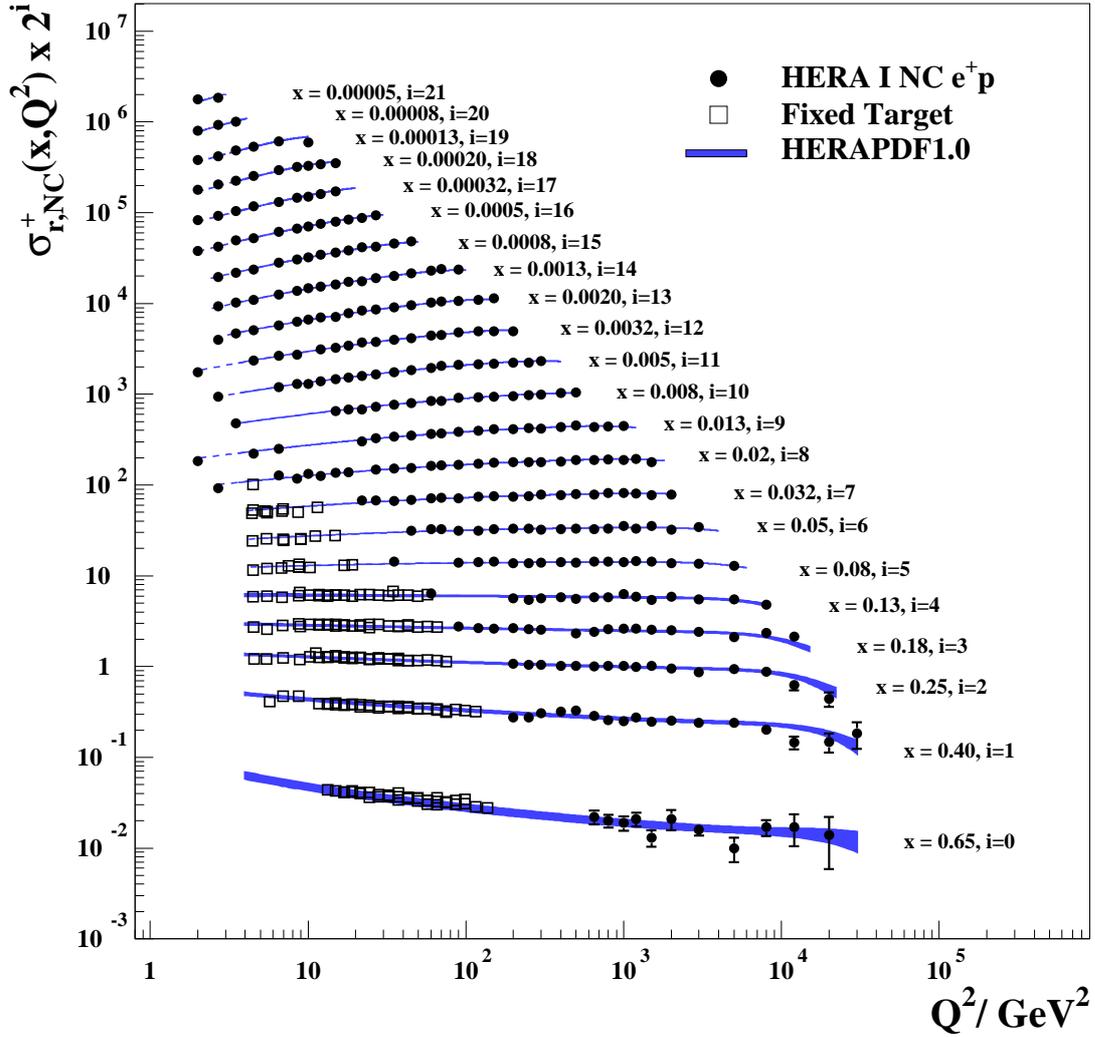 ,width=\linewidth}}
\caption {HERA combined NC $e^+p$ reduced cross section
 and fixed-target data as a function of $Q^2$.
The error bars indicate the total experimental uncertainty.
The  HERAPDF1.0 fit is superimposed.
 The  bands represent the total uncertainty of the fit.
 Dashed lines are shown for $Q^2$ values not included in the QCD analysis. 
}
\label{fig:scal}
\end{figure}

\begin{figure}[tbp]
\vspace{-0.5cm} 
\centerline{
\epsfig{figure=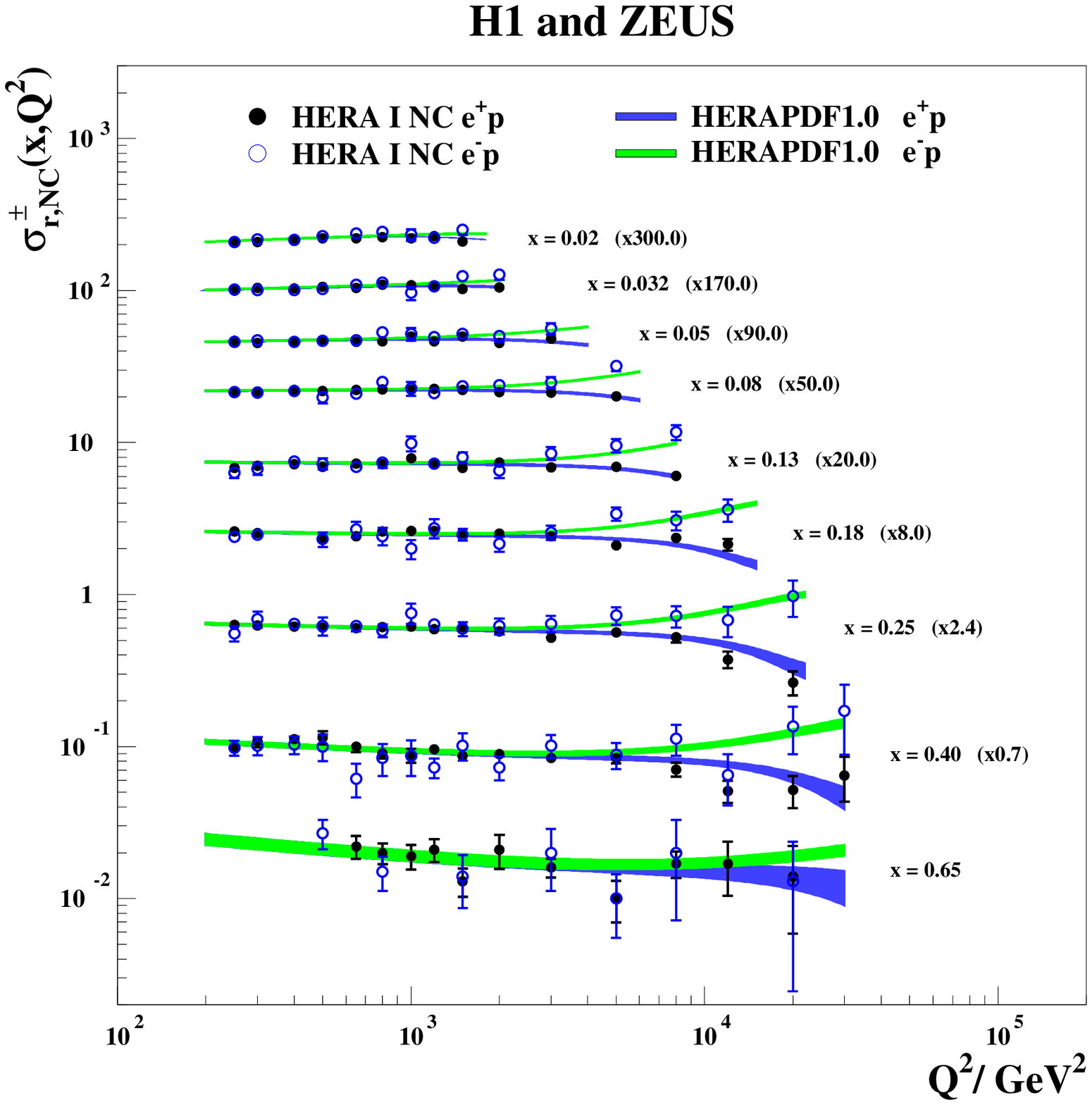 ,width=\linewidth}}
\caption {HERA combined NC $e^{\pm}p$ reduced cross sections
at high $Q^2$. 
The error bars indicate the total experimental uncertainty.
The  HERAPDF1.0 fit is superimposed.
 The  bands represent the total uncertainty of the fit.
}
\label{fig:ncepem}
\end{figure}

\begin{figure}[tbp]
\centerline{
\epsfig{figure=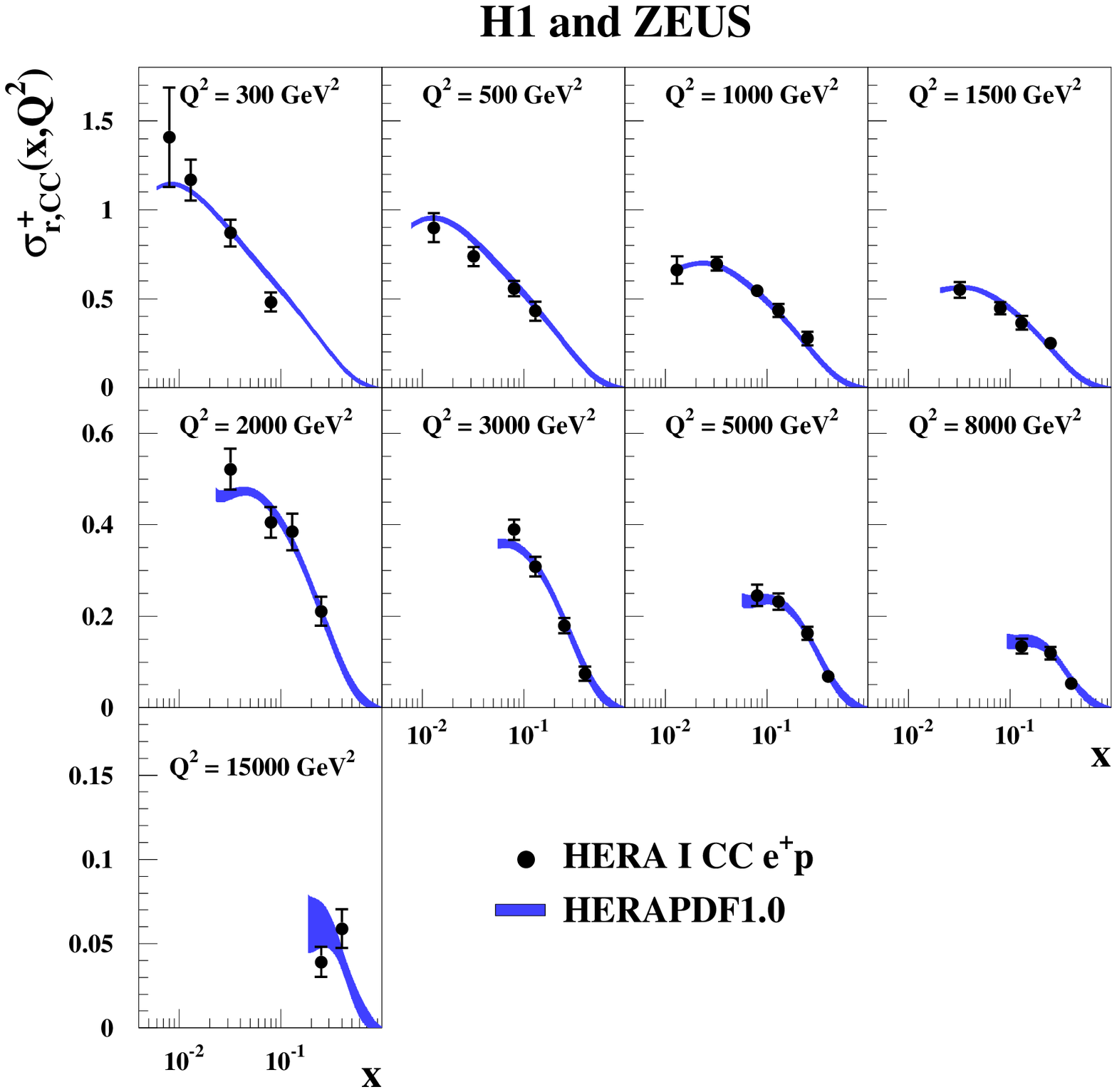 ,width=\linewidth}
}
\caption {HERA combined CC $e^+p$ reduced cross section. 
The error bars indicate the total experimental uncertainty.
The  HERAPDF1.0 fit is superimposed.
 The  bands represent the total uncertainty of the fit.
}
\label{fig:dataCCp}
\end{figure}

\begin{figure}[tbp]
\centerline{
\epsfig{figure=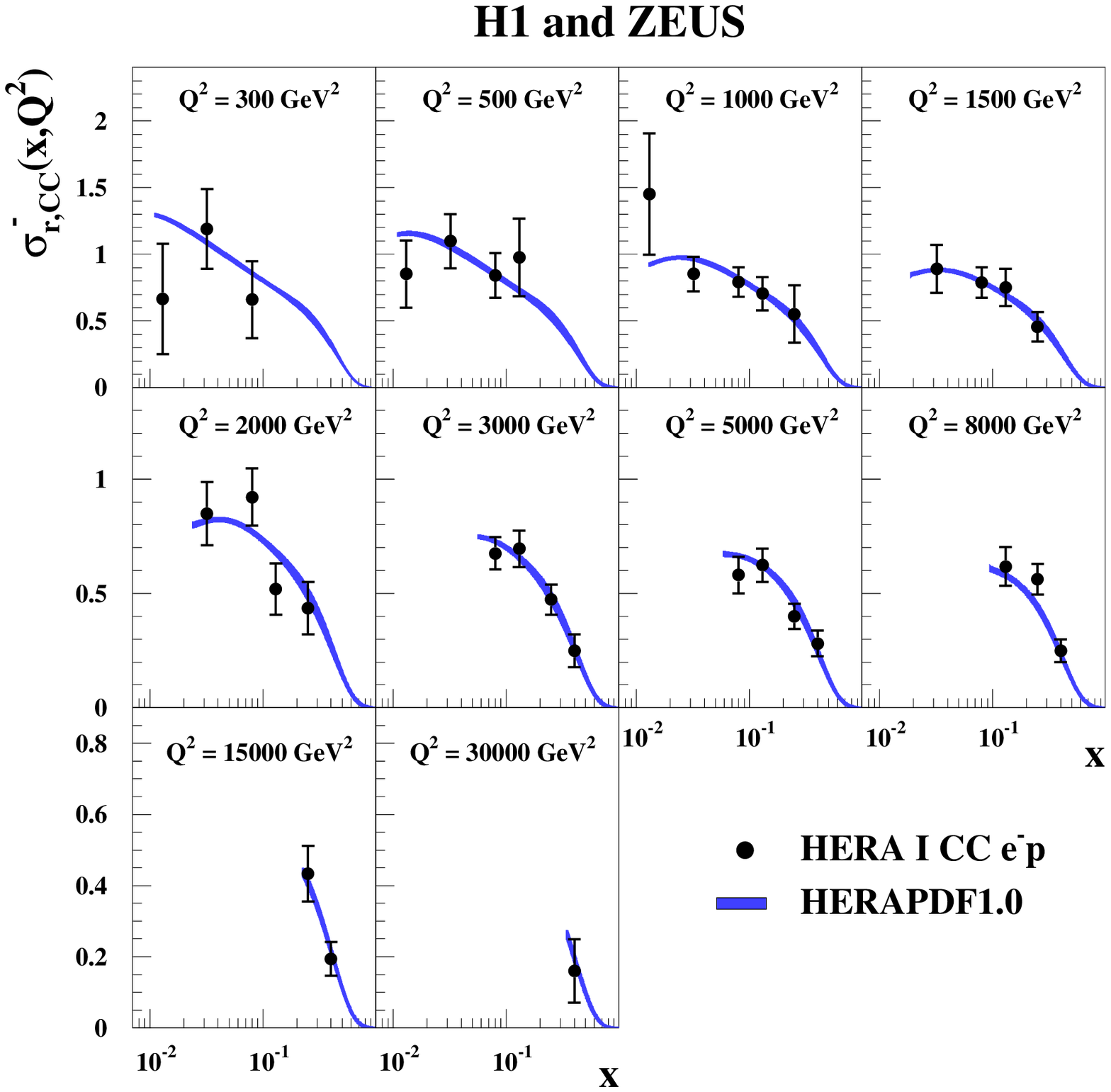 ,width=\linewidth}}
\caption {HERA combined  CC $e^-p$ reduced cross section. 
The error bars indicate the total experimental uncertainty.
The  HERAPDF1.0 fit is superimposed.
 The  bands represent the total uncertainty of the fit.
}
\label{fig:dataCCm}
\end{figure}

\clearpage

\end{document}